\documentclass[acmsmall,manuscript, research, nonacm]{acmart}

\AtBeginDocument{%
  \providecommand\BibTeX{{%
    \normalfont B\kern-0.5em{\scshape i\kern-0.25em b}\kern-0.8em\TeX}}}

\usepackage{multirow}
\usepackage{lipsum}
\usepackage{array}
\usepackage[normalem]{ulem} 
\usepackage{csquotes} 
    \DeclareQuoteStyle[american]{english}
        {\itshape\textquotedblleft} 
        [\textquotedblleft]
        {\textquotedblright}
        [0.05em]
        {\textquoteleft}
        {\textquoteright}

\usepackage{comment}

\usepackage{enumitem} 
\usepackage{paralist}
\newcommand{\lowHeading}[1]{\vspace{3pt}\textit{#1 --}}

\newcommand{\transitionGap}{\vspace{2pt}}  

\usepackage{graphicx}
\usepackage{color}
\usepackage{colortbl}
\usepackage{makecell}

\definecolor{greytext}{gray}{0.5}
\definecolor{DarkGreen}{rgb}{0.0, 0.5, 0.0}
\definecolor{KMGreen}{rgb}{0.4, 0.6, 0.6}
\definecolor{CadmiumGreen}{rgb}{0.0, 0.50, 0.24}
\definecolor{BrightGreen}{rgb}{0.0, 0.70, 0.24}
\definecolor{DarkKhaki}{rgb}{0.74, 0.72, 0.42}
\definecolor{DarkRed}{rgb}{0.7, 0.2, 0.2}
\definecolor{Purple}{rgb}{0.7,0.0,0.7}
\definecolor{DarkOrange}{rgb}{0.9,0.3,0.1}
\definecolor{Gray}{gray}{0.9}

\newcommand{\eg}{\textit{e.g.},\ }
\newcommand{\ie}{\textit{i.e.},\ }
\newcommand{\vs}{\textit{vs.}\ }
\newcommand{\etal}{\textit{et al}}

\graphicspath{
{figures/} 
}

\usepackage{ifthen}
\usepackage[normalem]{ulem}  

\usepackage{lineno}

\newboolean{showedits}
\setboolean{showedits}{false}  

\ifthenelse{\boolean{showedits}}
{

	\newcommand{\del}[1]{\textcolor{red}{\sout{#1}}}
	
}{

	\newcommand{\del}[1]{}
	
}

\newboolean{showcomments}
\setboolean{showcomments}{false}  

\ifthenelse{\boolean{showcomments}}
{
  \newcommand{\nbc}[3]{%
    {\colorbox{#3}{\bfseries\sffamily\scriptsize\textcolor{white}{#1}}}%
    {\textcolor{#3}{\sf\small$\blacktriangleright$\textit{#2}$\blacktriangleleft$}}}
}{
  \newcommand{\nbc}[3]{}%
}

\definecolor{kmcolor}{rgb}{0.0, 0.5, 0.0}    
\definecolor{btcolor}{rgb}{0.0, 0.5, 1.0}   
\definecolor{kicolor}{rgb}{0.98, 0.38, 0.5}   
\definecolor{ercolor}{rgb}{0.93, 0.53, 0.18}   
\definecolor{lkcolor}{rgb}{0.0, 0.55, 0.55}   
\definecolor{amvkcolor}{rgb}{0.7, 0.0, 0.7}    
\definecolor{rrcolor}{rgb}{0.28, 0.02, 0.03} 
\definecolor{rgcolor}{rgb}{0.0, 0.4, 0.8}    

\newcommand{\kmC}[1]{\nbc{KM}{#1}{kmcolor}}         
\ifthenelse{\boolean{showedits}}
  {\newcommand{\kmE}[1]{\textcolor{kmcolor}{#1}}}    
  {\newcommand{\kmE}[1]{#1}}                          

\ifthenelse{\boolean{showedits}}
  {}    
  {}                          

\ifthenelse{\boolean{showedits}}
  {}    
  {}                          

\ifthenelse{\boolean{showedits}}
  {}    
  {}                          

\ifthenelse{\boolean{showedits}}
  {}    
  {}                          
  
\ifthenelse{\boolean{showedits}}
  {}    
  {}                          
  
\ifthenelse{\boolean{showedits}}
  {}    
  {}   
  
\ifthenelse{\boolean{showedits}}
  {}    
  {}                          

\newcommand{\todo}[1]{\nbc{TODO}{#1}{DarkRed}}

\usepackage{subfig}

\newcommand{\citeESSWorkshops}{~\cite{essbotsWorkshops_anon}}

\newcommand{\ESSbots}{ESSbots}  
\newcommand{\ESSbot}{ESSbot} 
\newcommand{\ESS}{ESS} 

\newcommand{\FWname}{FoSTER}  

\newcommand{\theorSymb}{$_\tau$} 
\newcommand{\tST}{ST\theorSymb} 
\newcommand{\tGCP}{GCP\theorSymb} 
\newcommand{\tINT}{IN\theorSymb} 
\newcommand{\tAN}{AN\theorSymb} 
\newcommand{\tDC}{DC\theorSymb} 
\newcommand{\tMR}{MR\theorSymb} 

\newcommand{\affordSymb}{$_\alpha$}  
\newcommand{\aSTouch}{STouch\affordSymb}  
\newcommand{\aCoPres}{CoPres\affordSymb}  
\newcommand{\aCoordP}{CoordPart\affordSymb} 
\newcommand{\aEffortEng}{ElasticEffort\affordSymb} 

\newcommand{\hilite}[1]{\textbf{\textit{#1}}} 

\newcommand{\workshopform}{\textit{W0 \textsc{form}}} 
\newcommand{\workshopone}{\textit{W1 \textsc{concept}}} 
\newcommand{\workshopthree}{\textit{W3 \textsc{soc dyn}}} 
\newcommand{\workshopfour}{\textit{W4 \textsc{comm}}} 
\newcommand{\workshopfive}{\textit{W5 \textsc{generalize}}} 

\newcommand{\quotespc}{\vspace{3pt}} 

\newcommand{\tabASwid}{1.38in} 
\newcommand{\tabASmp}{1.53in} 
\newcommand{\tabASmpCor}{-0.30in} 
\newcommand{\icSp}{\kern-0.05in} 

\newcommand{\rSp}{-0.05in} 
\newcommand{\rSpRev}{0.05in}  
\newcommand{\lSp}{0.05in} 

\newcommand{\iconPlus}{\raisebox{\dimexpr-.60\height+.3\baselineskip}
   {\includegraphics[height=7pt]{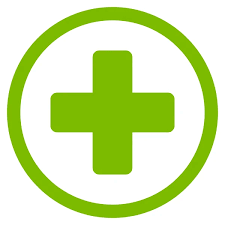}}}
\newcommand{\iconMinus}{\raisebox{\dimexpr-.60\height+.3\baselineskip}
    {\includegraphics[height=6pt]{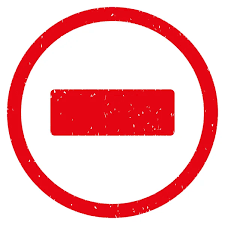}}}
\newcommand{\iconQuest}{\raisebox{\dimexpr-.60\height+.3\baselineskip}
    {\includegraphics[height=7pt]{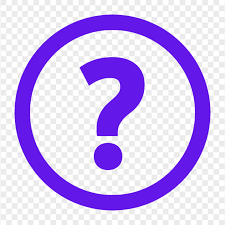}}}
    
\newcommand{\iconMeas}{\raisebox{\dimexpr-.60\height+.3\baselineskip}
    {\includegraphics[height=7pt]{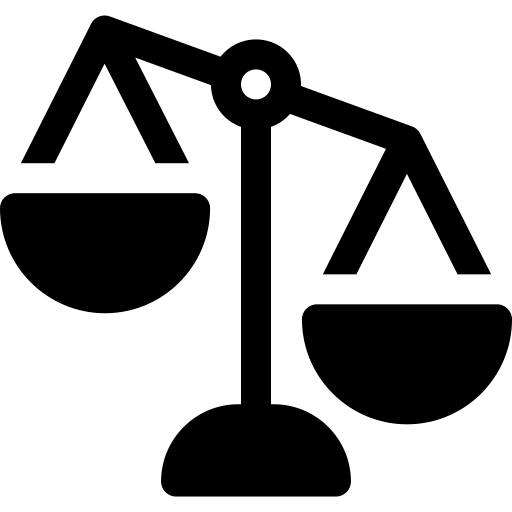}}}
    
\newcommand{\bulletPlus}{\par\noindent\makebox[3pt]{\iconPlus}}
\newcommand{\bulletMinus}{\par\noindent\makebox[3pt]{\iconMinus}}
\newcommand{\bulletQuest}{\par\noindent\makebox[3pt]{\iconQuest}}
\newcommand{\bulletMeas}{\par\noindent\makebox[3pt]{\iconMeas}}

\newcommand{\Erep}{\textit{SelfRep}$_e$}    
\newcommand{\Eobs}{\textit{Obs}$_e$}    

\newcommand{\DDdynsync}{\textit{DynSync}$_{dd}$}  
\newcommand{\DDcont}{\textit{Context}$_{dd}$}  
\newcommand{\DDint}{\textit{Interact}$_{dd}$}    
\newcommand{\DDscale}{\textit{Scale}$_{dd}$}    
\newcommand{\DDperf}{\textit{Perform}$_{dd}$}  

\begin{document}


\title[ESSbots, a Social Network that Jumps out of the Screen]
{ESSbots, a Social Network that Jumps out of the Screen: Could Embodied Social Swarms Foster Empathy and Connection in Remote Teen Groups?}


\author{Karon E. MacLean}
\email{maclean@cs.ubc.ca}

\author{Rebecca M. Todd}
\email{becket.todd@psych.ubc.ca}

\author{Elizabeth Reid}
\email{lizreid@uvic.ca}

\author{Rocco Tian Lang Ruan}
\email{ruan@cs.ubc.ca}

\author{Vasileia Karasavva}
\email{vasiakaras@gmail.com}

\author{Haley Foladare}
\email{haley.foladare@cw.bc.ca}

\author{Erin Chong}
\email{echong02@student.ubc.ca}

\author{Patrick Yung Kang Lee}
\email{patricklee@cs.toronto.edu}

\author{Fei Fan Yin}
\email{contact.feliciayin@gmail.com}

\author{Amori Yee Mikami}
\email{mikami@psych.ubc.ca}

\affiliation{
  \institution{University of British Columbia}
  \city{Vancouver}
  \state{BC}
  \country{Canada}
}

\author{Lawrence Kim}
\email{lawkim@sfu.ca}
\affiliation{
  \institution{Simon Fraser University}
  \city{Burnaby}
  \state{BC}
  \country{Canada}
}

\author{Katherine Isbister}
\email{katherine.isbister@gmail.com}
\affiliation{
  \institution{University of California, Santa Cruz}
  \city{Santa Cruz}
  \state{USA}
  \country{USA}
}

\renewcommand{\shortauthors}{MacLean, Todd, Reid, Ruan, Karasavva, Foladare, Chong, Lee, Yin, Mikami, Kim \& Isbister}

\begin{abstract}
Digital social media, alluring and consuming for today's teens, sidelines nonverbal communication and physical contact. Generally linked to well-being, these capacities are crucial as adolescents navigate new social landscapes, and their lack is often connected to the rise in youth loneliness. We envisioned \textit{Embodied Social Swarms} of small bots which embody a friend group in each member’s local space: through physical play and social gesture, ESSbots could be a conduit for feelings and support without performative pressure.  

This first investigation of the ESSbot conceptual and design space interweaves development of a theoretical underpinning with its examination via teen co-design of an early prototype. We contribute a crossdisciplinary review of foundational theory, the FoSTER (Forwarding Social Touch for Empathy with Remote Friend-groups) framework for studying and building ESSbots which connects theory, affordances and design requirements to social benefits, and an initial qualitative FoSTER assessment conducted within a longitudinal design workshop series with a Teen Advisory Board. We conclude with exciting use cases and a larger research plan.
\end{abstract}




     \begin{teaserfigure} 
      \centering
      \includegraphics[width=1.0\linewidth]{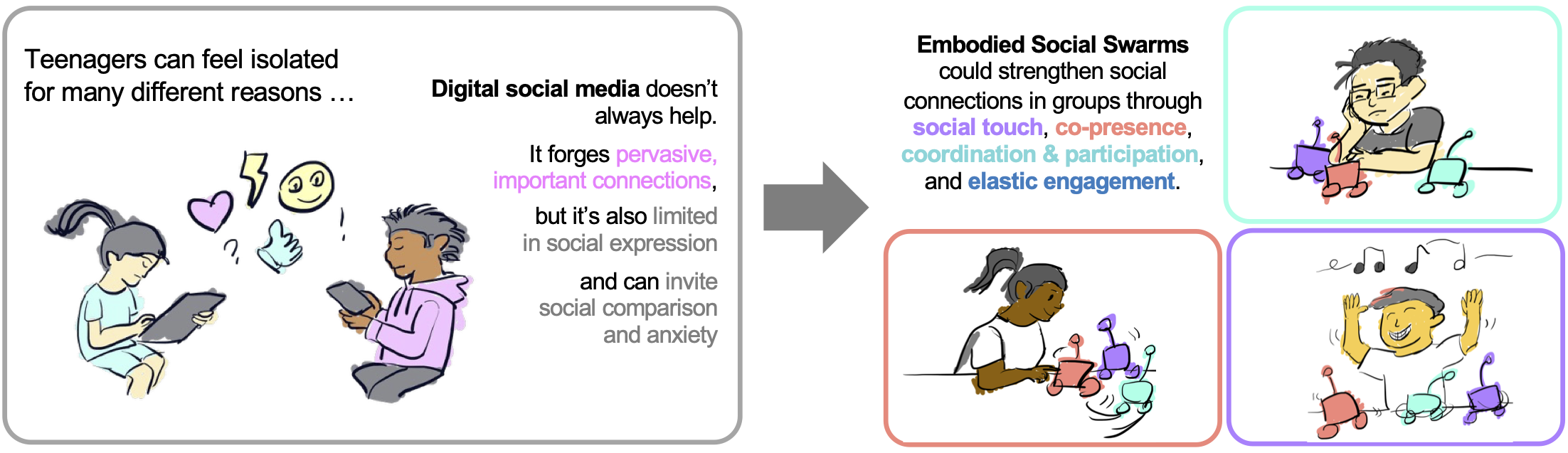}
      \caption{\ESSbots\ as a manifestation of \textbf{\FWname\ affordances}: teens who distributed across space and time \textbf{physically} interact with \textbf{multiple} friends through local, \textbf{socially interactive} robot swarms for \textbf{sustainable fun},  building empathy and connectedness by sharing emotional support and developing personal identity.
      }
      \label{fig:sec1:teaser}
    \end{teaserfigure}

\maketitle




\section{Introduction}
    \label{sec1:intro}

While  adolescents in every generation have felt alone, youth loneliness has reportedly doubled from 2012 to 2018 worldwide~\cite{twenge:2021:adolescentLoneliness}.
This unprecedented rise
paralleled growth in smartphone and digital social media (DSM) use
    ~\cite{twenge:2021:adolescentLoneliness,mcallister:2021:adolescentDepressionScreenMedia}, and suggests that
teens feel more isolated in the midst of DSM-lubricated social connections. 

DSM is purposefully designed to reformat social interaction through mechanisms like addictive incentive structures and feedback channel access
    ~\cite{thompson:2021:Atlantic:SocialMediaAlcohol,crouse:2021:NYT:instagramCesspool}.
This fosters visual performance: ``friends'' are audiences, and ``friendship'' is counted in likes and followers. 
While peer comparison is a normal human behaviour, DSM culture promotes immersing oneself in the highlights of others' lives. 
This displaces psychological development needed to navigate the social world as adults: 
    learning to listen, feeling heard, establishing and accepting intimacy, and taking emotional risks
        ~\cite{berndt:1982:featuresOfFriendshipAdolescence}. 
Girls may be most at risk for  negative emotional DSM impacts such as depression and self-harm --- sadly, in the direct context of close friendship~\cite{mcallister:2021:adolescentDepressionScreenMedia,twenge:2021:adolescentLoneliness,cohen:2020:socialMediaBodyImage}.

Today's DSM has weak to zero support for real-time nonverbal communication and physical contact, both key to  development~\cite{cascio:2019:socialTouchHumanDev} and mediating social perceptions~\cite{morrison:2010:skinSocialOrgan}. 
The COVID-19 pandemic of 2020-2023 and its proscription of physical contact highlighted our reliance on DSM's expansive, always-on allure
    ~\cite{hamilton:2020:teensSocialMediaCOVID}, 
its role in mitigating tension and loneliness amid stress and uncertainty
    ~\cite{vonMohr:2021:covidTouchDeprivationAdults,banerjee:2021:touchHungerBiopsychosocial}, 
and its potential emotional consequences
    ~\cite{hamilton:2020:teensSocialMediaCOVID,orben:2020:socialDeprivationAdolescent}. 
As enforced physical separation subsided, the new balance has become more digital than before.
    
Could \textit{physicalized} social media 
offer teens a new form of support for connection, communication, and co-presence? 

\subsection{Vision: \ESSbots\ as a Physical Social Medium} 
    \label{sec1:intro:vision}

We envisioned an \textbf{Embodied Social Swarm (ESS)} of tiny droids which together embody a friend group, transmitting feelings through gesture and physical (haptic) touch to each teen's local space (Figure~\ref{fig:sec1:teaser}). 
Small mobile bodies are linked to one other and to remote friends: each friend can author and jointly build behaviours that express their  feelings and respond to others, or simply play. 
The swarmbots we used are currently table-bound, but already delightful and evocative~\cite{kim:2019:swarmhaptics}. 
A future ESS could be an airborne cluster of bee-sized mites who travel with their friend to guide, play and comfort on the go,  then slip into a pocket when not needed.

This approach to social communication is hybridized over time and space.
It flips the dynamics of today's DSM from solo and performative to collaborative and empathic: instead of  visual depictions and comparisons, it privileges nonverbal, emotive, creative and playful screen-free expression. 
Its involvement of physicality is novel; haptic social mediation has generally focused on dyads, not groups
(\S\ref{sec2:foundationsRW:MST}, \ref{sec2:foundationsRW:CSCW}).

For this to succeed, teens must want to share their creations, feel free of oversight and  make their own rules. 
Using it has to be fun, rich, and support real  connection. 
A poorly designed platform could promote the same or new negative interactions as today's DSM; instead, we need a structure that values empathy, safety and collaboration, and does not easily lend itself to negative feedback loops and damaging comparisons.

There are open, interdisciplinary questions about using touch in a social medium for teens, which are beyond our present scope but which this novel social platform could assist in exploring~\cite{dahl:adolescent:2004}.
How does \textit{touch hunger}~\cite{field:2014:touchBook,banerjee:2021:touchHungerBiopsychosocial} apply to teens?
What is the larger importance of touch and embodiment to teen social development? 
These future questions are at the intersection of adolescent cognitive and affective development and their integration into society.

\transitionGap
\textbf{Our present objective} is a first investigation of this innovative conceptual and design space.
We have interwoven dual processes of 
(1) examining our ideas in a co-design process with teens, iteratively refining a first prototype; and 
(2) generating a robust theoretical underpinning, to assess and revise our initial ideas about the space's critical dimensions, identify exciting use cases, and establish a larger research plan. 


\subsection{Rationale}
    \label{sec1:intro:rationale}

We have targeted \ESSbots\ as a new digital social medium at the intersection of the under-explored {touch} modality,  an age group in need, and a specific {emerging technology}. 
We chose these for interlocking reasons.

     \begin{figure}
      \centering
      \subfloat[Zooid robots bumping a user's arm.  From~\cite{kim:2019:swarmhaptics} with permission.
          \label{figsub:sec1:zooids}] 
          {\includegraphics[height=0.28\textwidth]{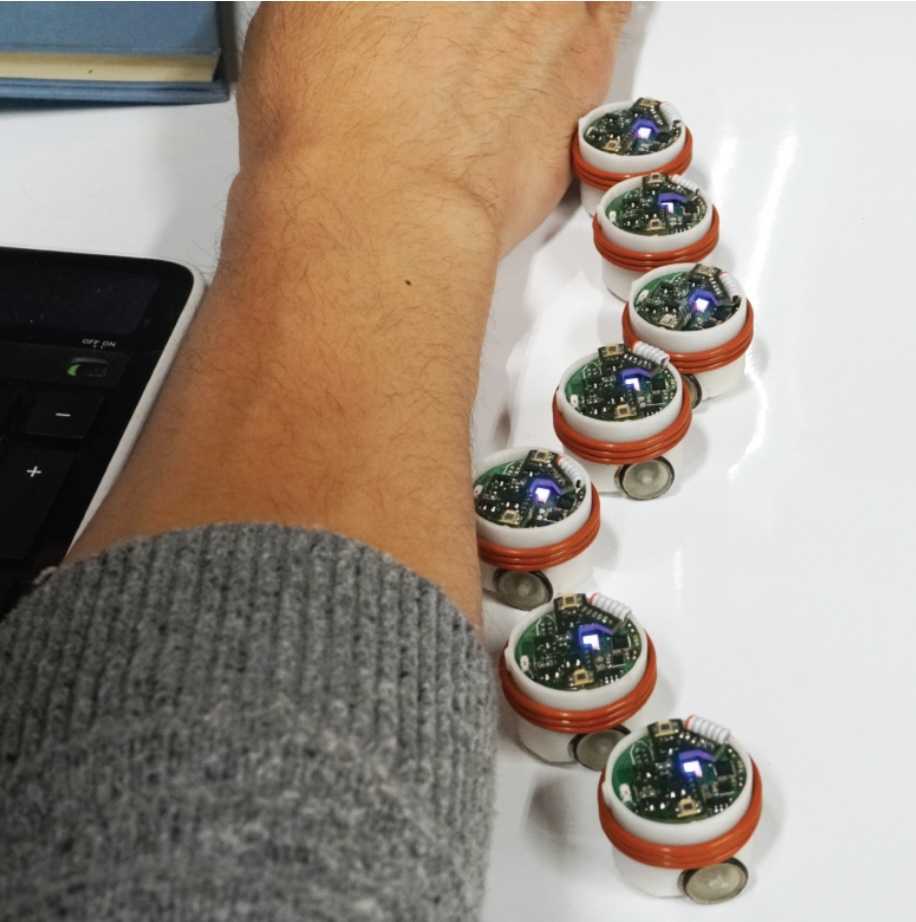}}\qquad
      \subfloat[Sony toio\textsuperscript{TM} robots on a position tracking mat~\cite{noauthor:toio:media:2018}. 
          \label{figsub:sec1:toios}]
          {\includegraphics[height=0.28\textwidth]{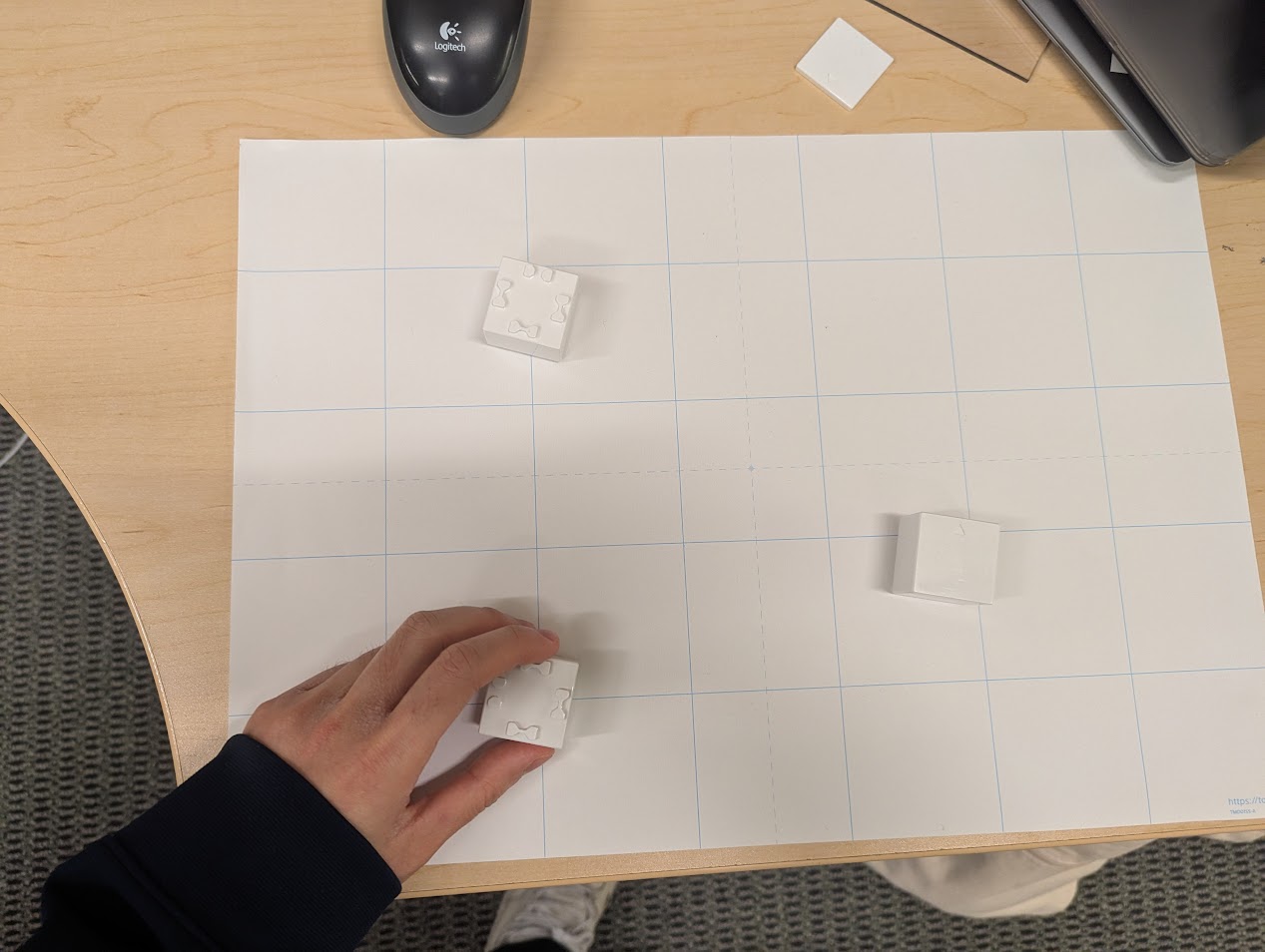}}
      \caption{Existing swarm  robot systems as tangible, haptic interfaces. We used toio\textsuperscript{TM} robots  in our workshop series.}
      \label{fig:sec1:existingSwarmBots}
    \end{figure}

\subsubsection{Why Touch and Embodied Movement?}

Fotopoulou \& Tsakiris (2017) argue the importance of embodied social interaction with other humans to development and the concept of self development~\cite{fotopoulou:2017:socHomeostasis}.
Reed \& McIntosh (2008) cite evidence that
    \textit{``\dots successful social interaction requires an enormous amount of information about the intentional relations of the self and other to be processed''  
    ~\cite{reed:2008:socialDance:chapter}}. 
To pull this off, human emotive modalities specialize by  social function. 
App \etal\ (2006) found that the body was preferred for both reading and conveying social-status emotions like embarrassment, guilt, pride, and shame; the face for anger, disgust, fear, happiness, and sadness; and touch for love and sympathy~\cite{app:2011:nonverbalEmotionChannel}.

Today's DSM  generally consists of text, still images and emoticons, and short audiovisual recordings. These can capture body language, but omit both  touch and embodiment (\eg proxemics).
Part of this gap is in realtime interactivity, a trait of broadcast-type social media: ``I do, you watch my recording'', in contrast to ``I do, you respond synchronously, and we move on together''.
For example, \href{https://www.tiktok.com/en/}{TikTok} and \href{https://www.instagram.com/}{Instagram} can be physically evocative, but in a performative rather than bidirectionally interactive way; as well as purely visual and without contact.
While ``broadly cast'', DSM (and their affordances) are also a core communicative medium for smaller, more personally-known social groups.

For children as well as adults, touch is well-documented as important to emotional health and belonging, relational reassurance and repair, and a way to involve  others in our  resilience    ~\cite{morrison:2010:skinSocialOrgan,fotopoulou:2017:socHomeostasis,app:2011:nonverbalEmotionChannel}. 
What if teens could  safely experiment with social touch even when connecting online?
 
\subsubsection{Why Teenage Groups?} 

One of the foundational developmental tasks of adolescence is the formation and maintenance of close peer relationships
    ~\cite{berndt:2002:friendQualSocDev}. 
Today's adolescents rely heavily on digital spaces to fulfill these social needs. 
A physical, embodied social medium would address a major gap in adolescents' choices for digital communications, allowing them to incorporate embodied proxemics and affective touch could help them create stronger and more solidly grounded social bonds with peers.  
As well, the current generation of adolescents (technology natives who came of age with vast Internet possibilities) may be open to adopting new forms of digital communication in their everyday life. 

\subsubsection{Relevant Properties of Social Swarmbots}

Embodiment and social touch suggest a robot form. 
However, today's social robots are often humanoid, and most research is about one-to-one interactions based on dialogue and facial expression, with minimal bodily, relational movement or physical contact in either direction.
In contrast, a swarm of little pucks rolling around on a table (Figure~\ref{fig:sec1:existingSwarmBots}) has unusual interactional and experiential qualities:


\lowHeading{Embodiment and tangibility} 
    Their small size, mobility (alone and relative to one another), and tangibility afford richly embodied haptic- and manipulation-based interaction with complex proxemics, and the relational and wellness benefits of touch which becomes social when a bot is perceived as invoking a remote friend. 

\lowHeading{Co-Presence, Coordination and Participation arising from multiplicity}
    A \textit{swarm} of robots supports natural mappings to members of a distributed friend \textit{group} (>1 remote interaction partner).
    With a guiding algorithm and narrative, users should be able to perceive their friends as embodied either in individual bots (physical avatars), or by the swarm as a whole.  
    When a member feels co-present in their local ``node'' with  remote friends,
    the swarm can facilitate new forms of communication as well as a safe place to try out social skills which  feel intimidating in the real world.
    
    Each bot in a node can interact  physically with the local member and  other bots, independently or swarm-coordinated, with fluid mappings and roles.  
    This embodied multiplicity invites active participation and social coordination.

\transitionGap 
Used together, these affordances could facilitate key needs that are missing or substandard in current DSM: a sense of 
    group connection, communication and mutual support when  members are physically distant.
Their scope is relevant across many dimensions of collaborative action;  time and space (synchronicity and physical distribution)~\cite{johansen:1988:CSCWmatrix}, 
and Lee \& Paine's proposal for scale, nascence and turnover~\cite{lee:2015:MoCA}.
While other suitable formats may exist, swarmbots' present availability and past examples of use as a haptic interface make them a practical exploratory platform.

\subsection{Approach and Contributions}    
    \label{sec1:intro:approach_contrib}

\begin{figure}[bt]
    \centering
    \includegraphics[width=1.0\textwidth]{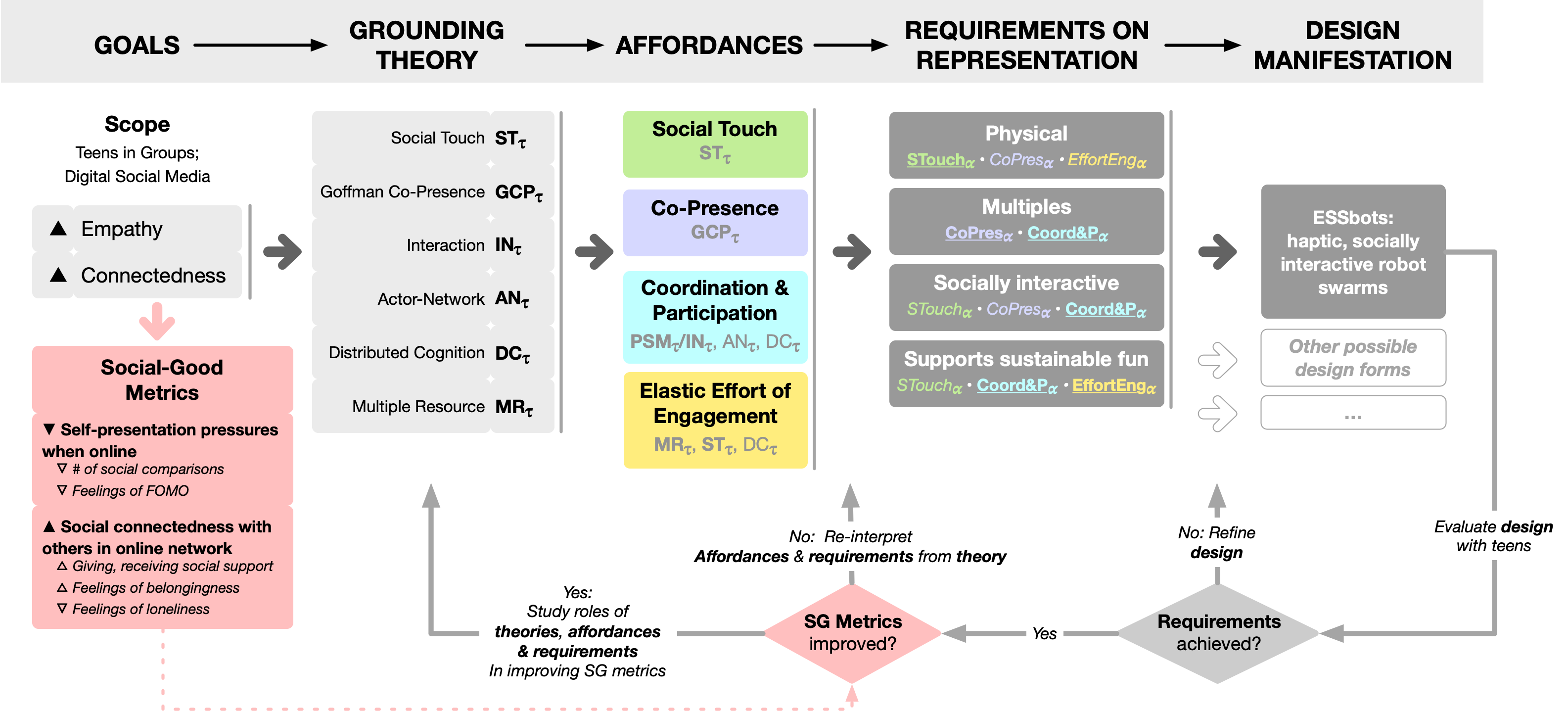}
    \caption{The \FWname\ Framework (\textbf{Fo}rwarding \textbf{S}ocial \textbf{T}ouch for \textbf{E}mpathy with \textbf{R}emote Friend-groups) 
    proposes theory-grounded affordances and design requirements for a physical teen social medium which could increase empathy and connectedness. 
    }
      \label{fig:sec1:Framework}
\end{figure}

Our entry point in this research was a notion --- informed by the team's own experience with pandemic-trapped teens --- that a robot system with not only \textit{physical}, haptic expressivity (which is also possible with a single robot) but also \textit{multiplicity}  (\eg linked haptic robot swarms) could make the jump from 1:1 interaction to engaging a  group.
However, we needed to ground this hunch in relevant cognitive, social and sensory theory, and explore it with teens.

Figure~\ref{fig:sec1:Framework} (the \FWname\ framework: \textbf{Fo}rwarding \textbf{S}ocial \textbf{T}ouch for \textbf{E}mpathy with \textbf{R}emote Friend-groups)  lays out our path, which begins with  
    scoped \textbf{goals}: to increase opportunities for empathy and connectedness for teen friend groups which are not well supported in the digital communication mediums which these groups predominantly rely on today, 
    and a set of \textbf{``social good'' metrics} by which we can test progress.
We first reviewed
    existing \textbf{theories} relating to social touch, co-presence, sense-making, attention management and others which might guide and ground a design in our goals.
As we gained theoretical confidence, we directly consulted teens who were co-designing a first implementation with us.
We iterated on early ideas to eventually propose a set of 
    \textbf{affordances} that could activate relevant theories, and then \textbf{requirements} for a representation that could capture those affordances.

Haptic swarmbots meet these requirements (not necessarily uniquely or optimally), so we used them to evaluate our idea: that these \textit{theoretically-based affordances} could have leverage on these \textit{social good metrics}. 
The \FWname\ framework provides us a pathway to examine this proposition.
%
Our primary contributions are: 

\begin{compactenum}
    \item \textit{An integrated, cross-disciplinary literature review and theoretical foundation} at the intersection of affective social touch, social swarmbots, and teenage development and social communication.
    
    \item \textit{The \FWname\ framework and design space} for studying and building tools for \textbf{\ESSbots}, a novel physical social medium \kmE{and platform} intended to support embodied, empathic group interaction.              
    
    \item \textit{An initial validation of \FWname} through qualitative analysis drawn from two  workshop series with teens, where we shared core \FWname\ and \ESSbot\ ideas to assess interest, practicality and  design directions.
\end{compactenum}

\section{Related Work}
    \label{sec2:relatedWork}

We review relevant research about 
    teens (their remote socialization, and what we know of social touch in their lives), and
    technological mediation of remote social touch in various forms and human-robot interaction specifically; 
    and situate this work within computer-supported cooperative work (CSCW) theory.

\subsection{Teens and Remote Socialization}
    \label{sec2:foundationsRW:teenSocial}

Adolescence is a developmental period where youth are strengthening social bonds with their peers and learning key communication skills. Having positive social connections with peers is crucial for enhancing well-being and social-emotional adjustment, and for preparing adolescents to successfully navigate the transition to adulthood~\cite{neel:longitudinal:2013}.
The brain undergoes extensive plasticity in early adolescence~\cite{dahl:adolescent:2004}, facilitated in part by pubertal hormones~\cite{laube:2020:relationship}, meaning that adolescence is a crucial time for the development of brain systems supporting social cognition~\cite{fuhrmann:2019:social}. 
In short, adolescence is a sensitive period for social development that can affect quality of life for many years to come. 

Over the past two decades, digital social media has transformed how people communicate and form social connections. Adolescents are among the most enthusiastic DSM users; up to 95\% of North Americans aged 13-17 may have access~\cite{anderson:teens:2018}. Increasingly, they  turn to DSM as a primary communication tool and to meet their social needs~\cite{smith:belonging:2021}. This pattern may have been accentuated in the COVID-19 pandemic lockdown period, with its restriction of in-person communication. 

However, prior reviews on adolescent DSM use have found a general correlation with negative mental health outcomes, particularly for factors like online social comparison and passive rather than active use~\cite{course-choi:socialMediaUse:2021, keles:systematicRev:2020}. 
Crucial features of face-to-face communication, including affective touch, are absent from most DSM. While some research has proposed tangible additions to social technology for children, few of these  are  fully realized systems~\cite{warren:exploring:2019}, or embody typical interactions of face-to-face friendships in their tangible components~\cite{abeyrathne:connectedChildren:2011}. 
It is  unclear how adolescents might engage tangible, embodied interaction in remote socialization, especially in comparison to traditional social media. 


While some scholars have raised concerns about the detrimental effects of DSM on adolescents, others argue that it provides important socialization opportunities~\cite{kross:2021:dsmWellBeing},
mirroring two DSM-evoked processes~ 
\cite{mikami:2025:dsmStrategies}.
On the negative side, DSM's visuality, permanence, and asynchrony can evoke \textit{self-presentation pressures}, or an intense awareness of one's online image and how it compares to others. 
As discussed above, this self-focus can detract from developing the social skills needed for healthy interpersonal relationships. 
On the positive side, DSM offers opportunities for \textit{social connection} with close friends and family, especially those that live far away.

    

\subsection{Affective Social Touch in Adolescence}
    \label{sec2:foundationsRW:touchAdolesc}

Affective touch is a powerful tool in communicating emotions, shaping social rewards, and establishing secure attachments for all ages~\cite{hauser:2019:physicalInteractionsAffect,cascio:2019:socialTouchHumanDev}. 
For example, nurses use comforting touch  to
personalize care and offer emotional or spiritual relief,  connection, security, peace, calm, confidence and self-esteem
~\cite{field:2010:review:touchWellbeing,chang:2001:conceptualTouchInCaring,fredriksson:1999:modesOfRelating,nist:2020:losingTouch,connor:2009:conceptualModelComfortTouch}.

These considerations are  central in adolescence, yet
most examination of touch's developmental role has focused on infants and young children~\cite{field:2010:review:touchWellbeing,mcglone:2014:affectiveTouchNeural,cascio:2019:socialTouchHumanDev}, where high levels of positive touch underlie lifelong social functioning~\cite{brauer:2016:maternalTouchDevelopingSocialBrain,bales:2018:socialTouchDevLTEffects}.
Relatively few studies directly support its importance in adolescence~\cite{bales:2018:socialTouchDevLTEffects,reed:2008:socialDance:chapter,tsachor:2017:LMAresiliency}:
\eg interpersonal touch reduces feelings of social exclusion~\cite{vonMohr:2017:TouchSocialExclusion}, and teens learn social skills in (tactile) pet relationships~\cite{purewal:2017:childrenPetsSocialDev}.

Neuroscience findings suggest evolutionary mechanisms that optimize affiliative touch.  Human hairy skin has receptors that respond preferentially to soft stroking touch associated with mammalian affiliation~\cite{mcglone:2014:affectiveTouchNeural}; 
pleasurable touch sensations are carried via dedicated unmyelinated fibers directly to brain regions that integrate affective information.  
These fibers function fully very early in infancy, in contrast to myelinated fibres for discriminatory touch, which mature more slowly. This system's early-maturing and dedicated wiring highlights its fundamental evolutionary as well as developmental importance in mammals~\cite{morrison:2010:skinSocialOrgan}. 

In adults, physically touching close others has been found to reduce patterns of neural activation associated with the stress response, and this relationship is stronger for those in better relationships
    ~\cite{coan:2006:lending,hauser:2019:physicalInteractionsAffect}. 
Social touch also reduces stress and provides reward as the young human brain develops. 
Rewarding affiliative touch from caregivers in early infancy is proposed to play a crucial role in development of the social brain, channeled through dedicated affiliative touch channels,  and this continues into the early childhood years~\cite{cascio:2019:socialTouchHumanDev}. 

In adolescence, a sexual component of social touch also emerges. However, 
in this sensitive and often tumultuous period of development, we argue that non-sexual affiliative touch in relational bonding may continue to be crucial to wiring the social brain and in stress reduction. 
However, there has been little study of teen social touch, perhaps in part because of its social complexity.


\subsection{Mediated Social Touch (MST), Groups, and Novel Perspectives}
    \label{sec2:foundationsRW:MST}

The notion of re-introducing social touch to online interaction is not new. 
Huisman \etal\ lists 
at least 32 prototypes for computer-mediated social touch (\textbf{MST}) between remotely-located people from as early as 1986~\cite{huisman:social:2017}, including vibrotactile objects, smartphone-based systems
    ~\cite{park:cheektouch:2010}, 
and wearables like a haptic vest~\cite{haans:2006:mediatedSocialTouch}. 
This existing body of work also shows desirable properties of in-person touch persisting in MST, such as capacity for emotional communication~\cite{smith:2007:IJHCS:affectiveLink}, increased likelihood of prosocial behaviour~\cite{haans:virtualMidasTouch:2009}, increased social connectedness between long-distance romantic partners~\cite{vanhattum:2022:longMSTcouples}, and attenuation of negative emotions~\cite{cabibihan:physiological:2017}. 

However, most of these touch-based systems are designed for one-to-one interactions. Group MST communication has focused more narrowly on communicating the same message to several people at once; \eg Lovebomb broadcasts singleton haptic or auditory emotive messages to nearby strangers~\cite{hansson:lovebomb:2001}). 

Moreover, many of these systems aim to replicate existing in-person touch sensations through MST, rather than exploring what \textit{new} interactions technology can afford. This may be due to an implicit assumption that social touch is a mere means of communication, so by replicating the {recipient's passive experience} 
of emotionally salient in-person touches like hugs, handshakes, and caresses, we can automatically recreate their meaning and positive effects. 

However, some  have recently pushed against this assumption. Schirmer~\cite{schirmer:2022:socTouchAsTool} and Huisman~\cite{huisman:2022:enactive} emphasize that MST requires conceptualization beyond the sender-and-receiver model, and the 2021 \textit{Manifesto for Digital Social Touch in Crisis}~\cite{jewitt:2021:manifestoDigTouch} calls for researchers and designers to "\textit{move beyond replication to forms of more innovative and creative digital social touch experiences}", while also emphasizing reciprocity, personalizability, and adapting to different contexts.
Della Longa \etal\  proposed the combination of in-person touch and immersive virtual reality as a means to address loneliness~\cite{dellaLonga:2022:virtualIATloneliness}, arguing: "\textit{virtual reality is not considered just a communication channel...but rather it has been recently reconceptualized as a social space... [giving] shape to interpersonal interactions and co-constructed relationships}"~\cite{galimberti:2010:intersubjectivity}. 

However, we are not yet aware of devices or social robots that create this kind of "social space" for touch interactions between remote groups of known others in the \textit{physical} world.

\subsection{Interactive Social Robots that Could Foster Connections}
    \label{sec2:foundationsRW:robotConnections}

Hegel \etal~\cite{hegel:understanding:2009} define a social robot as  physical robot plus metaphorical social interface, the latter encompassing the robot  attributes that a person would use to judge it as a social interaction partner. 
However,  expectations of this  interaction typically depend on the robot's form. 
We overview categories of social robots roughly based on Fong \etal's survey~\cite{fong:survey:2003}, then relate these experiences to \ESSbot\ design considerations. 
Research on group communication involving social robots appears limited to 
facilitating a single human operator's control over several social robots simultaneously (\eg Glas \etal's voice control~\cite{glas:teleoperation:2012}) rather than either bidirectional or haptic interaction.
    
\subsubsection{Anthropomorphic \vs Zoomorphic}
People tend to have higher expectations for anthropomorphic (humanoid) social robot capabilities compared to those for zoomorphic forms~\cite{fong:survey:2003, leite:social:2013}, including for the functions that particular social robots are intended to perform. 

Humanoid robots are typically intended to act in the place of actual humans, and may fulfill roles such as customer service representatives, tour guides, or playmates~\cite{zhao:humanoid:2006}. 
Interactions between humans and humanoid bots tend to be linguistic, relational, and normative: the robot communicates in familiar language using relational pronouns (``Can I help you with this?'') and follows human-human social constructs such as turn-taking
    ~\cite{zhao:humanoid:2006}. 
Humanoid robots sometimes emulate human expressions and emotions by manipulating their facial features, \eg eyebrow or mouth movement.

Attempts to create overly realistic robots (in appearance, movement or both) with any representational form can lead to the ``uncanny valley'' effect whereby it is perceived as unsettling or creepy~\cite{mori2012uncanny}. 
While zoomorphic forms may be less at risk of this since people often have less strict expectations for their appearance~\cite{fong:survey:2003}, some designers have deliberately taken a non-representational approach to avoid it --- \ie the caricature approach noted below.
    
People tend to expect social interaction with a \textit{zoomorphic} robot to resemble those with pet, companion  or therapy animals
    ~\cite{fong:survey:2003, yohanan:2012:humanIntentTouching}. 
Therapeutically-oriented examples include Paro~\cite{sabanovic:paro:2013}, a robot modeled on a baby seal and intended to support older adults with dementia, and the Haptic Creature~\cite{yohanan:2012:humanIntentTouching}, whose non-representational form was used to study how humans communicate emotion through their touch of a furry robot. 
Compared to the direct, literal nature of communication with humanoid robots, such robots facilitate a subtler connection via companionship or comfort, but also offer physical, tactile affordances (such as petting or holding) that might (today) seem socially awkward or inappropriate in interactions with a humanoid robot.   
We anticipate that \ESSbots' capacity for zoomorphism will come from emotional expression and social signaling in their individual and collective movement.


\subsubsection{Caricatured \vs Functional}

Fong \etal~\cite{fong:survey:2003} define \textit{caricatured} social robots as using unrestrained, potentially unrealistic aesthetic designs to draw attention to or distract from specific robot features, or to imply particular affordances
For example, the Haptic Creature~\cite{yohanan:2011a:AffectDisplay} is  zoomorphic in incorporating animal-like elements such as fur, ears, a tail and breathing motion (to encourage familiar ways of interacting with it) but it is also a caricature, because these elements are abstracted: they do not replicate any particular animal, and lack key mammalian elements like a face or eyes.

Finally, while most social robots have some kind of purpose, the physical attributes of  robots defined as \textit{functional} are specifically designed to accommodate the tasks the robot must perform. 
Huggable was created specifically for children in pediatric care, and thus takes on a toy-like teddy bear form that is familiar, would appeal to children, and could be used in a hospital setting~\cite{jeong:designing:2015}. 
Similarly, robots designed for autism therapy are often visually engaging so that the child does not lose focus, but not too bright or colourful lest the robot become over-stimulating~\cite{cabibihan:why:2013}.

\subsubsection{Telepresence}
Social robots of many forms can be an virtual agent, or alternatively  controlled by or serve as a mediator for a real person. 
Telepresence robots' communication with a remote participants is often facilitated by screen-based audiovisual interaction. 
In light of the COVID-19 pandemic-triggered shift to more online or hybrid activities, telepresence robots have been increasingly used in hospitals and nursing homes to allow remote visitors to visit patients, in high school and university remote graduation ceremonies, or in place of customer service workers at locations such as hotels~\cite{aymerichFranch:liaison:2022}. 
Many telepresence robots also allow the remote participant to pilot the robot throughout a physical space, of which several examples can be found in Kristoffersson \etal's review~\cite{kristoffersson:review:2013}. 

Some telepresence robots incorporate social touch, such as the tactile telepresence system for conveying touch patterns on the foreheads of isolated hospital patients developed by 
    Mostofa \etal~\cite{mostofa:tactile:2021} 
or the remote handshake system developed by 
    Nakanishi \etal~\cite{nakanishi:remote:2014}. 
However, this remote touch is most beneficial when both participants can experience it~\cite{nakanishi:remote:2014}, and such a setup may not be practically feasible in many cases since remote participants may not have the resources to acquire robots for themselves at home. 

\subsubsection{Insights for \ESSbot\ Design}
\ESSbots\ do not fit precisely into any of these categories.
They are neither anthropomorphic nor zoomorphic nor caricatured (they do not look like people or animals, even in abstraction, and their movement will not necessarily mimic a particular organism).  
They are not functional, aside from their social purpose. 
They are are telepresent, invoking a group of remote friends, but are not bound to a direct mapping between the remote other and a local avatar.
While the \ESSbots\ mapping could take many forms, by definition it cannot be exact since physical obstacles (like the local user's hand) will differ in each local node, forcing each node's momentary 
configuration to be unique. 
Moreover, we expect mappings to be most interesting when fluid, flexible and inexact.

That said, we have much to learn from all these social robot forms, including invoking people's willingness to attribute animacy and remote presence to a nearby robot that they see, hear and/or touch, and its evocation of a particular remote other. 
\ESSbots\ will  need to draw on biomimetic patterns of proxemics and emotive expression to find their own legible physical language, and learn how to deploy touch (both received and transmitted) to this end.

\subsection{Swarm Robots as Social and Physical}
    \label{sec2:foundationsRW:socialSwarms}

Swarm robots provide a natural platform for such group interactions. 
Interestingly, the field of social robotics began with the development and study of swarm robots --- 
communication between the robots themselves rather than between robot and human~\cite{fong:survey:2003}. 
Until recently, swarm robots have not generally been regarded as social, instead typically been designed and used for functional tasks such as collective manipulation, self-assembly, and distributed sensing
    ~\cite{rubenstein2014programmable,luo2011multi,werfel2014designing}. 
However, there is a growing literature on utilizing swarm robots for interaction with people, both in research and commercial contexts
    ~\cite{legoc:2016:zooids,kim:userDefined:2020,suzuki2019shapebots,kim2017ubiswarm}, 
including to provide haptic feedback in both real life and AR/VR environments
    ~\cite{kim:2019:swarmhaptics,suzuki2020roomshift,zhao2017robotic,suzuki2021hapticbots,li2022asteroids,kaimoto2022sketched,siu2018investigating,kim2023swarmfidget}. 

\textit{Zooids} are an open-source and open-hardware platform developed by Le Goc \etal~\cite{legoc:2016:zooids} as a tabletop swarm user interface (Figure~\ref{figsub:sec1:zooids}). 
Each Zooid is contained in a small, lightweight, cylindrical plastic shell, with wheels on the bottom to allow for motion across flat surfaces, and a touch sensor and colour LED for direct physical manipulation and additional visual feedback. 
The robots can move forwards and backwards up to 74cm/s, and can coordinate to produce complex movements via position tracking from an overhead projector~\cite{kim2017ubiswarm}. 
They can interact with other objects, \eg by pushing a phone across a desk towards the user. 
Most relevantly here, {Zooids} can be used as a haptic user interface, providing normal or shear force feedback by moving against the user's skin~\cite{kim:2019:swarmhaptics}.

Sony toio\textsuperscript{TM} robots~\cite{noauthor:toio:media:2018} (Figure~\ref{figsub:sec1:toios}) are a commercial swarm robot platform designed for children. 
While not quite as fast as Zooids (34 cm/s horizontal translation and rotation up to 1500deg/s), they use a mat rather than an overhead projector to track robot positions, and they are supported by an open-source Unity development kit~\cite{morikatron:toio:Unity:2022}. 
In addition to a built-in LED light, each has a sensor to detect magnetic force, cube posture angle, and collisions, which make it easy to integrate motion interactions such as double-taps and shakes~\cite{noauthor:toio:github:2021}. There is also a programmable button on the bottom of each robot. 

Research on swarm-human interaction has nevertheless been limited to interactions with one person or, at most, dyadic settings (\eg~\cite{kim:2019:swarmhaptics}).
None to our knowledge has explored using swarm robots to facilitate social interaction among more than two people, including use of the swarm to represent multiple other people, or multiple people coordinating control of the swarm at once --- intriguing social challenges which the \ESSbots\ platform allows us to take on.

\subsection{Situation within CSCW Theory} 
    \label{sec2:foundationsRW:CSCW}

Traditional CSCW theory posed a 2x2 matrix of space (same/different) and time (same/different) to understand contexts of software use  to support collaborative work~\cite{johansen:1988:CSCWmatrix}. 
However, a recent review noted that many current DSM applications sit in the single quadrant of different place \textit{and} time~\cite{zhang:2024:Form-From}, leaving an important gap that this  work helps to address. 

One key factor in assisting work communication and coordination is support of mutual/social ``awareness''~\cite{schmidt:2002:awareness,bodker:2006:socialAwareFexWork}.
CSCW researchers have pointed out that awareness is something formed between people in a complex give-and-take that includes many nonverbal signals. 
Though there has been some limited experimentation with awareness support in technologies such as vibrotactile belts~\cite{lopez:2017:awarenessReview}, there has been little focus on HRI tactics to support of ebb and flow of mutual awareness, according to a recent meta-review of the overlap between HRI and CSCW research~\cite{johansen:2024:cscwHRCreview}. 

The same review points out that there is also very little work in the space of groups engaging with robots.

\section{The \FWname\ Framework and its Theoretical Foundations}
    \label{sec3:framework}

In \S\ref{sec1:intro:approach_contrib}, we introduced the \FWname\ framework (Figure~\ref{fig:sec1:Framework}).
Here, we 
    link \FWname's \textit{grounding theories} to its prescribed \textit{affordances}, 
    propose a set of \textit{social good metrics} by which we can measure whether a resulting technology design is progressing towards our goals, and
    orient that design by deriving \textit{requirements} from the desired affordances.

\subsection{\FWname\ Affordances Grounded in Theory} 
    \label{sec3:framework:affordances}

\FWname\ technology should facilitate a specific combination of group behaviours, perceptions and responses: \textit{social touch}, \textit{co-presence}, \textit{coordination and participation}, and \textit{elastic effort of engagement}. 
We arrived at these affordances by examining theoretical perspectives and practical frameworks from cognitive science, developmental psychology, human-computer interaction and multidisciplinary team functioning; the most relevant are summarized in Table~\ref{tab:sec3:theorySummaries}.
Below we incorporate them into the motivation for the importance of each  affordance with respect to \FWname\ goals.

    

\begin{table}[tb]
\caption{Summary of psychology and social theories underlying the \FWname\ framework's Affordances. 
    Under References, [primary] sources are followed by other key [supporting] works mentioned in \S\ref{sec3:framework:affordances}. 
    } 
    \label{tab:sec3:theorySummaries}
    \vspace{-0.5em}
\footnotesize
\renewcommand*{\arraystretch}{1.1}
	\centering
	{
        \begin{tabular}{p{3.2cm}@{}p{1.9cm}@{}p{1.3cm}p{7.9cm}}
            \toprule
                \textbf{Theory} 
                & \textbf{References} 
                & \textbf{Affordances} 
                & \textbf{Key Details} 
            \\
            \toprule      
                {Social Touch Hypothesis}\newline
                    (\tST)
                & 
                    \cite{morrison:2010:skinSocialOrgan,
                mcglone:2014:affectiveTouchNeural,
                morrison:2016:affectiveSocialTouch};\newline   
                \cite{coan:2006:lending, cascio:2019:socialTouchHumanDev}  
                & \aSTouch \newline \aEffortEng 
                & Dedicated affiliative touch receptors are evolutionarily conserved to promote social bonding.
            \\
            \midrule[0.1pt]      
                {Goffman Co-Presence}\newline
                    (\tGCP)
                & 
                \cite{goffman:2008:copresence};\newline
                \cite{oh:2018:socialPresence,
                biocca:2001:networkedMinds,
                biocca:2003:socialPresence,
                vogel:2014:socialComparison}
                & \aCoPres 
                & Mutual awareness is facilitated through rich, embodied sensing of another, in a way that affords feedback. 
            \\
            \\
            \midrule[0.1pt]      
                {Interaction Theory}\newline
                    (\tINT) 
                & 
                \cite{froese:2012:IntTheoryXDynSystems, huisman:2022:enactive}; \newline
                \cite{thompson:2005:sensorimotor, dejaegher:ParticipatorySensemaking:2007, dePaolo:2012:interactiveBrainHyp, Gibson:Senses:1966}
                & \aCoordP 
                & Social cognition is understood in terms of embodied, co-regulatory dynamic processes spanning the participants and their environment.
            \\
            \midrule[0.1pt]       
                {Actor-Network Theory}\newline
                    (\tAN)
                & 
                \cite{Latour:1996}; \newline
                \cite{hutchins:2000:distributedCog}
                & \aCoordP 
                & Social processes are comprised of network interactions that include nonhuman/technological agents.
            \\
            \midrule[0.1pt]       
                {Distributed Cognition Theory}\newline
                    (\tDC)
                & 
                \cite{hutchins:2000:distributedCog}
                & \aCoordP \newline \aEffortEng 
                & Cognitive processes are distributed across agents.
            \\
            \midrule[0.1pt]       
                {Multiple Resource Theory}\newline
                    (\tMR) 
                & 
                \cite{wickens:2002:mrt,
                todd_implicit_2018};\newline
                \cite{sarter:2007:mrtDesign,
                    scolari_functions_2015,                        
                        todd_emotional_2020,
                        kryklywy_characterizing_2023}
                & \aEffortEng 
                & One task will compete with
another to the extent that it shares neural resources; affective/haptic systems compete less with key cognitive tasks.
            \\
            \bottomrule
        \end{tabular}
    }

\end{table}

\subsubsection{\textbf{Social Touch} (\aSTouch)}


The importance of affiliative social touch to social bonding and well-being is propounded in the \textbf{Social Touch Hypothesis} (\tST)~\cite{morrison:2010:skinSocialOrgan,morrison:2016:affectiveSocialTouch}: a specific class of mechanosensitive C-tactile (CT) afferents mediate affective rather than discriminatory aspects of touch,  signalling affiliative body contact primarily associated with positive social emotions such as comfort and affection, which in turn can promote social bonding.


CT afferents, 
    unmyelinated fibers present in mammalian smooth, and to a lesser degree hairy~\cite{cruciana:2021:CTglabrous,watkins:2020:CTglabrous}, skin 
respond preferentially to soft stroking touch~\cite{mcglone:2014:affectiveTouchNeural,kryklywy_characterizing_2023}. 
Fully developed perception of pleasurable touch  seems to rely on a combination of  affective and discriminative pathways ~\cite{mcglone:2014:affectiveTouchNeural}.
CT pathways are \textit{dedicated}, traveling directly to brain regions that integrate affective information, 
and \textit{mature very early} (full function in early infancy, in contrast myelinated discriminatory-touch fibers). 
These traits signify fundamental evolutionary importance of the affective touch system.

There is considerable empirical evidence of social and psychological benefits of social touch. 
These include stress reduction: 
as one of many examples, physically touching close others by adults reduces patterns of neural activation associated with the stress response, and this relationship is stronger for those in better relationships~\cite{coan:2006:lending}.
Studies of its widespread exclusion during the COVID-19 pandemic attributes touch loss specifically to increased anxiety and loneliness~\cite{vonMohr:2021:covidTouchDeprivationAdults} and conversely, its presence to resilience~\cite{dagnino-subiabre:2022:touchResiliance}  
and reduction in feelings of social exclusion~\cite{vonMohr:2017:TouchSocialExclusion}. 
Studies of early development suggests a reward mechanism; \eg from early infancy, rewarding affiliative touch from caregivers is proposed to play a crucial role in development of the social brain, channeled through dedicated affiliative touch channels,  and this continues into the early childhood years~\cite{cascio:2019:socialTouchHumanDev}.

Schirmer \etal~(2022) build upon these findings to argue for a ``nonverbal effecting model'', in which we recognize social touches as a way to regulate one's own state through the neurological and socially-mediated effects of interacting with another~\cite{schirmer:2022:socTouchAsTool}.
That is, while social touches can sometimes symbolically encode meanings or messages, they can access a much wider range of benefits to the touching agents, from immediate 
    (\eg regulating difficult emotions 
~\cite{hung:2019:benefitsSocialRobot,geva:2020:paroOxytocin,eckstein:calming:2020,sefidgar:2015:TAC:CalmingInteraction,vyas:2024:happraisalModel:EH,vyas:2026:passiveCalm:TAC})
to longer-term (\eg 
    strengthening an affective bond and ensuing emotion co-regulation support~\cite{uusberg:2023:reappraising, Sachs-Ericsson:2021:cogReapSocialSup,vyas:2024:happraisalModel:EH},
    or increasing likelihood of future positive interaction~\cite{haans:virtualMidasTouch:2009}.

Deeply reframing social touch as real-time \textit{co-regulation} between agents rather than mere disembodied communication highlights two \FWname\ design affordances. 
    \textit{Co-Presence} (\S\ref{sec3:framework:affordances:copresence}): 
    the system must assist users in locally recreating some extended notion of partners' embodied, animated selves.  
    Unlike a microphone for speech, socially inert touch ``sensors'' in the user's space might not suffice. 
    \textit{Coordination and Participation} 
    (\S\ref{sec3:framework:affordances:coordpart}): 
    It directs our attention away from communicative notions like information bandwidth, and towards affording the fundamental components of interaction. 
\subsubsection{\textbf{Variable Co-Presence} (\aCoPres)} \label{sec3:framework:affordances:copresence}

\FWname\ technology needs to facilitate varying degrees of co-presence: the \textit{``degree to which the observer believes he/she is not alone and secluded, their level of peripherally or focally awareness of the other, and their sense of the degree to which the other is peripherally or focally aware of them"}~\cite{biocca:2001:networkedMinds}. 

Co-presence is a subset of the broader quality of \textit{social presence}, loosely defined as a sense of ``being there with another''. Besides co-presence, social presence also includes psychological involvement and behavioural engagement~\cite{biocca:2003:socialPresence}. Social presence is often viewed as a fundamental goal of technology-mediated social interactions, measurable by Biocca \etal's Networked Minds Measure of Social Presence~\cite{biocca:2001:networkedMinds}. However, Biocca acknowledges that the concept  of social presence can blur the conditions required for it, the psychological state itself, and its effects~\cite{biocca:2003:socialPresence}. 

When designing for interactive social touch~\cite{schirmer:2022:socTouchAsTool,huisman:2022:enactive}, we need a specific definition of co-presence that emphasizes interactivity and embodiment.
\textbf{Goffman’s theory of co-presence} (\tGCP)~\cite{goffman:2008:copresence} identifies two key features required to achieve co-presence: 
    (1) mutual, synchronized sensing of the other's embodied self  (through the \enquote{naked senses}), 
    and (2) sufficient closeness to facilitate feedback 
        (\enquote{persons must sense that they are close enough to be perceived in whatever they are doing, including their experiencing of others, and close enough to be perceived in this sensing of being perceived}).

Both are continuous quantities.
Goffman emphasizes the importance of a range of closeness within which co-presence can be achieved, and its multi-dmensionality:
    \enquote{The physical distance over which one person can experience another with the naked senses –-- thereby finding that the other is `within range' --- varies according to many factors: the sense medium involved, the presence of obstructions, even the temperature of the air.}~\cite{goffman:2008:copresence} 

This definition directs our attention to technologies which, rather than simply passing disembodied messages like texts, extend some aspect of one’s embodied self so as to be perceivable through the naked senses of others.
Current DSM often accomplishes this by projecting one’s appearance to others (\eg video calls), but this emphasis may lead to negative social comparison processes that are detrimental to well-being~\cite{vogel:2014:socialComparison}.
We wonder how DSM might project an embodied version of the self through less easily-compared sensory channels. 
Meanwhile, embodying co-presence through touch may amplify social touch benefits beyond simply rendering sensations similar to in-person touch, because these stimuli confer full benefits only when attributed to a social agent~\cite{haans:2006:mediatedSocialTouch, huisman:social:2017}.  

Once co-presence is achieved, persons become ``\textit{uniquely accessible, available, and subject to one another}''~\cite{goffman:2008:copresence} in a way that can facilitate positive communication outcomes~\cite{oh:2018:socialPresence}. 
However, this does not mean it should be constantly maximized, just as many people find spending too much time in the physical presence of others draining or uncomfortable. 
DSM designs should allow people to modulate their co-presence over short or long periods; this also follows from the \tINT\ theory, which requires that agents retain autonomy over the nature of their participation in the interaction.


\subsubsection{\textbf{Coordination and Participation} (\aCoordP)} \label{sec3:framework:affordances:coordpart}

    
Increasing empathy and connectedness through embodied interaction require allowing for \textit{coordination} of action and reciprocal \textit{participation} among individuals in a group. 
Theories which point to the central importance of embodied coordination and participation to social cognition have their roots in Gibson's ecological psychology~\cite{Gibson:Senses:1966}: 
they propose that cognition does not happen within isolated agents but in the interaction between an organism’s embodied sensory systems and the opportunities for skilled action, or affordances, that the environment presents.  
For Gibson, cognition is embodied and happens through acting on the world. 

\textbf{Enactive theories} of cognitive science build on Gibson to argue that biological organisms, by definition autonomous and self-maintaining, make sense of objects and events in the world \textit{in relation} to whether interactions will support or threaten their ability to maintain their existence~\cite{thompson:2005:sensorimotor}. 
Most basically, in the process of chemotaxis a simple bacterium corkscrews its flagellae to swim up a chemical gradient to acquire nutrition and relaxes them to tumble away from noxious chemicals. 
This dynamically couples the bacterium's \textit{perceptions} of its environment (which guide its actions), and its possible \textit{actions} (\ie the ability to swim) which in turn guide the focus of its perception (whether to swim towards or tumble away).
Here, cognition is ``sense-making'' by which objects or events in the environment take on meaning in relation to whether they are to be approached or avoided (\ie good or bad for the organism's viability), \ie it emerges out of the relation with the environment.


Complex organisms such as humans have more complex and varied perception, and many more possible ways to interact with their environment. 
Enactivists~\cite{di_paolo_sensorimotor_2017} stipulate that simple sensorimotor loops, or actions, give rise to more complex concatenations of actions, called \textit{sensorimotor schemes}. 
Such schemes form to achieve ``norms'' of biological viability, whereby some outcomes of actions are more desirable than others. As schemes become solidified, the viability of the schemes themselves becomes a desired goal, giving rise to higher-level norms.

In the enactive view, sensorimotor schemes allow for the ability to perform such complex functions as social interactions. 
While these functions may seem far from the bacterium's life-or-death situation, the precariousness upon which sense-making depends can still exist even when built iteratively upon many cycles of experience and self-organization. 
For example, schemes for social interaction may seek to preserve precarious ``norms'' of an ongoing social bond, which can then gain importance through, \eg connections to physical safety in infancy.
In human development, a relational sense of self (self-as-betweenness) emerges from interactions with others~\cite{krueger_ontogenesis_2013}
beginning in the first days of life via embodied interactions that can be pre-reflective and non-verbal. 
It is through these interactions that children and adolescents learn the social affordances (behaviours permitted in a given social context) of their world~\cite{gallagher_socially_2013}.  
Here, the ability to participate fully in social interactions is crucial to the sense of having agency, or selfhood, as well as understanding others' intentions~\cite{krueger_selves_2024}.

By understanding social interaction as rooted in dynamic, layered sensorimotor schema, enactivism offers theoretical views into the importance of embodied touch interaction to social well-being. 
Since notions of social interaction begin forming even in pre-reflective, non-verbal stages, some underlying dynamic links of social cognition may best be stimulated through physical interaction, where meaning is conferred not through encoded representation but through mutual sense-making. 
\FWname\ relies specifically on three theories in this space: \tINT, \tDC\ and \tAN. 

\transitionGap

\textbf{Interaction Theory} (\tINT) explicitly extends the enactive concept of sensemaking to the social realm within cognitive science
~\cite{froese:2012:IntTheoryXDynSystems}. 
It challenges predominant theories that posit social cognition as a process that occurs in individual brains, which must send messages across a void to other brains, and in turn decode messages coming in.
Instead, it views social interaction as a situation where two or more cognitive, autonomous agents mutually and dynamically regulate their environments and each other, in a self-sustaining way~\cite{dejaegher:ParticipatorySensemaking:2007}.

Huisman \etal~\cite{huisman:2022:enactive} and Schirmer \etal~\cite{schirmer:2022:socTouchAsTool} connect \tINT\ directly to social touch, pointing out that standard sender/receiver views of social cognition fail to describe interactions outside the lab and thus fail to be useful in the development of mediated social touch technologies.  
In contrast, \tINT\ focuses on the interaction process itself as the site of social cognition. 
The dynamics of a given interaction take on an emergent life of their own that arises from and constrains the actions of each autonomous agent, as well as each agent’s evolving understanding of what is at stake. 

\tINT\ draws attention to two important qualities of the interaction: dynamics of \textit{coordination}, and a balance of \textit{participation} between agents and the interaction itself. Both of these are important affordances to FoSTER technology.

\transitionGap
This approach can further leverage concepts from theories of \textbf{Distributed Cognition} (\tDC), in which the system and not the individual brain is the unit of analysis
    (Hutchins~\cite{hutchins:2000:distributedCog}; also related to digital technology interaction design~\cite{hollan:2000:DistCogHCIFoundation})
and \textbf{Actor Network Theory} (\tAN)~\cite{Latour:1996}. 
\tAN\ focuses on social processes as networks of relationships that include non-human actors, including technology.
These ideas extend extend the concept of \tINT\ to include non-living agents (\eg sensemaking tools, environmental objects) as part of the network supporting social cognition --- a concept that is key to understanding \tINT\ as mediated by technology.

\lowHeading{Coordination} 
\tINT's focus is on understanding each agent’s evolving sensemaking in relation to the coordination dynamics of an interaction. 
We use De Jaegher and Di Paolo's definition of coordination as ``\textit{non-accidental correlation between the behaviours of two or more systems}'' which are related in some way~\cite{dejaegher:ParticipatorySensemaking:2007}. 
Rather than an individual’s intentional action, coordination refers to constantly-shifting properties of the interaction itself, which in turn entrains the actors. 
It can occur in emotion, physical movement, and intention; it can rise, fall, or change states over time.

\tDC\ describes coordination as distributing cognitive (sensemaking) processes across actors~\cite{hutchins:2000:distributedCog}.
Meanwhile, actors can intentionally maintain, shift or leave the coordinated interaction~\cite{dejaegher:ParticipatorySensemaking:2007}. 
Drawing on \tAN's proposal \tAN\ of non-human actors' roles in interactions, we can include robots as actors in these dynamics. 
 
In the context of social touch, \tINT\ posits that 
    ``\textit{social touch is a co-regulated process between two (or more) autonomous agents who actively participate in the generation of the meaning of the touch in interaction}''
        ~\cite{huisman:2022:enactive}, p. 3).  
 In most interactions, the degree of coordination between agents will rise and fall. 
 One hypothesis that arises from \tINT\ is that important shifts in sense-making will happen just before or after moments of high coordination~\cite{dePaolo:2012:interactiveBrainHyp}; 
     \kmE{see \S\ref{sec4:essbots:scenarios} for examples.} 
Empirically, measures of interaction coordination patterns can include patterns of high \vs low movement coordination between robots as well as coordination between brain and physiological measures in the teenagers operating them. 

\lowHeading{Participation} 
While coordination refers to correlations between the \textit{movement} or physiological processes of each cognitive agent, participation refers to the mutual extent with which each agent's \textit{sense-making} is affecting or affected by the interaction~\cite{dejaegher:ParticipatorySensemaking:2007}.
According to \tINT, a social interaction by definition occurs between agents, \ie  entities with \textit{agency}: 
self-organizing and self-maintaining, and able to modify the environment towards one's own goals~\cite{di_paolo_sensorimotor_2017}.

During a social interaction, a participant's agency may be at times reduced, but not eliminated. 
For instance, each participant's individual sense-making will be affected by other participants, as well as by emergent dynamics of the interaction itself. 
Their ability to manipulate the environment may also be shared with others. 
In a high-participation interaction, each participant is giving up a balanced amount of their own agency, lending to the emergent agency of the interaction itself, as a sort of ``super-agent'' made up of all participants and environmental factors involved. 
Here, \tDC\ and \tAN\ theoretically ground the idea that the entire system (including participants' brain and body, and physical objects like robots) can itself become a locus for emergent agency.
In a lower-participation interaction, each participant is less influenced by the interaction and retains more agency. 
If any participant's agency is reduced to the point of elimination, then there is no social interaction~\cite{huisman:2022:enactive}.
Thus, a balanced level of participation, even if varying over time, is a signifier of a design that is meeting the goal of fostering connection.


\subsubsection{\textbf{Elastic Effort of Engagement} (\aEffortEng)} 


Interacting with the technology should never \textit{demand} all of a user’s attentional resources. 
Affective communication, emotional support, fun and companionship  should be attainable with a scaleable level of engagement effort, ranging from none at a given time for a particular individual (\eg asynchronous interaction), to an easily legible parallel activity alongside other tasks, to full, synchronous engagement with others with  satisfying cognitive load. 

\transitionGap
\textbf{Distributed Cognition} (\tDC~\cite{hutchins:2000:distributedCog}) emphasizes offloading of cognitive processes from the individual to a distributed network of agents and technologies, reducing cognitive load of any given agent in the network.
Here, the robots and the individuals they embody act as a distributed cognitive system:
decisions about the content or nature of a given interaction are distributed amongst a friend group and the bots, and evolve from moment to moment. 
No single individual should be required to devote full attention to drive an entire  interaction in an effortful, planful way, and each should be able to respond to group dynamics intuitively and as desired.

\transitionGap
\textbf{Multiple Resource Theory} (\tMR)~\cite{wickens:2002:mrt,sarter:2007:mrtDesign} states that, in any multitasking context, one task will compete with another to the extent that it shares neural resources.  
For example, a task engaging tactile information not only activates different sensory pathways from one that engages auditory and visual information, but it also requires less linguistic processing, generating less competition in language networks.  
Here, the robots demonstrate (through expressive movement and social touch) remote agents' state and attitude towards the local user without each person having to process textual or verbally delivered information, or engage in conversation.  
Their physical embodiment, outside of a screen, enhances the absorption of this information through a low-cognitive effort,  `background' attentional process.

Research in cognitive neuroscience has firmly established a frontoparietal brain network that is consistently engaged during tasks that require effortful and explicit attentional engagement with the external world~\cite{scolari_functions_2015}.
Two tasks engaging this network will create high levels of competition.  
In contrast,  affective processes can function below the level of explicit awareness, and affective communication/touch can support rather than compete with other, attentionally more demanding, tasks. 
Affective information can guide attention implicitly, tuning sensory systems to relevant information via neural pathways that are distinct from those that are frontoparietally mediated~\cite{todd_implicit_2018}. 

Importantly, affective touch systems convey information about affective relevance from the point of contact with the world, before central brain resources are even engaged. 
Such valenced information is conveyed directly to brain systems that are sensitive to emotional information and  influence affective state, bypassing frontoparietal bottlenecks~\cite{todd_emotional_2020, kryklywy_characterizing_2023}. 
Thus, affective information conveyed via touch or via motion glimpsed through peripheral vision can allow for emotional communication and support without requiring a teen to turn their full attention from their homework.  
It allows for the embodied robot friends to work {with} each other, without hijacking other tasks.

\subsubsection{Summary}

Above, we have reviewed how psychology and social theories motivate a set of characteristics which we propose are central to a DSM technology that would increase empathy and connectedness. 
Specifically,  
(1)  Social Touch Theory (\tST) 
points to the importance of positive social touch (\aSTouch) for promoting closeness and reducing the stress response. 
(2) Goffman's Theory of Co-Presence (\tGCP) 
gives us insight into how the benefits of social touch can be realized even when remotely located from others. This is done by achieving \aCoPres; specifically, through extending one's embodied self in a way that facilitates attribution and feedback. 
3) Interaction Theory (\tINT), supplemented by Actor-Network Theory (\tAN) and Distributed Cognition Theory (\tDC), stress that social cognition arises from reciprocal and interactive dynamics, which can engage nonhuman actors and be distributed across them. They further emphasize that coordination and participation (\aCoordP) are fundamental components of healthy social interaction.
Finally, 
(4) Multiple Resource Theory (\tMR), 
alongside research in cognitive/affective neuroscience outlines ways in which embodied affective information, distributed across actors,  can \textit{manage} the effort of engagement (\aEffortEng), whether to reduce it in order to support rather than compete with other tasks, or  increase it for optimal interest and engagement, and allow these dynamics at times of the user's choosing.

\subsection{Metrics for Social Good} 
    \label{sec3:framework:metrics}




We expect that engaging with a DSM exhibiting the four \FWname\ affordances 
should yield two primary types of benefits for adolescents’ psychological well-being
 (listed in Table~\ref{tab:sec3:metrics} with metrics and assessment instruments).





\begin{table}[bt]
    \caption{Proposed desirable teen \textbf{outcomes} of theoretically-based DSM Affordances, 
        evaluation \textit{metrics} and 
        assessment \textit{instruments}.
    } 
    \label{tab:sec3:metrics}
    \vspace{-0.5em}
\footnotesize
\renewcommand*{\arraystretch}{1.1}
	\centering
	{
        \begin{tabular}{p{3.2cm}p{5.3cm}p{5.5cm}}
            \toprule
                \textbf{Social Good Outcome} 
                & \textbf{Metric} 
                & \textbf{Instrument}
            \\
            \midrule[0.1pt]      
                Reduced self-presentation \newline pressure online 
                &
                    $\searrow$ \# of social comparisons \newline
                    $\searrow$  reported feelings of FOMO
                &  $\circ$ Iowa-Netherlands Comparison Orientation Measure (\textbf{INCOM}), adapted for the social media context~\cite{cramer:2016:socialComparisonMeasure} \newline
                    $\circ$ 10-item Fear of Missing Out Scale~\cite{przybylski:2013:FOMOscale}
                    %
            \\
            \midrule[0.1pt]      
                Increased social connectedness with others  in online social \newline network
                &   $\nearrow$ observations of giving / receiving social support \newline
                    $\nearrow$ in reported feelings of belonging \newline
                    $\searrow$ in reported feelings of loneliness
                & $\circ$ Social support coding system such as \cite{mikami:2015:codingmethod} \newline
                $\circ$ General Belongingness Scale~\cite{malone:2012:generalBelongScale} \newline
                $\circ$  UCLA Loneliness Scale~\cite{russell:1996:UCLAlonelinessScale}
            \\
            \bottomrule
        \end{tabular}
    }

\end{table}
    
\subsubsection{Reduced Self-Presentation Pressure}

As discussed in \S\ref{sec2:foundationsRW:teenSocial}, acute awareness of one’s online image and how other people are perceiving it are known to accentuate social comparisons with others in one’s social network.
Strongly invoked by the design of existing DSM platforms~\cite{mikami:2025:dsmStrategies}, \FWname\ technology aims to avoid social comparison opportunities.        

Several kinds of comparisons and consequences can be measured via validated self-report scales. 
\textit{Upward} social comparisons (perceiving others in one’s social network to be better than oneself) are associated with poor self-worth and negative affect~\cite{vogel:2014:socialComparison}. 
\textit{Downward} comparisons (\ie perceiving oneself to be \textit{better} than others) can be detrimental because they may serve as temporary reassurance, fueling an anxious impetus to keep ensuring one’s place in the hierarchy. 
Finally, self-presentation pressures also give rise to fear of missing out (\textit{FOMO}) on rewarding experiences others in one’s social network are having, with implications for anxiety, negative affect, and distress~\cite{tandon:2021:fomo}.

\subsubsection{Increased Social Connectedness}

In general, we would hope to see people feel greater social connection with others in their social network. 
While existing DSM platforms could potentially evoke this~\cite{mikami:2025:dsmStrategies},  we argue that \FWname\ design should accentuate this tendency. 
This could be evidenced by adolescents showing more self-reported or observed social support exchanged with others in their online social network,
and in subjectively indicating 
    greater feelings of  belongingness (a sense that they belong in their online social network) and 
    reduced loneliness (a feeling that one’s social experiences are not the way one would like them to be).  
    
    
\subsection{From \FWname\ Affordances to Requirements} 
    \label{sec3:framework:requirements}

What does  facilitating
    \aSTouch, \aCoPres, \aCoordP\ and \aEffortEng\ 
need in a technological representation: what must it  
 be able to do, and what  characteristics must it have? 
A parsimonious set of {requirements} 
adds insight into what is unique about this form and allows us to navigate alternative design manifestations.

Our team iteratively distilled the following four (Figure~\ref{fig:sec1:Framework}) through a process that worked forward from our theories, and backwards from building and playing with system prototypes, observing and listening to the teens in our workshops, and extensively discussing both affordances and requirements from our varied disciplinary perspectives.  
While multiple affordances\affordSymb\ are involved in each, the dominant one(s) is/are \textbf{boldfaced} here and in Figure~\ref{fig:sec1:Framework}.

\subsubsection{Physical}
    \label{sec3:framework:requirements:physical}

To invoke \textbf{\aSTouch} over a distance, the technology first of all has to have at least an aspect which is physical and embodied.
The physicality must further be party to eliciting a sense of \aCoPres, \ie the remote network members captured or conveyed in the physical embodiment. 
Finally, by invoking an easily processed sensory modality and operating outside the crowded screen, this physicality should contribute to easing the minimum \aEffortEng\ while creating an appealing \textit{focus} of attention when users desire it.

An open question is the extent to which the representation should be \textit{solely} physical (\ie no graphical component), as well as co-located (given a graphical element, is it superimposed on the physical).

\subsubsection{Multiples}
    \label{sec3:framework:requirements:multiples}

For \textbf{\aCoordP}, the network of remote friends needs mobile identity elements which can be proxemically expressive with one another. 
While this might be achievable with a single linked entity with independently movable appendages (like  octopus  tentacles), multiplicity seems more apt for a group of remote agents who are autonomous but need to coordinate, and a much better bet for activating \textbf{\aCoPres}.
Multiple elements also lend themselves to a larger palette of movement, expressions and games. 
We see no need for the multiples to be identical in appearance or ability.  In fact, customization was a feature that teens frequently requested, and would likely enhance \aCoPres.

\subsubsection{Socially Interactive}
    \label{sec3:framework:requirements:socInt}

We anticipate that the primary modality of social communication will be coordination of movement, expressive individual movements, proxemic interactions with other bots; and the active participation of all the network members in this \textbf{\aCoordP}. 
The technology must support and encourage all of these.

\aSTouch\ occurs at the intersection of physical and social interaction. 
Touch should be possible between agents, observable by the user; as well as directly on or by the user's body when  touching objects controlled by others.

We see the media's social interactivity and by implication, expressive capacity as a primary conduit for the production of a sense of co-presence \aCoPres. 
These multiples must 
be able to move in a way that conveys socially legible meanings, and which (as expounded through \tINT) the network members dynamically interpret, build upon and elaborate.


\subsubsection{Supports Sustainable Fun}
    \label{sec3:framework:requirements:fun}

A system whose purpose is pleasure and comfort had better be fun: \ie have \textit{paratelic motivation}, the ability to sustain engagement in absence of the telic motivation of in-person presence.
Emergent behaviour (assisted by variability and unpredictability) can help sustain create the ``right'' level of hard, as well as sense-making (\tINT) interaction. 
The technology needs to support complex as well as simple behaviours and actions; and this will happen through  \textbf{\aCoordP}.
For many groups, game-like structure will be an engaging approach.


While fun brings people back, frustrating or unpleasant barriers to accessing rich behaviours will drive them away, making \textbf{\aEffortEng} critical.
Finally, physicality including \aSTouch\ opens opportunities for interesting, engaging interactions, including ``eyes-free'' interaction. What if you could play with the bots during a stressful or boring video call, much as you might play with your cat?

\section{\ESSbots\ as a FOSTER Design Manifestation}
    \label{sec4:essbots}
Embodied Social Swarms, present in and mirrored across  members' local spaces, meets these requirements. 
\ESSbots\ are
    physical, 
    based on multiples,
    and can move in socially communicative ways (\ref{sec2:foundationsRW:socialSwarms}).
Our design challenge is to turn them into a system that can support sustainable fun while building empathy and connectedness within a group. 

\subsection{\ESSbots\ Conceptual Model}
    \label{sec4:essbots:CM}

    \begin{figure}[bt]
        \centering
        \includegraphics[width=1.0\textwidth]{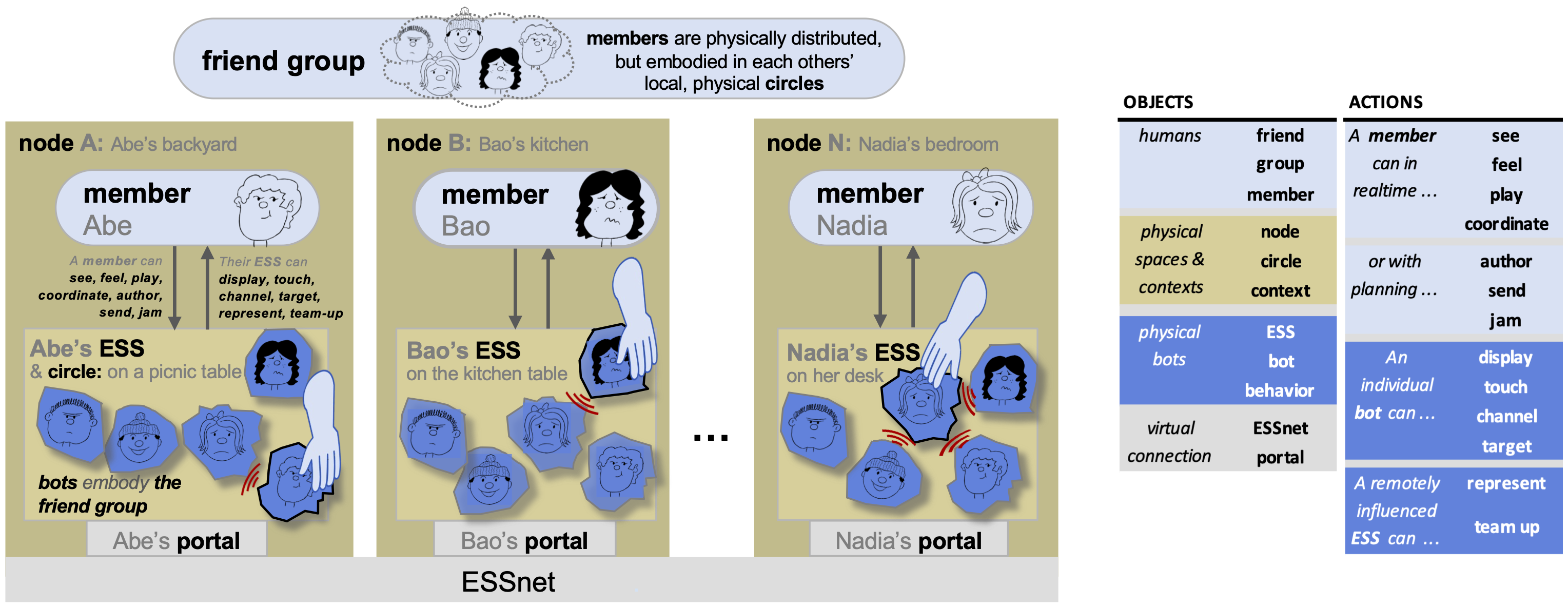}
    \caption{\textbf{The \ESSbots\ Conceptual Model (CM)}, portraying how \ESSbots\ work and how we would like users to think of them. 
        colour indicates physical space (tan) \vs virtual system structure (grey); and the interaction between humans (light blue) and physical robots (dark blue). 
        The bots' relative positions in the different nodes are related but cannot be perfectly mirrored because of differences in the local environment.
        \textit{\textbf{Use case:} Feeling sad, Nadia moves her bot slowly away from the others. 
        Abe and Bao notice this movement in their local swarms, and respond by gently, affectionately bumping Nadia's bot with their own. Nadia is able to feel these friendly bumps on and through her hand as it rests on her own bot in her local ESS.} 
        This depiction shows the individual bots embodying specific friends (an \textit{avatar} mapping); an alternative \textit{hive} mapping within this CM is offered in \S\ref{sec4:essbots:CM:fixity}.
        }
      \Description{}
      \label{fig:sec4:conceptualModelDiagram}
    \end{figure}

Figure~\ref{fig:sec4:conceptualModelDiagram} shows the \ESSbots\ conceptual model (\textbf{CM}) we adopt here;  
the full model is elaborated in 
    Table~\ref{tab:app:conceptsElaborated} (Appendix~\ref{app:CMtable}), 
    with detailed descriptions of its objects, their attributes and associated actions.
The CM consists of the minimal set of key ideas which the user needs to understand to make sense of the system and what they can do with it, and how these concepts relate to one another
    ~\cite{johnson:2002:conceptualModels}.
Primary concept objects include 
    the \textbf{friend group} as a whole;  
    group \textbf{members} each in a \textbf{node} with their local \textbf{Embodied Social Swarm (ESS)} composed of multiple \textbf{bots} that interact in that node's \textbf{circle} (a shared physical surface); 
    and  \textbf{portals} linking each node to an intelligent \textbf{ESSnet} of connected swarms.

\subsubsection{Conceptual Model Metaphor: Dance}
    \label{sec4:essbots:CM:metaphor}
%
This CM's language draws on metaphors in group dance, particularly breakdancing
    (\hyperlink{https://medium.com/cuepoint/9-dance-cipher-rules-no-one-ever-told-you-7ad4cd1817d8\%20Dancefloor}{dance circles}, 
    \hyperlink{https://en.wikipedia.org/wiki/Jamming_(dance)}{jamming}
and moves). 
While dance cultures can have performative and competitive elements, they are often more dominantly supportive, joyous, fun, funny, creative, emotionally expressive and celebratory.  
Crucially, they are deeply physical and epitomize many of the kinds of coordinated movement, building-upon, turntaking, and a fluid balance between solo and group movements that we seek here.  
Sports metaphors might also suit some groups and activities, but the teens in our workshops gravitated towards playful dance-like interactions. 

\subsubsection{Fixity of {bot} identity: Avatar \vs Hive}
    \label{sec4:essbots:CM:fixity}
%
Figure~\ref{fig:sec4:conceptualModelDiagram} depicts each bot as a manifestation of a specific group member;  we  call this the \textbf{avatar} mapping. 
It is an obvious interpretation, building naturally on the corresponding multiplicity and independence of bots and group members. 
Members could lean into this by decorating each bot to identify a  friend, as suggested by our teen advisors. 
They could exchange bots they have personalized (`signed') as a friendship bond. 

However, a fixed person:bot correspondence also raises issues.
Practically, if individual bots are permanently linked to a member's identity, then one bot's technical breakdown could knock its member out of that local ESS.
The group is arbitrarily constrained in size by the number of bots that each member has, and risks invoking a clique mentality. The latter problem might be addressed with high scalability, \eg if individual bots become inexpensive enough to have spares and a larger but still natural group size does not saturate the system computationally. 

Avatar mapping also complicates group-wide behaviours. 
These are better supported by a \textbf{hive} mapping: \eg an individual co-opting the whole ESS to express a dance they have designed, or many kinds of games.  
Futuristically, we envision large swarms of very tiny bots (picture bees), which can recombine atomically into member avatars in a meta-material manner~\cite{ion:2016:metamaterialMechanisms}, or act collectively as a hive. 

Figure~\ref{fig:sec4:conceptualModelDiagram}'s concepts are agnostic on this matter. 
They include both avatar actions (play, coordinate) and whole-hive moves (author, jam, represent, team-up). 
The corresponding implementation gives users license to explore both.

\subsubsection{Who's in charge?}
    \label{sec4:essbots:CM:ESSnet}
%
Some kind of intelligent system must connect the nodes and coordinate  swarm behaviours. The swarms cannot be perfectly mirrored across nodes, due to conflicts from local obstacles 
and transmission delays.
Since this coordination will be evident to users, it needs visual explanation. 
We chose to refer to it as a deliberately technical and non-animate \textbf{ESSnet}, 
in contrast to  an omniscient entity. 
We were concerned about the bots seeming like a wizard-controlled pet;
while this might be fun and lead to unexpected possibilities, it could also obstruct members' ability to perceive an ESS as representing their friend group.

\subsection{\ESSbots\ Use Scenarios}
    \label{sec4:essbots:scenarios}


Three scenarios convey some of the scope by which an \ESSbots\ swarm could be conceptualized,  structured and used, and illustrate potential opportunities and pitfalls.
The approaches are not necessarily exclusive, although they explore different mappings of the bots to the human group they represent, and moving between these mappings requires an increase in CM complexity.
The first two start the same way: 

\begin{quote}
    \textit{After a tough day at school, Nadia is miserable and feels she needs a hug. 
    She opens the door of a little hutch on her desk, and a swarm of sushi-sized \ESSbots\ roll out onto her node's ring. She squeezes the bot-avatar that represents herself long and steadily; she's trained it to recognize this signal for her low mood. 
    She sets it down and watches it sadly drift across the table. The bot-avatars representing her friends are still, so Nadia knows they are not home yet but that her bot-avatar will portray her feelings to her friends over the ESSnet once they are home. She turns her attention to her homework.}
    
    \quotespc
    \textit{Hutches across town start rattling.}

\end{quote}

\subsubsection{Scenario 1: \ESSbots\ as Avatars}
    \label{sec4:essbots:scenarios:1_avatars}

Each bot represents a specific group member (Figure~\ref{fig:sec4:conceptualModelDiagram}'s 1:1 CM mapping).

\begin{quote}
    \textit{Soon after they each arrive home, Abe and Bao pop open their rattling hutches, releasing several bots which each represent a specific friend. Bao has personalized their bots; the Nadia-bot has a blonde wig, and Abe's has a black skullcap like the one he wears. But she can tell who's who from their movement (Abe is always abrupt, and Nadia graceful and balletic). These `move-prints' are built into their avatars' personalities.}
    
    \quotespc 
    \textit{When the bots first come out, Abe remembers that he has some  news to share. He grabs his own avatar and shakes it like a dice to infect it with his excitement. When he sets it down, it spins around in the `happy dance' he's taught it.
    But then he notices Nadia's sad drifting. His bot's joy seems suddenly inappropriate; he calms it by trapping it under his hand, to save for later. 
    Instead, he pushes it gently towards hers to join in the hug that Bao's {bot} has already started.}
    
    \quotespc 
    \textit{Nadia looks up from her homework and sees Abe's and Bao's avatar-bots hugging her own. She rests her palm on hers, feeling the others' gentle rubs and taps. Then, she cups the whole swarm on her desk in both  hands and squeezes them in her own hug while they push and roll on her hands. 
    Their noticing helps her turn the corner in her mood. Done with sad drifting, she starts twirling her bot around. Noticing this change, Abe starts spinning his bot-avatar and chasing Nadia's. Suddenly the bots are playing tag.}
    
    

\end{quote}

With its reliance on proxemics, this scenario clearly invokes  \textbf{affordances} of 
    \aSTouch, \aCoPres\ and \aCoordP:
    Nadia's hug, others noticing her mood through embodied, nonverbal signals as if she were in their space, and responding by physically moving their own bot-avatars in coordination with the others. 
    This visibility allowed Abe to see the need to put his own feelings on hold, and put the brakes on an action which might set up Nadia for an unhappy \textit{social comparison}  --- 
    a sophisticated case of \textit{giving and receiving social support}
    (two of our \textbf{social-good metrics}).
    The teens' ability to make a choice about releasing the bots from their hutch to engage also supports \aEffortEng.
    
Opportunities for \textit{personalization} include 
    physical decoration to identify avatars;
    training a personally-linked bot to recognize the local human's moods through gestures;
    and training one's bot to respond to others' displays in particular ways. 
The first increases relational fixity, one of this mapping's restrictions (\S\ref{sec4:essbots:CM:fixity});  the others, being digital, do not.

\subsubsection{Scenario 2: Inhabiting a Hive}
    \label{sec4:essbots:scenarios:2_inhabitedHive}

Group members can alternatively interact with the swarm as a ``hive'', where all or most bots do not have an individual identity but move and touch in a coordinated way.

\begin{quote}
    \textit{When Abe and Bao each open their rattling hutches, one bot slips out.
    It drifts around their local rings, the picture of loneliness. 
    Bao guesses it isn't Abe, who doesn't do despondent; it seems more like Nadia. 
    For now, she  pulls out her phone, opens the ESS app and looks through several group-bot behaviours which she built earlier when she was bored. 
    She chooses one that triggers a gentle, physical game with one target bot; she touches and holds SadBot still for a moment, to set the focus on it.}

    \quotespc 
    \textit{At all nodes, the rest of the bots roll into the ring. They circle around SadBot, gently bumping and herding it into the center. 
    Two of the outer bots on opposite sides of the moving circle  suddenly dive towards SadBot, glancing off it and making it spin; then two more outer bots make it spin the other way, then all of them start circling again. 
    After doing this a few times, SadBot starts playing with the circle, moving around it in the opposite direction and bumping each outer bot as they pass. Whoever is in charge of SadBot has started to dance with the ring.
    Bao turns off the game she'd started, letting ApparentlyLessSadBot run the show.}
\end{quote}


Bao's  previous authoring of the semi-autonomous whole-hive behaviour requires an expressive \textit{composition and triggering tool}. 
Here, the movement is hybrid, with one bot being directly controlled to interact with and play off of the rest of the bots who are moving as a coordinated entity.  

The proxemic message conveyed differs subtly from the first example, because the non-target bots do not have specific identities. 
It is also unclear who invoked SadBot: Nadia can communicate her affect anonymously, and Bao could empathize anonymously. Co-presence can work even with ambiguous identity, at least in a small group.




\subsubsection{Scenario 3: Social Practice and Learning}
    \label{sec4:essbots:scenarios:3_practice}

Finally,  members can practice social communication and contact in an environment that could feel more safe due to acting through remote proxies. 
This scenario picks up a few days after one of the previous ones.


\begin{quote}
    \textit{Nadia had been sad because she was too shy and awkward to let someone --- know she likes them. So many things could go wrong! 
    The risk of disappointment and public humiliation seem high. She needs a subtle approach  and the confidence to pull it off. 
    Even though she sees her closest friends at school, this situation is so awkward that she hasn't had the stomach to ask them for help in person.
    }


    \quotespc 
    \textit{One day, Nadia, Bao, and Abe are playing Charades with their \ESSbots. Nadia privately orders Abe  to ``pretend he's trying to get the attention of someone he likes without showing he likes them''.
  Abe moves his bot with his hand and walks it by Bao's bot. Their bots have one marked edge, and they agree this is the `face', so they know which way each bot is looking. 
  Abe makes his `walk' clumsy. Bao and Nadia giggle. 
  Bao guesses the charade, and Nadia's secret is out. 
  They start discussing what else to try. Bao goes next. 
    }

    \quotespc 
    \textit{By the time it's Nadia's turn, she's feeling better about `Step 1' (getting Someone's attention) and starts working on an idea for upping the ante in `Step 2', by `accidentally' bumping into another bot. Abe and Bao try out different reactions, from offended to interested, and she figures out how to respond to each one. She thinks maybe she's ready to try it out for real next time she sees Someone.
    }
\end{quote}

We note that care must be taken so that the same quality which makes this scenario feel safe --- a remote proxy --- does not allow it to become unsafe through anonymity and lack of accountability.

\section{Teen Workshop Series: (I) Formative and (II) Co-Design}
    \label{sec5:workshops}


We conducted two  participatory design workshop series with teens to 
    explore this design space and its value propositions with teens, 
    get insight into how they conceptualize and interact with social swarm robots, 
    and direct feedback on the potential of swarm robots to facilitate embodied social connection for teens.
Over seven workshop sessions (each 2 to 3 hours), we discussed and co-designed  a wide range of themes and prototypes with 
26 different teenagers.

\begin{figure}
      \centering
      \includegraphics[width=1.0\linewidth]{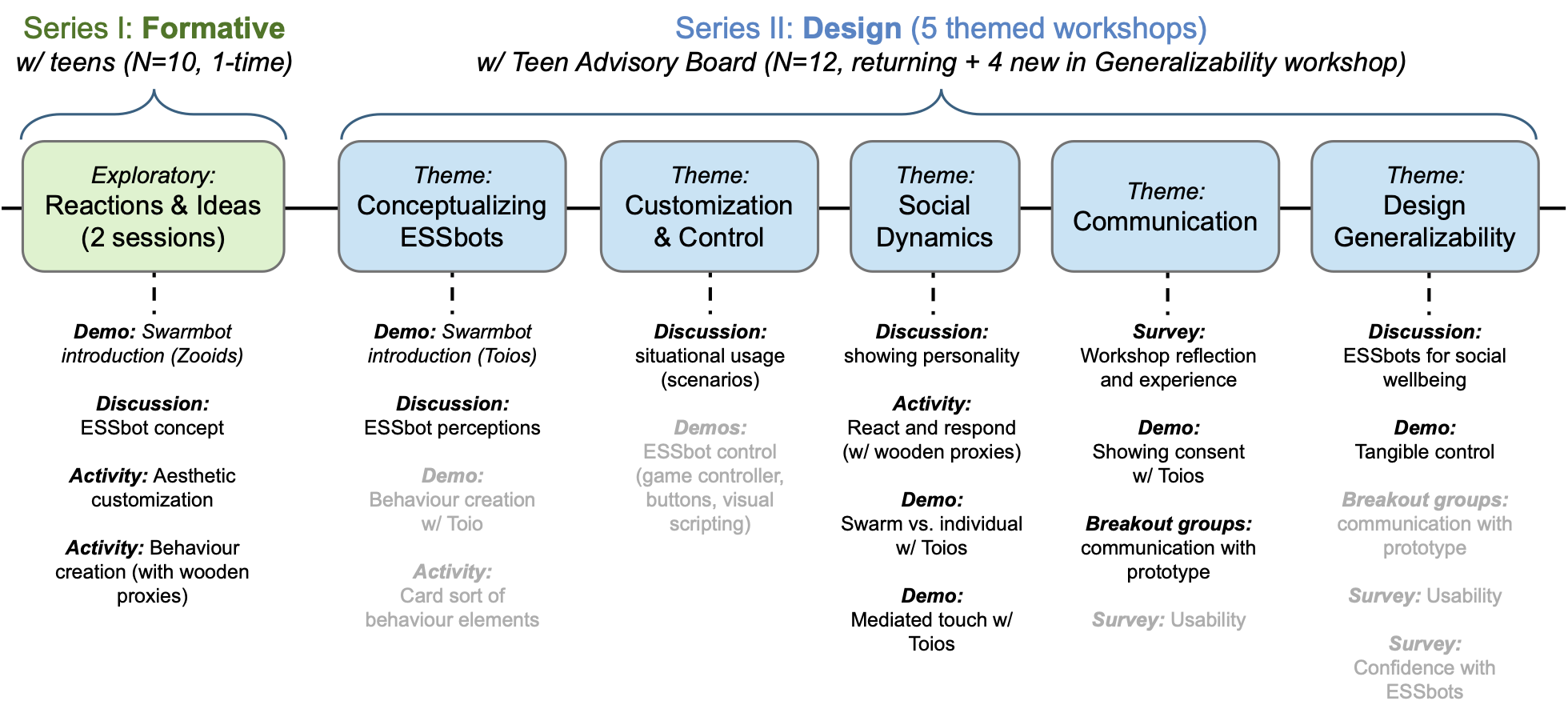}
      \caption{Workshop Series' contents, progression and analysis focus. Data analyzed in a separate paper is listed in gray.}
      \label{fig:sec4:workshopDivision}
    \end{figure}

Of this large corpus of data, in this paper we  have focused on the subset that addresses teen responses to \ESSbots\ concepts, how they perceive \ESSbots\ and anticipate using them, and how such a tool might relate to their own social and emotional wellbeing. 
A later paper analyzes data relating to how to make \ESSbots\ work, in a design sense, including the authoring-interface prototype~\citeESSWorkshops.         
This division is illustrated in  Figure~\ref{fig:sec4:workshopDivision}.

\subsection{Workshops Overview}
    \label{sec5:workshops:overview}

We conducted the workshops in the aftermath of the COVID-19 pandemic.
In our locale, secondary schools were fully in-person by Fall 2021, and social life had largely returned to normal (masking optional) by Fall 2022.

\subsubsection{Series I: \textbf{Formative} Exploration of Design Space}

These first two sessions took place in-person during Summer 2022.
We explored the design space for this innovative, embodied approach, through interactive activities and group discussions. 
We sought initial insights into how teens  conceptualized \ESSbots, their potential value to teens as a social technology, and how they would want to use the robots to communicate and express themselves. 
We had our participants use wooden bot proxies to mimic behaviours they would want the swarmbots to do, and asked  discussion questions about  design goals, practical use (including synchronicity and co-presence of use), and desired capabilities.

\subsubsection{Series II: \textbf{Co-Design} of \ESSbots\ Interaction}

The Design workshop series consisted of five thematically-focused participatory-design sessions, approximately every 3 weeks in Spring 2023.
We formed a Teen Advisory Board (TAB), and with this stable group as partners, progressively developed practical, engaging ways to support teens in working and playing with \ESSbots through hands-on activities, interactive demos, and group discussion questions. 

We required a \textit{behaviour-authoring interface} which was not only accessible and transparent to use, but also delicately balanced leveraging of \FWname\ affordances with support of what teens want to do. 
The early access it gave to bot behaviour helped us further identify what our TAB found most exciting; we would ask them to invent and describe behaviours and uses beyond or in other directions than what we had imagined or yet built.  

We first thought deeply about an \ESSbots\ conceptual model as informed by Series I, then for the first TAB meeting, constructed a skeleton tool which just began to expose the CM  and its \FWname\ affordances. 
Series II workshops then followed a cyclical pattern: TAB feedback in each session shaped our next  design steps (of system interactions and authoring-interface features), which we tested in the next workshop.
The TAB thus had increasingly direct control over the robots.
Ultimately, the authoring tool became a primary artifact of our collaborative exploration of this bot-behaviour design space with some balance {of} our theory-based affordances and teen input.

\subsection{Methods}
    \label{sec5:workshops:methods}

All workshops were conducted with teens aged 11--16, lasted 2--3 hours, and employed variations on three basic methodological components (\S\ref{sec4:workshops:procedures}).
In the following we describe the technology used,  basic procedures and specific variations for the workshop sessions analyzed in this paper, and the participants themselves, as
summarized in Table~\ref{tab:sec4:workshopSummary}.



\begin{table}[ht] \footnotesize
    \caption{\textbf{Methods Overview, Workshop Series I \& II.} 
            Primary goals, key activities, and data artifacts for \ESSbots\ concept explorations, by workshop, for activities within this paper's scope. 
            See Figure~\ref{fig:sec4:workshopDivision} for the full list of workshop activities.
        } 
    \label{tab:sec4:workshopSummary}
    \Description{Table with four columns:
        Workshop, Goals, Main Activities, and Data Artifacts. Each row lists the workshop title, the goals and main activities for that workshop, and the main data artifacts from each activity we were interested in investigating.}
    \centering{
        \begin{tabular}{p{2.0cm}p{3.7cm}p{4.0cm}p{4.0cm}}
            \toprule  
                \textbf{Workshop Theme} \newline 
                \textbf{\& Participants}   
                 &  \textbf{Goals} 
                 &   \textbf{Reported Activities}   
                 & \textbf{Data Artifacts}\\
            %
        \toprule
        \multicolumn{4}{l}{\textbf{SERIES I: Formative}} \\
        \toprule
        \textbf{FW1 \& FW2} \newline \newline
            FP1--FP7 (7) \newline
            FP8--FP10 (3)
            & Assess feasibility and openness to the \ESSbots\ concept with participants, including embodied and haptic communication with swarm robots.
            & \vspace{-.06in} \hspace{-3.0em}
                    \begin{minipage}{4.0cm} 
                    \begin{compactitem} 
                        \item Zooid demo
                        \item  Mediated touch demo
                        \item Discussion questions
                        \item Behavior creation activity
                        \item Customization activity
                    \end{compactitem}
                    \end{minipage}
            & \vspace{-.06in} \hspace{-3.0em}
                    \begin{minipage}{4.0cm} 
                    \begin{compactitem} 
                        \item Qualitative: first impressions
                        \item  Qualitative: haptic communication
                        \item Qualitative: \ESSbots\ concept
                        \item Qualitative: \ESSbots\ concept and perception
                        \item Personalized proxies
                    \end{compactitem}
                    \end{minipage}\\
        \toprule 
        \multicolumn{4}{l}{\textbf{SERIES II: Co-Design}} \\
        \toprule    
            \textbf{W1 \newline Conceptualizing \ESSbots} \newline \newline
                P1--P12 (12) 
                & Familiarize participants with the concept of \ESSbots\ and swarm robot features, learn how they want to express themselves with the robots and how they might want to control them.
                & \vspace{-.06in} \hspace{-3.0em}
                    \begin{minipage}{4.0cm} 
                    \begin{compactitem} 
                        \item Toio demo
                        \item Discussion questions
                    \end{compactitem}
                    \end{minipage}
                & 
                \vspace{-.06in} \hspace{-3.0em}
                    \begin{minipage}{4.0cm} 
                    \begin{compactitem} 
                        \item Qualitative: first impressions
                        \item Qualitative: \ESSbots\ concept
                    \end{compactitem}
                    \end{minipage}\\
            \cmidrule[0.4pt]{1-4}   
            %
            \textbf{W2 \newline Customization \& Control} \newline \newline
                P1--P3, P5--P7, \newline P9--P11 (9) 
                & Provide participants with multiple robot control methods and discuss their preferences and design suggestions, explore how situational context might affect how they use and control \ESSbots. 
                & 
                \vspace{-.06in} \hspace{-3.0em}
                    \begin{minipage}{4.0cm} 
                    \begin{compactitem} 
                        \item \ESSbots\ scenario discussion
                    \end{compactitem}
                    \end{minipage}
                & 
                \vspace{-.06in} \hspace{-3.0em}
                    \begin{minipage}{4.0cm} 
                    \begin{compactitem} 
                        \item Qualitative: situational usage
                    \end{compactitem}
                    \end{minipage} \\
            \cmidrule[0.4pt]{1-4}  
            %
            \textbf{W3 \newline Social Dynamics} \newline \newline
                P6, P10, P11 (8) 
                & Explore how in-person group dynamics are translated through \ESSbots\ (reacting and responding, interruptions or co-opting of robots, mediated social touch, and showing personality). 
                & 
                \vspace{-.06in} \hspace{-3.0em}
                    \begin{minipage}{4.0cm} 
                    \begin{compactitem} 
                        \item Reaction and response with proxies activity
                        \item Swarmness and individuality demos (`robot dance party', co-opting)
                        \item Mediated touch demo
                        \item Showing personality discussion
                    \end{compactitem}
                    \end{minipage}
            & 

            \vspace{-.06in} \hspace{-3.0em}
                    \begin{minipage}{4.0cm} 
                    \begin{compactitem} 
                        \item Qualitative: dynamic group usage
                        \item Qualitative: level of agency
                        \item Qualitative: touch preferences
                        \item Qualitative: showing personality
                    \end{compactitem}
                    \end{minipage}\\
            \cmidrule[0.4pt]{1-4}  
            %
            \textbf{W4 \newline Communication} \newline \newline
                P1--P6, P11 (7) 
                & Test remote communication in breakout groups with a working \ESSbots\ prototype, reflect on the workshop series and major takeaways for participants (what they learned, how their perceptions about \ESSbots\ changed). 
                & 
                \vspace{-.06in} \hspace{-3.0em}
                    \begin{minipage}{4.0cm} 
                    \begin{compactitem} 
                        \item Showing consent demo
                        \item Workshop series reflection activity
                        \item Survey about workshop experience
                        \item Breakout group communication
                    \end{compactitem}
                    \end{minipage}
                & 
                \vspace{-.06in} \hspace{-3.0em}
                    \begin{minipage}{4.0cm} 
                    \begin{compactitem} 
                        \item Qualitative: showing consent
                        \item Qualitative: perception comparison
                        \item Qualitative: workshop outcomes for participants
                        \item Qualitative: \ESSbots\ concept and perception
                    \end{compactitem}
                    \end{minipage}
                    \\
            \cmidrule[0.4pt]{1-4}    
        %
            \textbf{W5 \newline Design \newline Generalizability} \newline \newline
                P1--P3, P5, P6, \newline P11, P13--P16 (10) 
                & Evaluate the remote communication prototype and \ESSbots\ concept with a new group of participants, reflect on how workshop participation and influence on the prototype design could affect participants' perception of the \ESSbots\ system. 
                & 
                \vspace{-.06in} \hspace{-3.0em}
                    \begin{minipage}{4.0cm} 
                    \begin{compactitem} 
                        \item  Demo and discussion: tangible control
                        \item \ESSbots\ for social wellbeing discussion
                    \end{compactitem}
                    \end{minipage}
                & 
                \vspace{-.06in} \hspace{-3.0em}
                    \begin{minipage}{4.0cm} 
                    \begin{compactitem} 
                        \item Qualitative: tangible control
                        \item Qualitative: social wellbeing
                    \end{compactitem}
                    \end{minipage}\\
            \bottomrule
            \vspace{0.1em}
        \end{tabular}
    } 
\end{table}


\subsubsection{Technology Used}
    \label{sec4:workshops:methods:tech}

We structured our workshops around two existing swarm robot systems (described in \S\ref{sec2:foundationsRW:socialSwarms} along with their capabilities), to provide participants with thought-objects as they conceptualized what swarm based communication might look like and how they could work.
Series I employed \textit{Zooids} for demonstrations, and wooden pucks for hands-on activities.
In Series II, we used  Sony toios\textsuperscript{TM}; as  a commercial product that uses  a watermarked mat instead of a projection system, it proved more usable and scalable.

\subsubsection{Custom Authoring Interface}
    \label{sec4:workshops:methods:interface}

The custom authoring interface which evolved over the course of Series II was not formally evaluated for usability, but the final version technically supported our cumulative observations of how participants tried to use the system to communicate with one another, and  the comments they made in discussion groups. 
While we primarily discuss the interface design evolution and functionality and analyze its use in a separate paper, we note that participants could choose between 1) creating custom new behaviours by dragging and dropping simple visual scripting actions for each robot into the main workspace, and 2) clicking on buttons to play pre-made emotion or touch-based behaviours, which were developed based on earlier participant feedback (Figure~\ref{fig:sec5:scriptingInterfaceFinal} in Appendix~\ref{app:behaviour_interface}).

\subsubsection{General Procedures for Workshop Components}
    \label{sec4:workshops:procedures}

Each workshop consisted of one or more 
	\textit{demonstrations} of the robots and design tools and (as the series progressed) their updated capabilities, 
	\textit{activities} in which participants used and/or designed various \ESSbot\ aspects, 
	and \textit{facilitated group discussions}. 
While all tasks were tailored to the specific goals of each workshop, we summarize the structure of each task type below to clarify our terminology and illustrate the general procedures followed during our workshops.
The specific focus for each element is overviewed in Table~\ref{tab:sec4:workshopSummary}, with further detail on each activity  in Appendix~\ref{app:detailedMethods}.

\lowHeading{Demonstrations} 
Participants observed or (in later sessions) interacted with actual swarm robots in researcher-defined tasks, with increasingly direct control.
In breakout sessions in the last two Series II workshops,  participants used the \ESSbots\ prototype to pilot robots for one another in two separate rooms, including sending haptic messages.

\lowHeading{Activities} 
Participants did interactive tasks relating directly or indirectly to the robots.  
Examples of the latter were using robot proxies to communicate movement or touch-based messages, card-sort tasks to organize types of behaviours, crafting customization of robot proxies, and answering surveys about the \ESSbots\ prototype or concept.

\lowHeading{Facilitated Discussions} 
As a follow-up to each workshop task, we asked targeted questions  about both perceptions of the broad implications of \ESSbots\ as a technology, and specific experiences with our prototype and participants' design recommendations.

\subsubsection{Data Analysis}
    \label{sec4:workshops:analysis}

We analyzed  Series I (Formative) exhaustively, then took a lighter-weight approach to Series II (Design) textual analysis while attending closely to design artifacts 
(analyzed in depth in \citeESSWorkshops).   

\lowHeading{Corpus Analyzed}
We recorded videos of all Series I and II workshop activities, and transcribed participant responses to both facilitated discussions as well as general comments made during design activities.  These transcriptions were the primary source for all data analysis in the present paper. 
We also collected and analyzed written responses to the workshop reflection and experience activities in \workshopfour{}, and took photographs to analyze how participants customized their robot proxies in \workshopform{}.




\lowHeading{Codebook Generation}
Three co-authors collaboratively coded this corpus. 
We followed the hybrid approach of Fereday \& Muir-Cochrane~\cite{fereday:demonstrating:2006} and Proudfoot~\cite{proudfoot:inductivedeductive:2023}, \ie combining inductive and deductive coding.
Specifically, we  rigorously generated an initial inductively-generated codebook from  Series I workshops, then used these codes for both series' analysis.
Series II material that did not fit our initial codebook was inductively coded. 

We engaged in both latent and semantic coding to generate our initial codebook; but given Series II's reflective and fast-paced nature, we primarily coded semantically and did not iterate or reduce our initial set of low-level codes during collection.
Following the final workshop, we collaboratively reviewed the entire set of codes, removed redundant codes, and added latent codes specifically related to how participants conceptually perceived the swarm robots. 
For example, low-level codes related to how teens perceived their interactions with the robots included ambivalent, negative, and positive, as well as the latent conceptual construct codes (\eg viewing the robots as proxies for other people \vs as inanimate objects). 
We further organized these into the `Perception' category, \eg `Perception.ConceptualConstruct'. 
For a complete list of codes and when they occurred in the workshop series, see Appendix~\ref{app:codebook}.

We organized and discuss our codes as domain summaries, to reflect on how our own ideas for \ESSbots\ were interpreted and co-opted by teens. 
We focus on how participants conceptualize \ESSbots, participant-described usage scenarios, social dynamics, negotiating consent, and \ESSbots\ for social wellbeing. 
These domains are not fully realized themes~\cite{braun_reflecting_2019}, but  structure our discussion around the diversity of responses to relevant \ESSbots\ concepts. 


\subsection{Participants}
    \label{sec5:workshops:participants}

The study was approved by the university ethics board; recruitment took place on social media. Eligibility was restricted to respondents aged 11--16 who could attend the sessions. 
Assent and consent forms were collected from participants and  guardians.  
Participants were compensated for each session attended (participation details in Table~\ref{tab:sec4:workshopSummary}). 
    

\subsubsection{Series I: Formative --- 2 workshops; N=10, one-time} 
The first workshop included 7 teens (self-reporting as 2M, 3F, 2 chose not to disclose);
3  attended the second (1M, 1F, 1 undisclosed). 
All were between 11--15 years old.
In the first session, one pair were previously friends and 3 others were siblings; the rest did not know each other beforehand. 

\subsubsection{Series II: Co-Design --- 5 workshops; N=12 (TAB) + N=4 (final session})

Our goal was to design \ESSbots\ as a technology that teens could eventually co-opt for themselves; we needed partners who over time could build experience and knowledge about working swarm robots.
We therefore recruited a Teen Advisory Board (TAB) who engaged and provided ongoing feedback, in a role between  
    participatory design partners~\cite{druin:role:2002,guha:cooperative:2013} and protagonists~\cite{iversen:child:2017}, and encouraged them to attend as many workshops as possible. 
Three TAB members had also taken part in a Series I Formative
session: P1 (FP4), P5 (FP8), and P8 (FP1).

Twelve TAB members attended at least one of the first 4 workshops (6M, 5F, 1 non-binary);
four more (2M, 2F) joined the final Generalizability workshop.
All Series II participants 
    were 13--16 years old,  
    lived in North America at the time, 
    and identified as Mexican (n=3/16), Latino/a/x (2), Asian (5), European/White (3),  Asian and European/White (1); or did not disclose ethnicity (2). 
Five  indicated that they did not previously know other participants (3/12 and 2/4 respectively); 9/14 were related or already friends with at least one other.

\subsubsection{Participants' Context of Social Media Use (Summer 2022--Spring 2023)} 
    \label{sec5:workshops:participants:social_media_use}



\begin{figure}[tb]    \centering
        \subfloat[\textbf{Reasons participants used social media, and their frequency of reporting them.} Reasons are ordered from most to least commonly reported, based on length of the light blue (\textit{always)} bar.
            \label{figsub:sec5:social_media_use}]
            {\includegraphics[width=0.8\textwidth]{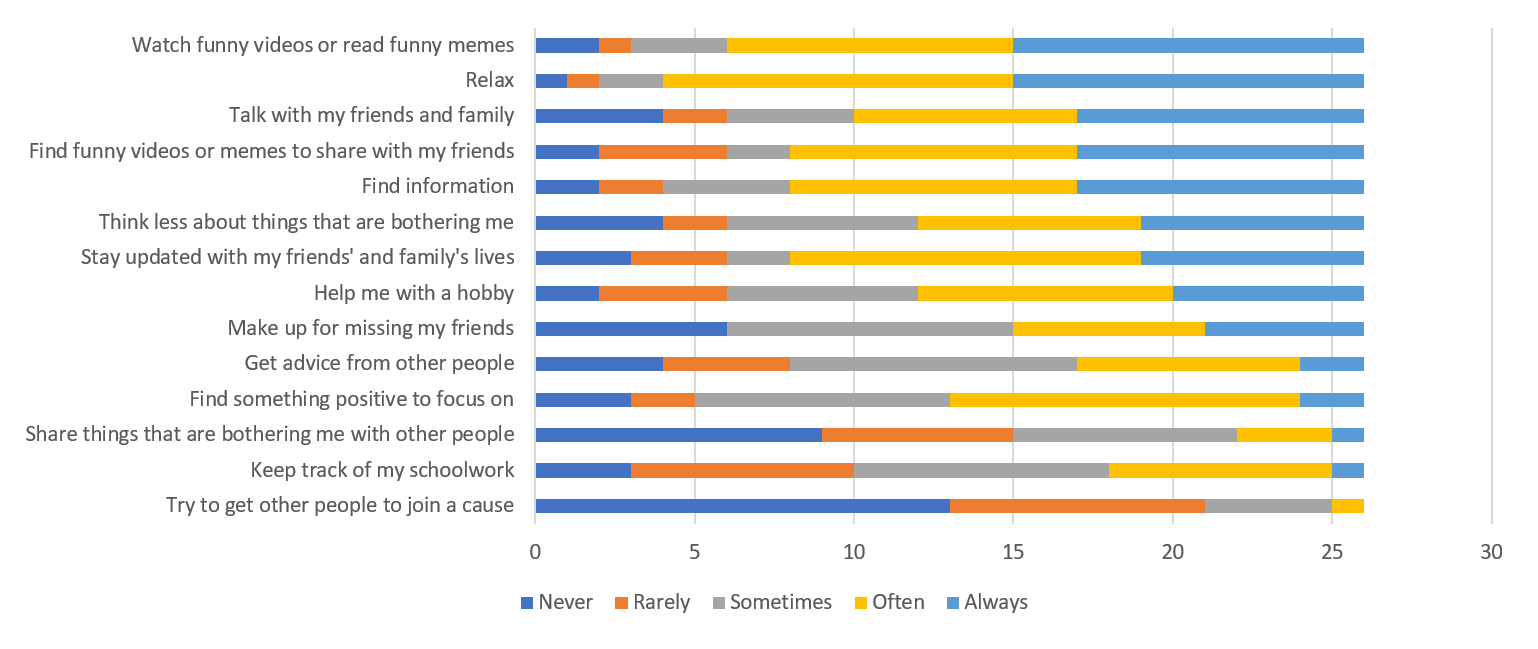}}
        \hfill
        \subfloat[\textbf{Frequency relative to COVID-19 lockdown period:} Comparison of social media use in 2022 to 2020. 
            For each reason of use, most participants reported using social media to a similar extent now as during the lockdown period (orange bars). Reasons where they used social media more during COVID-19 are bolded; of these, the most relevant to  social connection are also in black (\eg \textit{make up for missing my friends}).
            \label{figsub:sec5:social_media_covid}]
            {\includegraphics[width=0.8\textwidth]{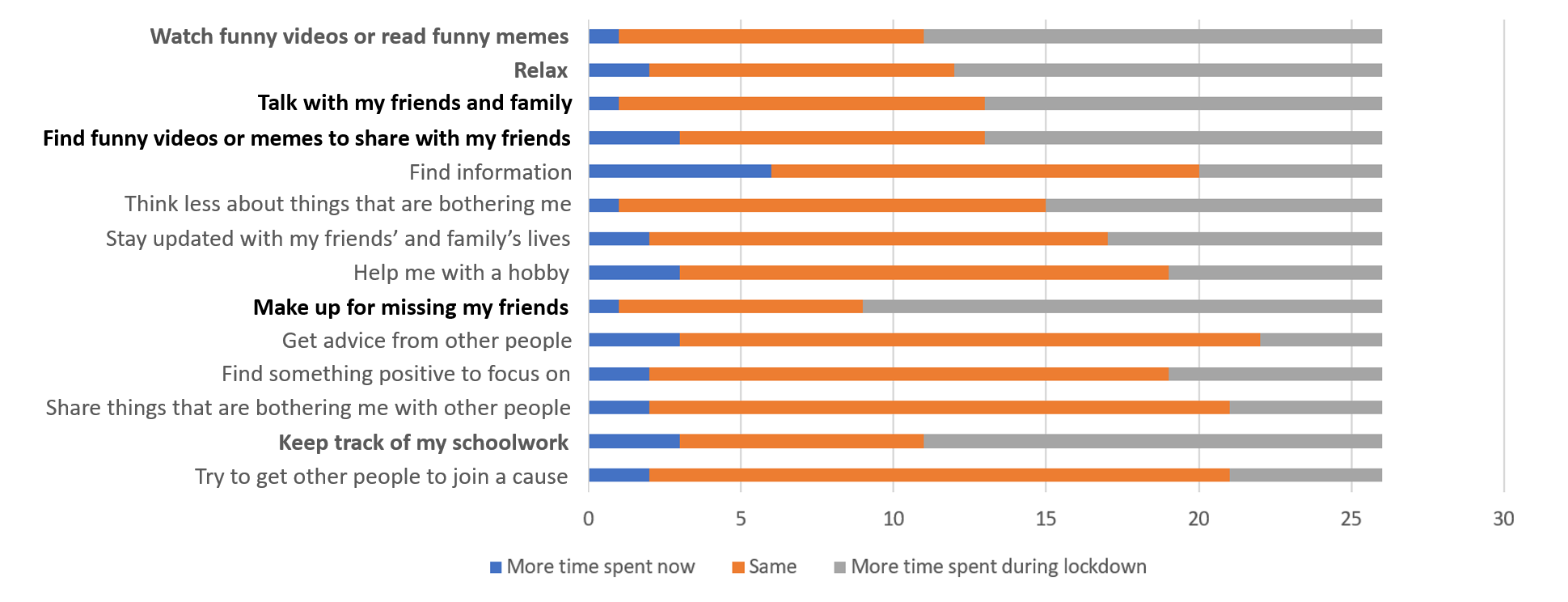}}
        \caption{\textbf{Teen participants' self-reported social media use (N=26).} (a) Frequency of different reasons participants use social media; (b) whether or not they used social media more, less, or the same during COVID-19.}
        \Description{Two stacked bar charts. The first shows participant responses to a 14-item Likert scale, indicating whether they never, rarely, sometimes, often, or always used social media for the given reasons. The most common reasons for social media use included to relax, watch funny videos or read funny memes, and talk with friends and family. The least common reasons included getting advice from other people, keeping track of schoolwork, and trying to get other people to join a cause. The second bar chart shows the same 14-items, this time indicating the frequency with which participants engaged in each item more, less, or the same during the COVID-19 pandemic. Most participants engaged in the activities the same or more during COVID-19.
        \kmC{put into appendix? is it adding enough here?}
        }
        \label{fig:sec5:social_media_use}
    \end{figure}

As our work explored how teens would react to a new kind of remote social technology, we asked participants in both series about social media use both at the time of the workshops and during the recent COVID-19 pandemic (when many relied more on remote interaction). 
Their most common reasons were to relax, have fun (videos and memes), and socialize with close others 
    (Figure~\ref{figsub:sec5:social_media_use}). 

Participants reported using social media 4 hours/day on average; most
0--7, and two >11 hrs (Formative Series).
65\% reported increased use in the COVID-19 lockdown period, especially to stay connected with friends or family (bolded rows in Figure~\ref{figsub:sec5:social_media_covid}), 
highlighting a connection between social media reliance and obstructed in-person socializing.

\subsection{Workshop Findings: \ESSbot\ Design Values}
    \label{sec5:workshops:findings}

In the following, we denote: 
        workshops as \workshopone, and
        evidence of specific Affordances as \hilite{\aSTouch},  
    The full codebook used to reach these linkages is in Appendix~\ref{app:codebook}. 

    

    

\subsubsection{How Participants Conceptualize \ESSbots}
    \label{sec5:workshops:findings:conceptualize}

\begin{figure}[tb]
      \centering
      \includegraphics[width=\linewidth]{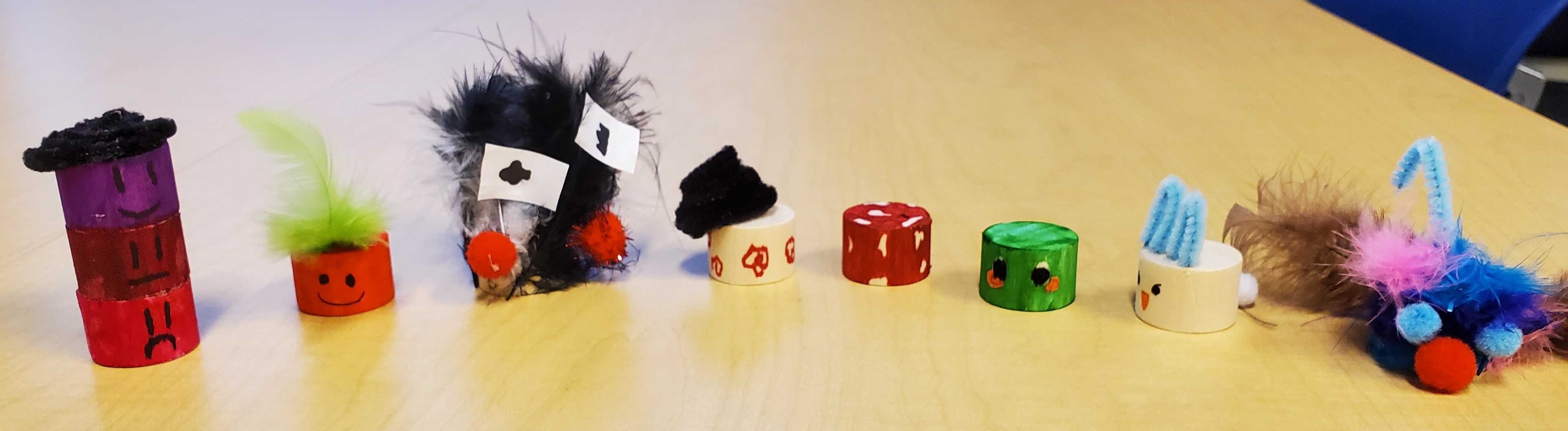}
      \caption{Decorated robot proxies from the crafting activity in Workshop 2.
      }
      \label{fig:sec5:crafting_results}
      \Description{This figure shows an image of eight decorated small wooden robot proxies from workshop 2. Some robots include emoji-like faces, others emulate animals (including a frog and a rabbit), while a few are colourful and decorated with abstract shapes.}
    \end{figure}

In general, participants conceptualized \ESSbots\ flexibly but stably across the series: as objects, representatives of themselves or others, and at times as having their own agency or emotions.  
Notably, when using the robots for games or tasks, they tended to describe them as objects  (\eg chess pieces or race cars); but  when discussing robot behaviours or communication, they used more personal language, sometimes identifying specific robots as representations of themselves or their friends: 
    \enquote{First the other person's robot [would flash yellow], and then mine [would move] around it} 
        (FP10, \workshopform{}; 
        \hilite{\aCoPres}; 
    \enquote{They're comforting us!} 
        (P11, \workshopfour{}; 
        \hilite{\aCoPres}, 
in reaction to a robot behaviour sent from other participants). 
Similarly, the \workshopform{} crafting activity produced customizations of faces or zoomorphism, but also abstractions (Figure~\ref{fig:sec5:crafting_results}).

Some participants seemed to attribute 
    feelings to the bots: 
    \enquote{it's trying to make [the other robot] more comfortable} 
        (P1, \workshopfour); 
    \enquote{[the other robots] are protecting it} 
        (FP6, \workshopform{};
        \hilite{\aCoPres}).
How the robots themselves moved and interacted led to higher perceived agency. 
    For example, breakout group participants in \workshopfour{} responded to a message before the other group had finished sending their own message, causing two robots to run into each other. 
    This in turn made the participants re-interpret the overall message as a disagreement between robots, rather than the initial comforting response the other group had tried to send
    (\hilite{\aCoordP}).

Over time, participants increasingly conceptualized the robots as social compared to their use as objects in games (possibly encouraged by the workshops' focus on remote connection and communication).
When asked to compare their perceptions of the robots from \workshopone{} to \workshopfour{}, TAB members noted that they initially saw the robots as more task-oriented (\eg to ``\textit{move}", ``\textit{fight}", ``\textit{jump}", ``\textit{play}"), but by \workshopfour{}, they were more communication-focused (\eg to ``\textit{communicate}", ``\textit{comfort}", ``\textit{express emotion and feelings}")
    (\hilite{\aCoPres}). 
 
Finally, direct descriptions illustrate how teens viewed \ESSbots\ as a tool for affective interaction despite varying levels of familiarity
    (\hilite{\aCoPres}). 
When \workshopfive{} teens were asked anonymously to describe \ESSbots\ to someone new, a TAB veteran wrote: 
    ``\textit{They are little robots that show a new way to communicate with your friends, by using emotions and commands}''; 
and a newcomer: 
    ``\textit{White, mini cube shaped robots that you can control/program to move, make sounds, light up colours and express emotions. These robots can be used to communicate your current state to a friend (\eg make the bots move rapidly and flash red when you feel angry)}''.

\subsubsection{Participant-Described Scenarios}
    \label{sec5:workshops:findings:PScenarios}

In both series we heard a range of potential \ESSbot\ uses,
including sending short messages, playing games or pranks, and as a notification system. 
Here, \ESSbots'  physical and embodied nature was viewed as an asset. 
FP3 (\workshopform{}) suggested that robot notifications would be harder for friends to ignore, and that robots could interact both with a friend and that friend's local environment, \eg by hiding objects around their room or jumping out to surprise them. 
Series II participants echoed this: ``\textit{I think that it’s different as in, it’s obviously physical, it’s not just text on a screen or something like that. So it’s cool to have a new way of communicating other than just reading. It’s actually physical, it’s in front of you}" (P13, \workshopfive{},
    \hilite{\aCoPres}).

Participants felt that the robots would not be well suited to convey complex messages, instead emphasizing their ability to facilitate closeness and convey emotions more impactfully than other modalities, especially text-based mediums. 
FP8 said: ``\textit{I might use like texting or whatnot to convey like, direct information, but use these to incorporate, you know, that feeling of touch, or you know, that more closeness or intimate feeling}"
    (\hilite{\aSTouch, \aCoPres}).
Some  described using the bots to try to evoke a particular emotion in others, as opposed to simply expressing an emotion themselves. 
Evocation was typically tied to symbolic meaning.
\eg when trying to come up with a behaviour for the phrase \enquote{that's not okay}, one participant suggested the robots could turn slowly while moving along the person's hand, \enquote{like tears falling} (P1)
    (\hilite{\aCoordP}).

We observed this view of ``bots for emotion expression'' during  \workshopform{}'s behaviour-guessing activity, when participants tended to guess about \textit{emotions} rather than \textit{specific phrases}. For example, when the demonstrated behaviour involved direct touch, guesses more often involved relational phrases or  evoked closeness and social group dynamics, such as ``I'm here for you", ``I'm happy for you"
    (\hilite{\aSTouch, \aCoPres}).

Participants suggested that some emotions or expressions, like hugs, might translate better through swarmbots than text: ``\textit{[when] I think they need a hug or something, and I can't [hug them], if I text `hugs' it'll be kinda weird. I think the robots can be a bit clearer what I meant to do.}" (P5,\workshopone{})
    (\hilite{\aSTouch, \aCoPres}).
However, while reporting mediated touch as a highly important aspect of expression, they did not see it as completely replacing in-person social touch; \eg P2, \workshopfive{} stated that ``\textit{When I’m using the robots, I try to make [my friend] feel good by, I don’t know, tapping them on the hand, but it’s like an electronic, so I’m not physically touching them... so it’s more like 50\% [effective] for me}."
    The incomplete \hilite{\aSTouch} expression in this \ESSbots\  implementation underscores its broader importance.

\subsubsection{Social Dynamics}
    \label{sec5:workshops:findings:socialdynamics}



In both series, participants indicated that \ESSbots\ use would be influenced by personal comfort, existing relationship norms, and  communication partner identity, especially in their willingness to use mediated touch.

Perceived interaction benefits  appeared dependent on existing social ties. 
\workshopfive{} participants using \ESSbots\  with strangers felt less connected compared to using the prototype with fellow returning TAB members: ``\textit{If we’re sending each other messages, like [the people from the original workshops], I feel like it helps me. But I don't know [the new group] as well, so I don't feel a connection really.}" (P2, \workshopfive{};
    \hilite{\aSTouch, \aCoPres}).
Similarly, participants felt that the importance of expressing their individual personality through the robots depended on who they were communicating with: ``\textit{If it’s for distance friends, I would say it’s important that it has her personality, and they don’t see me daily, so they need to know it's me.}" (P5, \workshopthree{}).
FP6 said they may be more willing to pay attention to the robots or return playful messages knowing they represented close friends \vs strangers. 



The influence of existing social dynamics and norms was especially evident regarding \hilite{\aSTouch}. 
Most expressed they would only use touch messages with close friends, comparing using the \ESSbots\ prototype with unfamiliar participants to the in-person experience of ``\textit{randomly grabbing a stranger's hand that you just met}" (P14, \workshopfive{}) --- itself a strong indicator of \hilite{\aCoPres}. 
Notably, those who  did not normally engage in social touch with in-person friends were most hesitant about \ESSbots\ social touch, \eg finding it too intimate or personal (P10, \workshopthree{}).
Further, the specific ways participants would use remote swarmbot touch directly reflected their in-person use of touch. 
For example, P11 (\workshopthree{}) preferred playful touches like gentle smacks and liked the demo where robots bumped against their hand, linking it to the ways they already interact in their male friendships
    (\hilite{\aCoordP}).

Others viewed touch as a way to express support and comfort: ``\textit{I feel like it just makes it more personal, like I’m here, not just saying it, which is why the robots are sometimes better than just messaging them, or like DMing them, cause it’s got touch}" (P13, \workshopfive{};
    \hilite{\aSTouch, \aCoPres}). 
P5 also described how social touch could help them express support indirectly, especially towards a friend who may not normally want to discuss their feelings openly: ``\textit{I want to tell them like everything is okay, without needing to like, text them}" (P5, \workshopfive{};
    \hilite{\aEffortEng}).


\subsubsection{Negotiating Consent}
    \label{sec5:workshops:findings:consent}

Participants 
recognized the importance of setting boundaries for whom should be allowed to manipulate bots at a given time, whether one or a whole swarm.
Some concerns were practical.
P11 (\workshopthree{}) and FP3 (\workshopform{}) mentioned distraction if a friend made the swarm move when they were trying to sleep, 
and P1 (\workshopthree{}) was concerned about the swarms falling off the table if they were out of their room.

More substantively, bot co-opting  was closely linked to personal consent and expression (\hilite{\aCoordP}).
P10 (\workshopthree{}) and P11 (\workshopfour{}) wondered about  communicative integrity if their bot automatically joined a group behaviour:
    it might misrepresent their current emotion, and then inadvertently mislead  friends.
P10  wanted to feel acknowledged before co-option, especially if it felt like an interruption: ``\textit{If it was trying to like block off your original comment as insignificant, then that’s a problem. That’s not a good friend}" (P10, \workshopthree{}). 
In general, participants wanted to maintain some level of input on their robots' actions, even for potentially autonomous tasks like shared music-listening or a dance party: 
    ``\textit{It’s you and your friends. So obviously it should have some spark of you and your friends in it, not just [be] the robots decision.}" (P10, \workshopthree{}).

To support boundary-setting, participants proposed features like a do-not-disturb mode, and consent-based usage for group behaviours such as indicating co-opting refusal by activating a red light or making a sound.
When we added a requirement of explicit receiver consent for a touch-based message to be sent (\workshopfour{}), they became more open to using mediated touch, even those who were initially hesitant
    (\hilite{\aSTouch}). 
This change in openness suggests that expressing consent may be especially important for remote social touch, and is relevant beyond this type of configuration.

\subsubsection{ESSbots for Social Wellbeing}
    \label{sec5:workshops:findings:wellbeing}

Overall, participants felt like using the robots with their friends would help them feel closer;
especially for mediated social touch
    (\hilite{\aSTouch, \aCoPres}), 
because ``\textit{they know you’re there}" (P15, \workshopfive{}). 
The robots might help them comfort others when they couldn't be there in person: 
    ``\textit{I really think that would be really helpful, cause sometimes I know my friends are feeling a little bit bad and I want to help them and I can't, because I’m not there}" (P5, \workshopfive{}). 
They might enhance empathy by showcasing diverse emotional expressions and personalities: 
    ``\textit{It makes you see other’s personality and the way they show their emotions}" (P5, \workshopfive{}). 


Participants also foresaw misunderstandings, and 
suggested mitigating strategies misunderstandings like adding layered levels of meaning with different robot features.
They also noted the need to incorporate respectful \textit{in-person} interaction strategies
here, such as thoughtfulness, consent, and active listening. 
P13 (\workshopfive{}) pointed out that the thoughtful creation and interpretation of messages actually added value to \ESSbots\ interaction: ``\textit{The fact that you actually have to like, program it [is better than texting or calling]. They actually had to do something to tell me this}". 
In this respect, a slightly \textit{increased} \hilite{\aEffortEng} was potentially valued as an indicator of caring.

\section{Discussion: Early Insights into \FWname\ Potential}
    \label{sec6:discuss}

We can now revisit the full \FWname\ framework (Figure~\ref{fig:sec1:Framework}) in light of observations from the teen workshops.
We 
    (1) review where this specific \ESSbots\ implementation and evaluation might impact generalizability; 
    then (2) address satisfaction of framework \textit{requirements}, a necessary precondition of 
    (3) capturing \FWname\ affordances.


\subsection{Project Focus, Constraints and Emergent Design Dimensions}
    \label{sec6:discuss:focusConstraints} 

    
The perspective on \FWname\  documented here is 
based on 
    specific swarmbot technology,
    an interface co-designed with teens, 
    3-bot swarms,
    just 2 nodes in adjacent rooms in a lab, 
    teens of varying familiarity and desire to connect with one another,
    and longitudinal themed workshops.
%
The larger \textit{vision} is more typified by  groups with 5 nodes and 5 bots/node connected across larger distances, with groups who either start with or come to know one another well and have an intrinsic motive to socialize.

More broadly, our approach was thematically focused, rather than fully open-ended: 
W1 \textit{Conceptualizing \ESSbots} targeted early reactions, but Series II (W2-W5) delved into how teens wanted to use them, to support future design.
Examining each theme with \FWname\ in mind exposed an evolving conceptualization across varied topics.
In Series II, theme specificity also helped us maintain momentum and  evolution, where a full, bottom-up analysis of all collected material was infeasible and would have produced diminishing returns.
That we discovered things we were not looking for (\eg the issue of consent to touch one's physical avatar, or how interpersonal and activity context consistently shaped participants' perceptions of the robots),
is an indicator that we may have found a reasonable  structural balance.

In the following, we generalize findings cautiously,  acknowledging  many tradeoffs, \eg in: 
    continuity with a small number of teens rather than a larger-scale study;
    rapid iteration over deep analysis in Series II;
    small scale in node and swarm size and  physical separation;
    and the potential influence of a specific platform.

As our analysis and reflection progressed, we identified several \textbf{emergent design dimensions} along which it will be productive to investigate the impact of implementation  sophistication and variety, and mention them where relevant:
    \hilite{\DDdynsync} (dynamic synchronization),
    \hilite{\DDint} (modes or features), 
    \hilite{\DDscale} (more bots and/or people), 
    \hilite{\DDperf} (strength, miniaturization, expressive movement) and 
    \hilite{\DDcont} (context of interaction).

\subsection{Can \ESSbots\ Meet \FWname\ Requirements?}
    \label{sec6:discuss:requirements} 

We proposed four \FWname\ system requirements   
 (\S\ref{sec3:framework:requirements}), 
then identified how \ESSbots\ as a general approach seemed in principle to satisfy them (\S\ref{sec4:essbots}). 
To what extent were those requirements indeed met with this initial implementation?

\lowHeading{Physicality \iconPlus\ and Multiples \iconPlus}
Both are objectively met by the use of swarm technology reproduced on each user's local space. 
Teens physically touched the robots, and typically used and observed them acting as multiples rather than individually. 
While this behaviour was not always spontaneous (many activities were directed or constrained), teens' comments about conceptualization and engagement indicate that they recognized these properties, and also appreciated and could make use of them.
Conversely, although we learned a great deal here, larger and more capable swarms 
    (\hilite{\DDscale, \DDdynsync, \DDperf})
would unlock possibilities and support larger friend groups.

\lowHeading{Socially Interactive \iconPlus}
Regarding the ability to
    \enquote{move in a way that conveys socially legible meanings} (\S\ref{sec3:framework:requirements:socInt}),
We have ample evidence that teens were able to infer social meaning in individual and collective bot movement, both as intended by a sender or otherwise.
Technically, the bots were able to move briskly and with low system latency; teens did not mention lag or slowness, and such factors did not appear to impact on social interpretations.

Importantly, the ways teens expressed themselves through the robots appeared closely tied to pre-existing social dynamics between senders, reflecting the norms and expectations they already have of communication partners. 
Future work should explore how teens both interpret and respond to the robots in more sophisticated and sustained  interactions, to assess how social interactivity is sustained through the bots over time
    (\hilite{\DDdynsync, \DDperf, \DDcont}).
    

Teens also pointed to ways to improve social expressiveness, mainly through customization of bot physical appearance and capability.
Decoration could move beyond glue and tape with snap-on sites or draw-on surfaces. 
Similarly, teens often wanted bots to convey orientation for social clarity, \eg with a front or face edge. 
This requires bots to be aware of its `front'.
Both needs are easily addressed
    (\hilite{\DDint}).

\lowHeading{Supports Sustainable Fun \iconQuest}
Regarding \enquote{the ability to \textit{sustain} engagement in absence of the telic motivation of in-person presence} (\S\ref{sec3:framework:requirements:fun}): 
teens appeared to and reported having fun, and believing in  telic potential.
We  observed sufficient engagement for participants to continually ideate and provide valuable insights on their perceptions, feelings, and desires for the robots over multiple  sessions. 
In particular, teens were eager to interact with the robots, envisioning use-case scenarios that were both communicative and playful. 
However, structured activities and artificial 
group composition and context restrict generalization on its sustainability
    (\hilite{\DDint, \DDcont}).


\subsection{Can \ESSbots\ Support \FWname\ Affordances?}
    \label{sec6:discuss:affordances} 

Our teen workshops provided feedback regarding each of the proposed \FWname\ affordances, which
Table~\ref{tab:sec6:affordanceSummary} summarizes as evidence relating to their importance, how the \textit{current} prototype (with its specific design focus and limitations) captures that affordance, and how it and its potential benefit could be further amplified and examined. 

%

\begin{table}[ht]  \footnotesize 
    \caption[FoSTER Affordances Findings Summary]{\textbf{\FWname\ Affordances Findings Summary.}
        \textbf{\textit{Evidence}} is self-reported (\Erep) or observed (\Eobs).
        Each key item is noted as
        %
            \textit{supporting} \iconPlus,
            \textit{countering} \iconMinus, or
            \textit{inconclusive / uninvestigated} \iconQuest,
        with next steps for {development} (amplification) or {evaluation} (inquiry about its role).
        Emergent \textbf{\textit{Design Dimensions}} meriting further development/study (\eg whether they should be requirements) are
            \DDdynsync\ (dynamic synchronization),
            \DDscale\ (more bots and/or people), 
            \DDperf\ (strength, miniaturization, expressive movement), 
            \DDint\ (modes or features), and
            \DDcont\ of interaction.
            \iconMeas\ indicates prerequisites for more targeted inquiry.
    } 
    \label{tab:sec6:affordanceSummary}
    \Description{\todo{describe}.}
\centering{    
        \begin{tabular}{
            @{\kern0.01in}p{0.70in}  
            @{\icSp}p{\tabASwid}
            @{\icSp}p{\tabASwid}
            @{\icSp}p{\tabASwid}
            @{\icSp}p{\tabASwid}
            @{\kern-0.1in}}
        \toprule  
                \textbf{Affordance}   
                 & \textbf{Is Important to Users} 
                 & \textbf{Is Captured Here} 
                 & \textbf{Could be Amplified} 
                 & \textbf{Further Inquiry}  
            \\ \toprule 
        \vspace{\rSp}
        \textit{\textbf{\aSTouch}}
        \vspace{\lSp}\newline 
        \textbf{Social Touch}
            & \vspace{\rSp}\hspace{\tabASmpCor} 
                    \begin{minipage}{\tabASmp} 
                    \begin{compactitem}
                        \item[\bulletPlus] \Erep: \ESSbot\ touch preferred over non-touch media for regulatory and affiliative functions.
                        \item[\bulletMinus] \Erep: Social touch (bot-skin or bot-bot) confirmed as non-optimal for communicating detailed information.
                    \end{compactitem}
                    \end{minipage}\vspace{\rSpRev}
            & \vspace{\rSp} \hspace{\tabASmpCor}
                    \begin{minipage}{\tabASmp} 
                    \begin{compactitem} 
                        \item[\bulletPlus] \Eobs: Facilitation of two key  affiliative touch components: attribution of the touch to another, and affiliative interpretation. 
                        \item[\bulletQuest] 
                        Affiliative \textit{benefits} not measurable in this study design.
                    \end{compactitem}
                    \end{minipage}\vspace{\rSpRev}
            & \vspace{\rSp} \hspace{\tabASmpCor}
                    \begin{minipage}{\tabASmp} 
                    \begin{compactitem} 
                        \item \DDscale, \DDperf: Social touch could be more convincing and varied with bots that are more numerous, strong and/or expressive. 
                        \item \DDdynsync: Dynamic synch might  strengthen affiliative and attribution tendencies.
                    \end{compactitem}
                    \end{minipage}\vspace{\rSpRev}
            & \vspace{\rSp} \hspace{\tabASmpCor}
                    \begin{minipage}{\tabASmp} 
                    \begin{compactitem} 
                        \item[\bulletMeas] Devise means of quantifying \textit{affiliative benefits}.
                        \item[\bulletQuest] \DDdynsync: Will dynamically synchronized \ESSbots\ touch behavior strengthen observed affiliative associations?
                    \end{compactitem}
                    \end{minipage}\vspace{\rSpRev}
        \\ \midrule
        \vspace{\rSp}
        \textit{\textbf{\aCoPres}}
        \vspace{\lSp}\newline 
        \textbf{Variable \newline Co-Presence}

            & \vspace{\rSp}\hspace{-0.29in} 
                    \begin{minipage}{\tabASmp} 
                    \begin{compactitem}
                        \item[\bulletPlus] \Erep: Desire both for high (\eg live bidirectional interaction) and low (\eg emotion button) co-presence features. 
                    \end{compactitem}
                    \end{minipage}\vspace{\rSpRev}
            & \vspace{\rSp} \hspace{\tabASmpCor}
                    \begin{minipage}{\tabASmp} 
                    \begin{compactitem} 
                        \item[\bulletPlus] \Eobs: Current ESSbots attributed as partner ``extensions''. 
                        %
                        \item[\bulletQuest]  
                        This prototype's `message-exchange' style might be constraining this affordance.
                    \end{compactitem}
                    \end{minipage}\vspace{\rSpRev}
            & \vspace{\rSp} \hspace{\tabASmpCor}
                    \begin{minipage}{\tabASmp} 
                    \begin{compactitem} 
                        \item \DDdynsync: Co-presence (especially its ``feedback'' component) may be facilitated with dynamic interactivity.   
                        \item \DDdynsync, \DDcont: Tendency to attribute agency to  bots (not people) might be countered by greater synchronicity, and contexts ingraining bot-human links.
                    \end{compactitem}
                    \end{minipage}\vspace{\rSpRev}
            & \vspace{\rSp} \hspace{\tabASmpCor}
                    \begin{minipage}{\tabASmp} 
                    \begin{compactitem} 
                        \item[\bulletQuest] \DDdynsync: Will more directly interactive social touch enhance \ESSbot's ability to achieve and vary co-presence? 
                    \item[\bulletQuest] \DDdynsync, \DDscale, \DDint,  \DDcont: How can we encourage attribution towards remote users rather than agency of the bot itself?  
                    \end{compactitem}
                    \end{minipage}\vspace{\rSpRev}
        \\ \midrule      
        \vspace{\rSp}
        \textit{\textbf{\aCoordP}}
        \vspace{\lSp}\newline 
        \textbf{Coordination \newline \& Participation}
            & \vspace{\rSp}\hspace{-0.29in} 
                    \begin{minipage}{\tabASmp} 
                    \begin{compactitem}

                        \item[\bulletPlus] \Erep, \Eobs: Desire to use \ESSbots\ to coordinate emotionally with others (\eg communicate feelings, ``I'm with you'').
                        \item[\bulletPlus] \Eobs: Reactions and comments demonstrating participants' valuing of autonomy and consent.
                    \end{compactitem}
                    \end{minipage}\vspace{\rSpRev}
            & \vspace{\rSp} \hspace{\tabASmpCor}
                    \begin{minipage}{\tabASmp} 
                    \begin{compactitem} 
                        \item[\bulletPlus] \Eobs: Ability to communicate (via message exchange).
                        \item[\bulletPlus] \Eobs: Co-created  norms.
                        \item[\bulletPlus] \Eobs: Mutually suspended disbelief (ascribing agency to \ESSbots) indicated high participation; 
                        ~\iconMinus~unchecked,  could weaken co-presence.  
                        \item[\bulletQuest] 
                        Small groups, messaging style and fixed contexts and group composition made \aCoordP\ hard to assess.
                    \end{compactitem}
                    \end{minipage}\vspace{\rSpRev}
            & \vspace{\rSp} \hspace{\tabASmpCor}
                    \begin{minipage}{\tabASmp} 
                    \begin{compactitem} 
                        \item \DDdynsync, \DDscale: Coordination opportunity and bot-member associative strength should grow with synchronicity, group size and activity variety.
                        \item \DDcont, \DDint: New contexts, \eg game environments, offer ways both to motivate and contextualize use, and to study users' dynamic behavior.
                    \end{compactitem}
                    \end{minipage}\vspace{\rSpRev}
            & \vspace{\rSp} \hspace{\tabASmpCor}
                    \begin{minipage}{\tabASmp} 
                    \begin{compactitem} 
                        \item[\bulletMeas] Improve measures for both \textit{coordination}, \textit{participation}.
                        \item[\bulletQuest] \DDint, \DDscale, \DDcont: How do different interaction modes, system scale and designs, and context of use affect coordination and participation? 
                    \end{compactitem}
                    \end{minipage}\vspace{\rSpRev}
        \\ \midrule      
        \vspace{\rSp}
        \textit{\textbf{\aEffortEng} } 
        \vspace{\lSp}\newline 
        \textbf{Elastic Effort \newline of Engagement} 

            & \vspace{\rSp}\hspace{-0.29in} 
                    \begin{minipage}{\tabASmp} 
                    \begin{compactitem}
                        \item[\bulletPlus] \Erep: Desire both for features that allow investment for some benefit (\eg programming a custom behavior) and low-barrier usage (\eg pre-programmed behavior). 
                    \end{compactitem}
                    \end{minipage}\vspace{\rSpRev}
            & \vspace{\rSp} \hspace{\tabASmpCor}
                    \begin{minipage}{\tabASmp} 
                    \begin{compactitem} 
                        \item[\bulletPlus] \Eobs: Participants successfully implemented and used features with different levels of required effort.
                        \item[\bulletQuest] 
                        Message-passing style and constrained contexts limit generalization of interaction effort  and engagement.
                    \end{compactitem}
                    \end{minipage}\vspace{\rSpRev}
            & \vspace{\rSp} \hspace{\tabASmpCor}
                    \begin{minipage}{\tabASmp} 
                    \begin{compactitem} 
                        \item \DDint, \DDdynsync:  Increasing effort range with high/low-effort \textit{synchronous} (as well as asynchronous) interactions can meet user needs and allow study of its impact.
                    \end{compactitem}
                    \end{minipage}\vspace{\rSpRev}
            & \vspace{\rSp} \hspace{\tabASmpCor}
                    \begin{minipage}{\tabASmp} 
                    \begin{compactitem} 
                        \item[\bulletMeas] Improve quantification of \textit{engagement effort} and benefits.
                        \item[\bulletQuest] \DDdynsync: Can high-synch social touch afford greater {range} in interaction effort? 
                        \item[\bulletQuest] \DDint, \DDscale, \DDcont: How do interaction style, group size  and context impact  effort?  
                    \end{compactitem}
                    \end{minipage}\vspace{\rSpRev}
    \\ \bottomrule
    \vspace{0.1em}
        \end{tabular}
    } 
\end{table}

    %
            %



\subsubsection{Social Touch (\hilite{\aSTouch)}} 
    \hspace{1ex} 

\lowHeading{\ESSbots\ for Feelings \vs Information}
Participants' suggestions about \ESSbots' potential roles align with the theoretical conception of social touch as regulatory \cite{schirmer:2022:socTouchAsTool} and affiliative~\cite{cascio:2019:socialTouchHumanDev} rather than  conveying information. 
They pointed out that existing technology (texting, calling) is suited to communicating ``\textit{direct information}'' (FP8) and felt that social touch for that purpose would lead to misunderstandings. 
Conversely, they recognized limitations in existing technology for communicating empathy (P5), closeness, or intimacy (FP8), and the suitability of social touch for this role. 

\lowHeading{\ESSbot's Achievement of Social Touch; and in an Affiliative Form}
Individuals' ability  to attribute touch to an embodied other seems to unlock  affiliative benefits of social touch (\S\ref{sec3:framework:affordances}), such as 
    social bonding~\cite{morrison:2010:skinSocialOrgan,morrison:2016:affectiveSocialTouch} 
    and stress reduction~\cite{coan:2006:lending}.
We found evidence that \ESSbots' social touch affordance is sufficient for this attribution to occur, \ie that users could achieve a sense of social touch with their remote partners. 
For example, participants discussed issues of consent and bodily autonomy for the robots as they would if they \textit{were} ``physically touching''(\S\ref{sec5:workshops:findings:consent}), and also tended to interpret robot behaviours involving direct human-robot touch as affiliative.

\lowHeading{Would a more powerful form of touch be more affiliative?} 
In this implementation, social touch took the form of one or more robots tapping or moving against participants' hands in simple coordinated patterns. 
While participants described  these touches as evocative and expressive of playfulness (P11) or social support (P5, P13), they also recognized that mediated touch was not the same as in-person touch and at this point perhaps not as effective (P2).

Systems which allow for more nuanced MST might amplify affiliative benefits. 
Future \ESSbots\ studies could test whether more directly bidirectional, immediately responsive social touch, and richer interactions and context  strengthen potential affiliative associations 
    (\hilite{\DDdynsync, \DDperf, \DDint, \DDcont}).

\subsubsection{Variable Co-Presence (\aCoPres)}
    \hspace{1ex}
    
\lowHeading{Interest in Varying Co-Presence, and Ability of \ESSbots\ to Support it}
Participants repeatedly confirmed desire to vary co-presence. 
At the outset (Figure~\ref{fig:sec5:social_media_use}), they reported using existing DSM for interactions both high in co-presence (\eg talking with friends and family) and low (sharing videos and memes, staying updated with others' lives). 
In  workshops, they expressed desire for more, and appreciated that the \ESSbots\ seemed to help their friends (``ˆ'', \ie `thumbs-up', P15), especially when comforting was needed (P5). 
However, they also wished to regulate it, \eg when they emphasized the importance of consent, and suggested context-dependent interaction modes (such as those captured in \S\ref{sec5:workshops:findings:PScenarios}; (\hilite{\DDcont, \DDint})).

\lowHeading{Attribution of Robot Touch to a Remote Other} 
Participants' interpretation of bot identity was flexible, 
sometimes referencing them as objects, and other times 
appearing to attribute \ESSbots' physical multiples to  remote group members' physical selves to some degree  (\S\ref{sec3:framework:affordances:copresence}); 
This increased over time, with growing use of possessive pronouns and especially when interpreting behaviours sent by others (\S\ref{sec5:workshops:findings:conceptualize}). 
They reported that the identity of the robot's human controller was important: their perception of a touch would change based on if that person were a stranger or friend, implying  specific and emotionally salient associations (\S\ref{sec5:workshops:findings:socialdynamics}).
When attribution occurred, its form seemed to be of \ESSbots\ as \textit{extensions} of others' bodies. 
It seems likely that the physical multiples, via coordinated, evocative activity, added to the sense of co-presence (a single shared object would not elicit these verbal references).

Future studies with more nodes (\hilite{\DDscale}) and continuously synchronized bidirectional interaction between groups (\hilite{\DDdynsync}) would help validate the extent to which participants view \ESSbots\ as embodied extensions of their friends. 
Likewise, \tST\ theory and our findings there point towards a tangible medium further deepening attribution over a graphical one (\hilite{\DDcont}), a premise that needs future empirical validation.


\subsubsection{Coordination \& Participation (\aCoordP)}
    \hspace{1ex}

\lowHeading{Basic Inter-User Coordination}
Even in this initial \ESSbots\ implementation 
(which supported message-exchanging rather than continuous node synchronization), we still observed some basic coordination, \eg in activities where emotional messages were sent and interpreted.

We wonder what spontaneous, non-conscious and dynamically shifting  patterns could tell us about \aCoordP,
further motivating a bidirectionally responsive implementation which would enable them to emerge 
    (\hilite{\DDdynsync}).


\lowHeading{Perception of Inter-Robot Coordination and Agency}
Participants \textit{perceived} a degree of coordination in swarm behaviour which was not user-intended;
\eg interpreting a case of bots colliding (due to a programming error) as a ``disagreement'', thereby shifting their own sense-making about the entire interaction. 
Such examples indicate that at least under some conditions, the \ESSbots\ themselves can play a social role outside of what any group member intends.
    
We also saw users more deliberately co-creating an interpretation of agency in the \ESSbots\, \eg when describing the robots' as ``playing'' or ``protecting'' each other.
This hints at a perception that the robots are tending towards internal norms (to seek fun and avoid danger; \S\ref{sec5:workshops:findings:conceptualize}), 
despite awareness that the robots do not actually ``care'' about either.
In suspending that belief, users allow the robots to impact (\ie participate in) the group's joint sense-making.

Perception of robot agency may undermine user ability to fully attribute the bots to remote users. 
The balance in this attribution (between robots as their own selves, and as avatars, extensions or embodiments) could likely be influenced through design, \eg balancing direct user control and autonomous behaviours, or occasionally using the voice of a user to reinforcing the attribution between themselves and their robot
    (\hilite{\DDint, \DDcont}).

\lowHeading{Co-Creating ESSbots Social Norms through Participatory Sense-Making}
Participants' preferences and feedback make it clear that they experienced their \ESSbots\ use as participatory interaction, beyond simple communication. 
For example, their use case of comforting others (\S\ref{sec5:workshops:findings:PScenarios}) indicated awareness that they might impact others. 
This behaviour is motivated by an external norm --- the desire to offer support to someone in a difficult real-life situation.

However, users sometimes also respected ``norms'' that do not exist through \ESSbots\ in the same way that they do in regular life ---
for instance, in  respecting bodily autonomy via proper consent. 
With physical co-location, this norm stems from physical and psychological safety concerns; the former is largely absent with \ESSbots.
A \textit{psychological} concern implies a strong social interpretation of who is touching, who is being touched, and what the touch means.

This transfer of a real-life norm to a virtual environment exemplifies co-creation through participatory sense-making. 
When faced with the new environment posed by \ESSbots, participants emergently and jointly realized that their robot's bodily autonomy matters, and that their human boundaries should be respected by others' robots despite altered context. 
This joint sense-making is key to what makes an interaction participatory and social under \tINT.

\subsubsection{Elastic Effort of Engagement (\aEffortEng)}
    \hspace{1ex} 
Comments indicated desire to control the effort of interacting through a \FWname\ system; and how some \ESSbots\ features afforded this. 
On one hand, social touch offers a low-effort avenue for interactions like emotional support, especially for those unaccustomed to speaking about emotions (P5). 
This aligns with \tINT\ and Schirmer's ``nonverbal effecting model'' of touch: certain kinds of interaction (like sharing a social touch) cannot easily be rendered as information, and \aSTouch\ eases that burden. 
Conversely, participants also appreciated occasional high-effort engagement (\eg time spent programming a complex behaviour), seeing it as more meaningful than texting or calling because \enquote{they actually had to do something to tell me this} (P13). 

It seems clear that access to a range of interaction modes will be important for fulfilling \aEffortEng, from supplied, pre-programmable and triggerable (here), to (future) live, responsive behaviours 
    (\hilite{\DDdynsync}). 
We further conjecture that burden and its control could also be managed when this technology is used in balance with others, like texting and calling. 
Future work could examine this more directly by comparing ways of integrating or moving between interaction modes and quantifying relative cognitive and physical workload 
    (\hilite{\DDint, \DDcont}).

\section{Next Steps}
    \label{sec7:next}

Above, we used our design and workshop experience to assess two key \FWname\ propositions, working backwards towards the start: the degrees to which
 (\S\ref{sec6:discuss:requirements}) this initial \ESSbots\ \textit{design manifestation} meets our proposed \textit{requirements on representation};
and  (\S\ref{sec6:discuss:affordances}), those requirements provided four \textit{proposed affordances}, and teens made use of them.
We found successes and shortfalls at both steps. 

What comes next?
Here, we examine 
    the \FWname\ technological and social design space (of which this first platform inhabits just a corner), including how future \ESSbots\ versions might better satisfy requirements;
    practical considerations;
    priorities in \FWname\ validation; 
    and how all of the above can relate to the many research communities who are needed to participate in this approach.

\subsection{Fostering Embodied Social Swarms:
    Expanding the \FWname\ Design Space} 
    \label{sec7:next:expanding}


The \ESS\ technology examined here  supports expressive swarm-behaviour-exchanging  interactions within a small group. 
It is an evocative instance of a larger design space of embodied and tangible interactions for group communication and online connection. 
For the same small group of friends, greatly altering any of many dimensions could upturn the interaction; and  group composition will likely influence use of a given implementation. 
We explore a few of these.

\subsubsection{Technological Form \& Capabilities (\DDperf, \DDscale)}
    \label{sec7:next:expanding:tech}

What if an \ESS\ were a literal hive of airborne, bee-sized elements that follow each member,  buzzing around them and disappearing into small wearable hive when inactive? 
Such a version would become a {portable}, physical, augmented-reality presence of the distributed social group.
Such a vision has many dimensions to unpack, which we consider just for swarmbot-type technology; there  may be other hardware approaches that can also support \FWname\ requirements.

Meanwhile, putting technical aspirations in context: highly realistic abilities are probably not the right standard. 
We saw many instances of teens ``filling in the gaps'' between \ESSbots' abstract movements and the social touch interactions of real life; expressive, fun and inexpensive will count more than perfection. 

\lowHeading{Size and capability of bots and swarms}
What is the  ``ideal'' bot size?
If  tiny bots could fly and exert tiny pressure or impact forces, perhaps we would want many at a node (regardless of human group-size), and interactions which engaged them more as a cloud. 
For bots that are table-bound and with numbers in parity to humans, smaller size will generally mean reduced generative force and weaker interactions. 
Current swarmbot platforms typically support $\sim$20–30 units.
Small, numerous bots would enable different kinds of mappings and interactions, but require advances in coordination, communication and cost.
Larger bots could be stronger, but need more space and power. 

We also imagine new ways for bots to 
    \textit{move} (fly, jump) 
    with greater \textit{awareness} of their local space, and the user's body; 
    and \textit{push} (much harder and in more coordinated ways).
Small bots might join in more powerful conglomerates.

\lowHeading{Bot form and uniformity}
Our participants found both tangible embodiment and ability to express identity particularly beneficial to feel close to remote friends;
and we note that individualization of bots is better afforded at  sushi than bee size. 
New designs could relax {constancy and uniformity of  physical form}, in appearance, shape or materiality.

\lowHeading{Haptic readiness}
For small, desktop-scale mobile robots, mechanical trade-offs emerge between speed and force output: achieving higher speeds often requires higher gear ratios, which in turn reduce force output. 
Moreover, the bots’ small size and low mass limit traction, constraining their ability to deliver significant haptic feedback. Prior work has explored magnet-based approaches to increase effective force~\cite{kim:2019:swarmhaptics}, but such solutions add  environment constraints and reduce deployment flexibility. 
Meanwhile, greater omnidirectional mobility will support more fluid, responsive interactions through rapid, orientation-independent movement, but  may also increase mechanical complexity and typically require larger form factors to accommodate additional components.

\lowHeading{Sensing} 
%
\ESSbots\ should operate robustly in unstructured, everyday environments, supporting both collision avoidance and context-aware interaction (\eg leveraging environmental features during engagement).
This requires the ability to perceive the human user's body, other robots and surrounding objects. 
Eventually, we imagine \ESSbots\ that can infer their user’s emotional state.

\lowHeading{Supporting a larger (human) group size}
Interaction styles will naturally change as a friend group grows; we imagine important jumps from dyads to triads, to small and then larger groups, with drops in intimacy and shared understanding at each but also in the technical complexity of the system.
For example, group size might tend to shift the balance between users wanting to engage in live interaction 
and broadcast messaging, if the former becomes incoherent.
Conversely, shared-manipulation of a swarm might become more fun and  rich with groups of 10 to many more. 
At an extreme,  {broadcast} communication of swarmbot behaviours to {a large number} of global users who {do not know one another at all} would be more akin to current social media platforms --- an embodied swarmbot TikTok or Instagram.

\lowHeading{Realtime networking} 
The network layer connecting group members must support low-latency, high-throughput communication to enable real-time synchronization across distributed deployments. 
Minimizing lag is critical for maintaining the temporal coherence of interactions and preserving a seamless, shared experience among users.

\subsubsection{Interaction Synchronicity (\DDdynsync)}
    \label{sec7:next:expanding:synchronicity} 

We often think of synchronous interaction as meaning simply ``at the same time''. 
Our current prototype is {synchronous} in the sense  of immediate message-exchange (as is texting in current DSM).
At even this level, there are architectural and usage-related  decisions; and there are nuanced levels beyond it.

\lowHeading{Asynchronous communication}
A system's practicality could be increased by relaxing the need for everyone to be online and available at the same time.
It also gives more latitude for crafted `media'. 
While we imagine that dynamic synchronization (below) will increase the potential for a sense of connectedness, this might be compensated for with greater offline creative attention.

\lowHeading{Dynamic synchronization}
    \label{sec7:next:expanding:responsivity} 
To fully activate most \FWname\ affordances and leverage the potential of a physical robot swarm, we may need  physical, bidirectional interactivity across nodes, maintained through continuous synchronization.
Instead of  User A simply watching the same robot behaviour at roughly the same time as User B, or interacting with local, non-synchronized copies,  
A feels the robot move \textit{as B moves it}, and B feels A directly push back on them.
This is a higher technical bar; what happens when it's met?

%
However, exact temporal-kinematic matching of all local nodes is inherently impossible:
local users and environments will constrain physical avatar movements differently in each location.
This imposes \textit{physical} elasticity to co-presence, \eg  a local bot  connected by a virtual string to its virtual intended location. 


\subsubsection{Motivation: Context and Subversion (\DDcont)}
    \label{sec7:next:expanding:motivating}

Our teens used the \ESSbots\ because we asked them to, in imposed groups and structured activities. 
We guessed and teens confirmed that this medium is not ideal for conveying high information density, and won't replace more semantically suited channels.
Why will they use it in everyday life?

\lowHeading{Games: for fun \ldots\ and a study sandbox}
\ESSbots\ are supposed to afford fun.
Games offer a motive and narrative for interaction.
Furthermore, games could be a natural 
way to design towards most \FWname\ affordances: they generally direct \aCoordP,  and could be thoughtfully constructed to evoke others.
In fact, games may be an ideal mechanism through which to \textit{study} the impact of varying these affordances on the social-good outputs we are interested in~\cite{allen:2024:games}.

\lowHeading{Designing for subversion}
%
We imagine that teen acceptance and adoption of a new DSM technology would be greatly amplified if seen as teen-only territory.
This dynamic can be seen in systems like MineCraft~\cite{ringland:2016:minecraft}, whose high and persistent engagement with young users and associated online communities likely stems not only from the engaging activity itself (with a low barrier to access creative, fun construction), but also as a place to build and share without grownups around. 
We envision \ESSbots\ tapping into similar creative and adult-free-zone urges, but with greater capacity for emotional connection. 
Achieving this will depend on the interface by which users control the swarm and the context by which users are introduced to and access it (as noted below), as well what the system can do.


\lowHeading{Group diversity and familiarity}
In this first cut, we recruited teens independently and with age and local availability as the only eligibility factor, then asked them to work with strangers over multiple workshops. 
There are, however, many kinds and purposes of groups, as well as individual experience within a group.

Our current system allows teens to draw on their joint history in creating an embodied language, and perhaps to increase their comfort level in the intimacy of physical social touch.
It will be important to understand how interaction and commensurate support of emotional connection changes as  group composition varies (\eg in many axes of diversity, shared history, size and specific preferences, styles and tendencies).

\lowHeading{How individuals navigate within a group}
Individual identity will influence both personal and group expression.
For example, what happens when a robot that one member identifies with becomes co-opted by another in a group action), communicating the need for distance or privacy?
When does consent matter?
What is the role of synchrony or asynchrony in communication?

\subsubsection{Interfaces: Providing Access to the Technology (\DDint)}
    \label{sec7:next:expanding:interface}

The  many potential ways teens might control an \ESS\ generally fall into two interface categories: 
    visual (behaviours can be programmed and sent via a phone or computer); 
    or tangible (bots are controlled by direct manipulation
    or piloting, \eg with a 
    video game controller). 
We built a two-mode visual interface: participants could
    (1) click buttons to send system-made (or their own)  behaviours, or 
    (2) create custom behaviours via simple visual scripting.
While more details regarding the interface design are reported elsewhere, the results presented above give insight into what it supported.
This interface allowed teens to express themselves and explore the system's potential, in that they could communicate emotions and establish co-presence behaviour sharing.
Participants chose different control methods based on scenario, and wanted to vary effort of engagement.

We infer that interfaces that are extendable, support different control methods and encourage teens to customize and personalize them will support the goal of \ESSbots\ as a system for subversion.
Future work should investigate alternative 
    \ESSbots\ interfaces and control methods, and 
    the use-cases and affordances that they enable.

\subsubsection{Situating Within Teen and Other Communities}
    \label{sec7:next:expanding:otherCommunities}

Moving from a platform that supports good things to one that gets  adopted could happen by design or serendipity or both (if at all). 
Beyond providing the right contextual and motivational furniture for a given community within an \ESSbots\ platform, and addressing nontrivial technical obstacles, groups need to learn about it, be able to access it and want to share it with their friends. 
Our instinct is to avoid schools and other `adult-run' environments, and to find a model more like Minecraft (\S\ref{sec7:next:expanding:motivating}) for dropping into a teen (or other) ecosystem.

We imagine that \ESSbots-like technology could find a place in other groupings; in fact, many places where distance and social norms breed isolation. 
What would \ESSbots\ \textit{within} a senior centre look like, versus a system oriented at connecting seniors to their distant, younger family members?

\subsubsection{Steering away from Potential Negative Effects}
    \label{sec7:next:expanding:negatives}

Making some of the jumps outlined above will be technically nontrivial, and are not guaranteed to be interesting, engaging or practical. 
It is also important to consider their social impact and strive to minimize negative effects:  
we do not want to reproduce the pitfalls of existing DSM by inadvertently promoting artificial or performative expression (a risk of broadcast formats) or negative self-judgment and peer comparison, and undermining genuine connection. 

We must also consider what \textit{new} forms of harm social swarm robots could afford. 
For example, the physicality of tangible robots could make purposefully negative communication feel more visceral.
Violations of established physical boundaries may be more upsetting for  a physical, embodied presence in a person's space.

\subsection{Future Viability and Practicality of \FWname\ Technology}
    \label{sec7:next:practicality}

We did not focus here on making \ESSbots\ practical,
but can  speculate about pathways and obstacles to practical systems.

\lowHeading{Technical and commercial} 
Swarmbots' dependence on physical bodies with sensors and actuators mean that they will always cost more than a purely graphics-based system.
Swarms with simplified physical capabilities leveraged by complex intelligence may be most likely to breach these barriers.

\lowHeading{Social}
Along with many positive qualities, strong mappings of local physical avatar to a specific remote friend impose  limitations, not least a potential for social exclusion. 
More flexible mappings and low-cost (possibly low-capability) individual bots will contribute to more flexible human-group membership.
We imagine a  social practice of gifting a physical, perhaps personalized bot as a symbolic offer of membership to a friend group.

\lowHeading{Fitting into users' lives}
Adoption  will be facilitated by congruence with  how users live.
Practical access spans convenience and contextual availability as well as cost: ``always with you'' \vs ``at home by table''.
The surface or space on/in which the \ESSbots\ operate matters, \eg restricted to a specific location, or varying degrees of portability.

\lowHeading{Ecological sustainability}
A rapidly-advancing physical technology routinely creates waste. 
in contrast to a graphical app (typically adding a marginal  energy burden as it leverages an existing physical device, \eg a phone or laptop), acquiring then updating robots often means putting the old ones into landfill.
We are acutely aware of this conflict. 
We hope that an idea like personal swarmbots will eventually take hold within very simple, re-usable devices that have a chance of respecting crucial ecological requirements.

\subsection{Further Validating the FOSTER Framework}
    \label{sec7:next:FWvalidation}

We first consider what it will take to comprehensively validate the proposed \FWname\ affordances,   
then step back to consider next steps based on the theories which we drew on to postulate this framework. 

Table~\ref{tab:sec6:affordanceSummary}'s \textit{Is Important to Users} column offers preliminary validation to our proposed affordances: every affordance was appreciated by teens (\iconPlus), including in ways we did not predict, such as teens' desire for \ESSbots\ to protect autonomy (\aCoordP). 
We also observed (the \textit{Is Captured Here} column) that all the affordances were observed being utilized to some extent (\iconPlus).
This suggests that this affordance set is reasonable to retain as we further explore this path.
It does not conclusively tell us if other affordances might supplement them, although our analyses did not unearth obvious gaps.

A comprehensively validated affordance set will require more investigation, such as for more targeted manipulation and measuring longer-term hypothesized benefits (\iconMeas).
Furthermore, there are hints that the proposed affordances were \textit{incompletely} activated:
the question marks (\iconQuest) in Table~\ref{tab:sec6:affordanceSummary} indicate gaps that obstructed what we could conclude.
On one hand,  stronger effects should be possible; 
but on the other, we have few observations about a large part of the design space and cannot predict influence without a more performant \ESSbots\ version.

The gaps stem from three primary limitations highlighted in Table~\ref{tab:sec6:affordanceSummary}'s \textit{Could be Amplified} column, which collectively involve  design dimensions associated with ongoing technical implementation and study design, including support of realistic social context 
(other design elements, including \DDperf, can certainly be improved but are not obviously impeding affordance analysis at this point):
\begin{compactitem}
    \item \DDdynsync: Better support for live (continuously synchronized) interaction     (especially important for \aCoPres);
    \item \DDscale: Larger group size and more natural composition; 
    \item \DDcont: Compelling motivation and context for interaction.
\end{compactitem}

\noindent 
We see \aCoordP\ as perhaps the most critical priority for better support, 
both because it is probably the most hampered (due to the current message-exchange platform limitation) 
and because once available, alongside \aSTouch\ it may be most powerful in unlocking unique \FWname\ benefits.

Our theoretical foundations, especially \tINT, also support the importance of these three design dimensions. 
A central \tINT\ claim is that social cognition goes beyond passing information back and forth; \ie social interaction cannot be captured as message-exchange alone (calling for \DDdynsync). 
\tINT\ and \tDC\ challenge us to consider cognitive processes which exceed the boundaries of a single agent, and at the next step, those which 
cannot be fully explained by observing \textit{two} agents (\DDscale). 
Finally, \tINT\ and \tAN\ advocate for the consideration of elements aside from humans in social cognition; this can include contextual rules and goals that organize and motivate behaviour (\DDcont).

\subsection{Implications for Relevant Research Communities}
    \label{sec7:next:resCommunities}


This  mixed  team joined up because we each found something exciting about \FWname\ and \ESSbots\ --- different things. 
Having established that there \textit{is} something  here, we can build from this starting point to further explore many of those directions, both back in our own fields and as a team.

\subsubsection{Engineering and Computer Science} 

\S\ref{sec7:next:expanding} details many technical innovations that  would both expand \ESSbot\ capability, and facilitate research in \FWname\ potential. 
These in particularly include swarm robot and haptic mechatronic engineering, novel sensing, network architecture and realtime systems; ideally conducted in coordination from sociology and HRI/HCI to advise on implications of form, scale, capabilities and communications.


\subsubsection{Cognitive Science}

Access to \ESSbots\ research platforms may expose new implications 
for the grounding theories that inspired them. 
For instance, \tAN\ and \tDC\ indicate that physical objects like \ESSbots\ can play a role in social cognition. 
Our workshop findings support this: users saw individual bots both as extensions of  remote humans, and sometimes as ``themselves'' (socially-active agents). 
Controlled experiments with  manipulated mirroring could elucidate when users can distinguish computer-driven \ESSbots\ behaviour from human-controlled movements. 
If so, what are the socio-cognitive implications? 
Can co-presence (as defined by \tGCP) be achieved with a computer agent?

\tINT\ posits that in a group interaction, emergent group-level processes can take on a life of their own and influence behaviour beyond the influence of  individual participants.  
\ESSbots\ can support further inquiry into this hypothesis:
mathematical analyses like integrated information decomposition~\cite{rosas:2020:iid,mediano:2019:iid} applied to physiological measures of participant biometrics in response to  swarm-level motion characteristics can quantify the presence and impact of emergent group-level social processes.

Finally, in the domain of social cognition as construed by \tINT, there are ongoing empirical efforts to bridge externally-measurable data, and subjective experience only accessible through self-report~\cite{berkovich-ohana:2020:neuropheno}. 
\ESSbots\ could be used to reduce complexity here, \eg by looking for  recurring trends in \ESSbots\ users' self-reported experience, and their correspondence to measurable patterns in swarm movements. 






\subsubsection{Developmental Psychology}
%

By centering embodied, affective interactions and meeting teenagers where they are (\ie in digital spaces), \ESSbots\ are a promising avenue for both research and clinical practice. 
Broadly, they can provide a controlled environment for isolating and examining components of social communication. 
This could help identify the elements most critical for social development and clarify the ways adolescents build relationships. 

For example, compared to face-to-face interactions, social media may be more accessible to adolescents with internalizing psychopathology (\eg social anxiety) because it reduces demands for immediacy and real-time integration of multiple and often ambiguous or even contradicting communication cues ~\cite{Angelini:2024}.
\ESSbots\ could be  be used to investigate social functioning outcomes in response to differences in input complexity and immediacy demands, in adolescents with and without internalizing psychopathology. 
Clinically, \ESSbots\ may represent a ``middle-anxiety'' context and could be used as a step in exposure interventions for social anxiety. 

Finally, digital and in-person relationships are not as separate as they are often conceptualized in policy and research~\cite{Evans:2025}. 
Adolescents may be particularly sensitive to cultural shifts, and especially receptive to new technologies~\cite{Blakemore:2014}. 
\ESSbots\ could represent a vehicle to facilitate the integration of the physical and digital world.

\subsubsection{Human-Computer and Human-Robot Interaction} 

We anticipate that HCI/HRI researchers will continue to be needed at the centre of a rich disciplinary soup.
They will ultimately determine the future of this new form of telepresence and connection, so different from the current commercial product landscape: 
embodied, group-focused, and about social-emotional support rather than task-based.

They will help and learn from psychology investigations and push strategically on technology boundaries, meanwhile working closely with users themselves in the differently-formatted evaluation forms noted throughout, from controlled studies exploring the impact and need for different levels of synchrony to longitudinal examinations of longer-term evolved behaviour. 
They will follow up on our teens' excitement about personalization, customization, boundaries and consent, and how to give users access to both choreography and spontaneous fun.
Importantly, they will investigate \textit{context}, learning how to fill the hybrid \ESSbots\ ``playground'' with engaging environments and structures.

\subsection{Conclusion}
    \label{sec7:next:Conclusion}

We began by
    proposing a link between two observations: 
    (1) reduced opportunity for social touch among friends in the predominant social environments of young people;
    and 
    (2) documented growth in teen isolation and loneliness.
This link is based on a large body of research that ties affiliative social touch to empathy and connectedness, and thence social functioning and longer-term resilience, but which has mostly overlooked adolescents. 

We wondered if \textit{integrating social touch into group digital media} could use this same link in a positive way.
Drawing from a multidisciplinary team's respective theoretical backgrounds, we distilled design affordances that should access healthy outcomes predicted by those theories.
With teen co-designers, we developed an engaging swarm robot platform and interface, and in this collaboration found evidence of this \ESSbots\ implementation exercising the affordances, and user behaviour aligned with what the founding theory predicts.
This evidence \textit{begins} to validate the ``flow''  posited by the \FWname\ framework: from affordances to representation to social benefits from group use.

Further validation is needed.
An exciting design terrain has been revealed, and  deeper explorations require platform advances, some achievable in months,  others more difficult. 
The technology we employed is far from either commercial or ecological readiness. 
But we found evidence among our teen partners of both a hunger for touch, and signs that these little robots might be able to fill the gap to some degree. 

Finally --- teens are not the only ones who tend to be isolated today, and who might not be getting enough hugs.

\begin{acks} 
This work was substantially funded by the New Frontiers in Research Fund (Award \#  NFRFE-2021-00592021-0059), as well as student fellowships provided by 
the Natural Sciences and Engineering Research Council of Canada (NSERC, AWD-010773).
Human user research was conducted under UBC Ethics \#H22-01554. 
\end{acks}

\bibliographystyle{ACM-Reference-Format}
\bibliography{bib_EssbotsDesignValues.bib}


\newpage
\appendix
    \label{sec:appendix}

\section{Appendix: Conceptual Model Terms and Descriptions}
    \label{app:CMtable}


\begin{table}[bh]
\footnotesize
\centering

    \caption{\textbf{Concepts in an ESS-supported Social Network:} 
        key system elements as we need members to think of them in order to facilitate a perception of social group embodiment, and which the system's technical design needs to support.
        Attribute notation loosely follows UML, \eg [1..*] means 1 or more, [1] exactly one. 
        We specify downward associations but omit reverse associations when fairly obvious (\eg listed: \textbf{groups} have [3:n] \textbf{members}; implied: \textbf{members} belong to one \textbf{group}).
    } 
        \label{tab:app:conceptsElaborated}
    \vspace{-1em}
    \includegraphics[width=0.84\textwidth] 
        {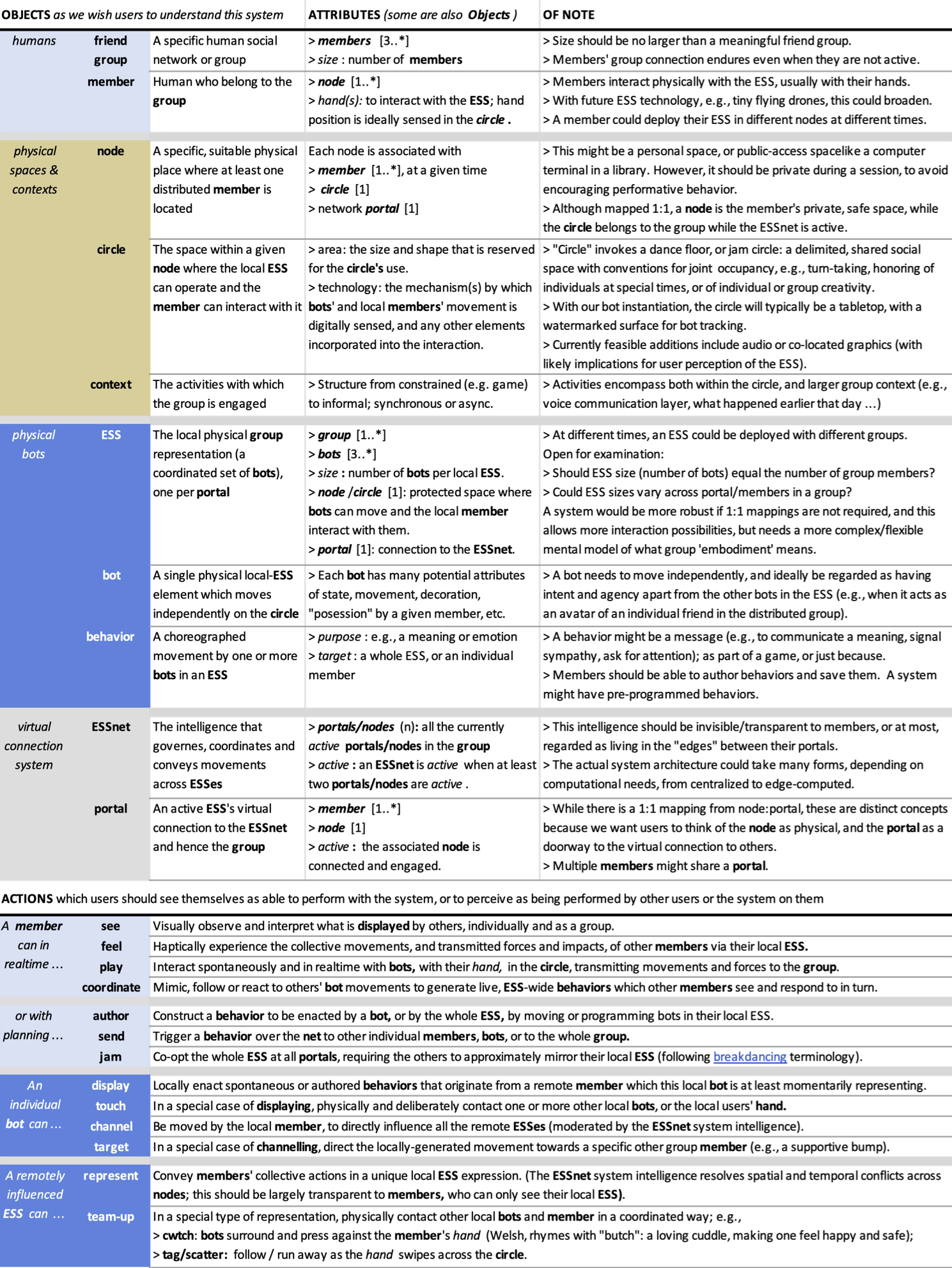} 

\end{table}

\newpage
\section{Appendix: ESSbots behaviour Authoring Interface}
    \label{app:behaviour_interface}

    \begin{figure}[h]
        \centering
        \includegraphics[width=0.9\textwidth]{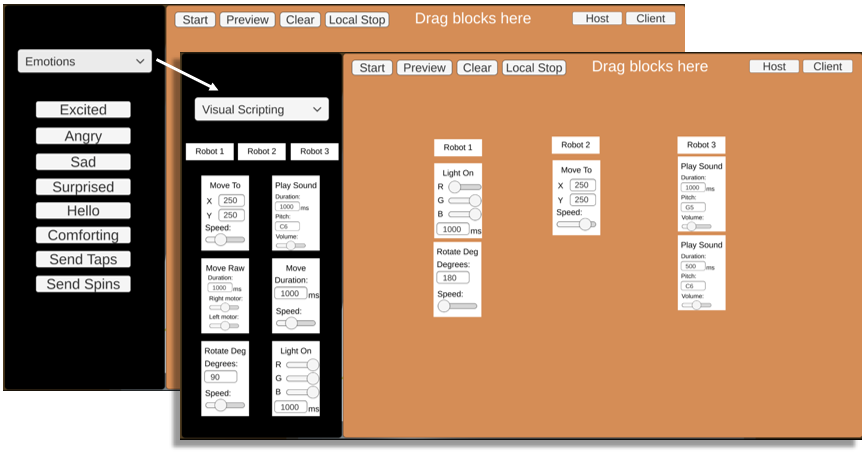}
        \caption{\textbf{Emotion Button and Visual Scripting Interface.} Users can work in two modes: Emotion Buttons or Visual Scripting, selected by a pulldown menu at upper left, to control the robots and send messages to others remotely after connecting over a local network with the host and client buttons.}
        \Description{ESSbots behaviour interface. There is a dropdown menu on the left, where users can select from ``Emotions'' or ``Visual Scripting''. When ``Emotions'' is selected, 8 buttons with pre-made behaviours become visible on the left sidebar. These include excited, angry, sad, surprised, hello, comforting, send taps, and send spins. When ``Visual Scripting'' is selected, draggable objects become visible on the left sidebar. This includes control blocks for each robot, and action blocks, such as various movement blocks, rotate, play sound, and light on.}
        \label{fig:sec5:scriptingInterfaceFinal}
    \end{figure}

\newpage
\section{Appendix: Detailed Methods: Workshop Series I and II}
    \label{app:detailedMethods}


\begin{table}[ht]
\caption{\textbf{Summary of all workshop activities analyzed in the paper.}}
        \Description{TODO}
        \label{tab:appendix:activitySummary}
    \footnotesize
	\centering
	{
        \begin{tabular}{p{2.0cm}p{2.5cm}p{8.5cm}}
            \toprule  
                \textbf{Workshop Theme}   
                 &  \textbf{Activity} 
                 &   \textbf{Activity Summary}\\
            %
        \toprule
        \multicolumn{3}{l}{\textbf{SERIES I}} \\
        \toprule
        \textbf{Formative Workshops 1 \& 2}
            & Zooid demo
            & Live demo showcasing single Zooid robot movement, followed by demo with two robots moving together. Follow-up discussion questions about participants' first impressions of the robots (e.g., \textit{What is your first impression of the robots?}; \textit{How do you think you could communicate by using the robots?}).\\
        \bottomrule
        & Mediated touch demo 
        & Live demo, where each participant felt a single robot bump and then rotate against their hands. Follow-up discussion question: \textit{How did it feel, and what are your overall thoughts about this aspect of the robots?} \\
        \bottomrule
        & Behaviour creation activity
        & In breakout groups of 3--4 people, participants took turns illustrating how they would use the robots to express simple emotions or phrases to their friends. Participants were given wooden robot proxies to move around the table to illustrate these robot behaviours. As a follow-up, participants repeated the activity after being instructed to think of one robot specifically representing them in the group. Participants then returned to the full group, and demonstrated one emotion and phrase to the other breakout group, who guessed what was being communicated.\\
        \bottomrule
        & Discussion questions
        & Group discussion about participants' experiences communicating messages with the robot shells, what aspects about the robots seem most interesting, and how this form of communication differs from the ways they currently communicate with distant friends (e.g., \textit{How would the robot communication style be different or similar to existing ways you communicate online? How would you use the robots with your friends? Do you see yourself using it for similar or different reasons than existing social media?}). \\
        \bottomrule
        & Customization activity
        & Participants were given several crafting materials (e.g., markers, pipe-cleaners, paper, feathers) and asked to customize one or more wooden robot proxies. They were then asked follow-up discussion questions about how they personalized the robots and how that might affect they way they would use them to communicate (e.g., \textit{How do you think that the design and the aesthetics of the bots would affect the way you use them to communicate with friends?}; \textit{Since your friends are distant, how would you want them to personalize their robot that represents you?}). \\
        \toprule
        \multicolumn{3}{l}{\textbf{SERIES II}} \\
        \toprule
        \textbf{W1 \newline Conceptualizing ESSbots}
        & Toio demo
        & Live demo with one robot, then four robots, showcasing different movement patterns, sound, light, rotation, and vibration/rattle. Examples of more complex and haptic behaviours were also shown (e.g., robots touching someone's arm, moving in a torus pattern). Follow-up discussion about participants' first impressions (e.g., \textit{What are your first impressions of these robots? (Anything you like or dislike? Do they remind you of anything?)}; \textit{Imagine you are using these robots to communicate with a group of friends who aren’t with you in person. What do you think you could communicate using these robots?}). \\
        \bottomrule
        & Discussion questions
        & Group discussion about participants' attitudes towards the robots, using the robots to communicate, and their conceptualization of atomic behavours. Example questions included: \textit{We talked a lot about using the robots to communicate emotions or relate to your friends. After using the robots and creating some possible behaviours, how do you feel about this goal? (e.g., was it easy to do? Why or why not?)}; \textit{Aside from trying to communicate emotions, are there other things you would use the robots for with your friends?}). \\
        \toprule
        \textbf{W2 \newline Customization \& Control}
        & ESSbots scenario discussion
        & Participants were given several scenarios related to friend or time availability, type of interaction, type of message, and physical location of use. For each scenario, participants were asked how they would use the ESSbots and why (e.g., to send a message or play a game), and how they might want to control them (e.g., programming or via a remote control). Participants were asked follow-up discussion questions about using the robots in different scenarios (e.g., \textit{Do you think ESSbots would be particularly useful for any of these
        scenarios? Why?}, \textit{Are there any scenarios here where you would prefer to use other tech? Why?}). \\
        \bottomrule
       
            \vspace{0.1em}
        \end{tabular}
        }

\end{table}

\begin{table}[ht!]
    \caption{\textbf{Summary of all workshop activities analyzed in the paper, continued.}}
    \label{tab:my_label}
\footnotesize
    \centering
    \begin{tabular}{p{1.8cm}p{2.0cm}p{9.2cm}}
            \toprule  
                \textbf{Workshop Theme}   
                 &  \textbf{Activity} 
                 &   \textbf{Activity Summary}\\
    \toprule
        \textbf{W3 \newline Social Dynamics}
        & Reaction and response with proxies activity
        & After splitting into two smaller breakout groups, participants took turns using robot proxies to demonstrate simple prompts (emotions or simple phrases). The other participants used their own proxies to respond or join in on the behaviour, and were told they could vary the behaviour at any point during the activity. Participants were then asked what behaviour they thought they were responding to, what they tried to show with their own proxy, and what they thought the final group behaviour represented. Participants were also asked other follow-up discussion questions (e.g., \textit{Do you think this way of communicating (reacting and responding) is similar to what happens in person when you’re with a group of friends? Why or why not?}).\\
        \bottomrule
        & Swarmness and individuality demos (`robot dance party', co-opting)
        & Participants responded to different use cases where they would have less direct control over the robots. This included a `robot dance party', where we demoed live robot coreography to music, and different co-opting scenario demos (e.g., one robot is sad, but gets co-opted into a happy group behaviour; one robot is happy, but gets co-opted into a sad group behaviour). Follow-up discussion questions after each scenario included: \textit{How would you feel if you did have less control over the robots?}, and \textit{Are there any instances where you wouldn’t want your robot to get co-opted?}. \\
        \bottomrule
        & Mediated touch demo
        & Live demos where participants felt different haptic patterns with single or multiple robots. Examples include simple taps and spins against participants' hand, as well as symbolic behaviours with multiple robots (hug, attention, notification). Participants were asked follow-up discussion questions about their experience with the haptic touches (e.g., \textit{How did you feel when the robots touched you?}; \textit{Are there any benefits or downsides to being able to use the robots as a proxy for something like an in-person hug?}).\\
        \bottomrule
        & Showing personality discussion
        & Participants given paper and markers, and asked to write down ``What makes you, you''. Without having to share what they wrote, participants were asked several discussion questions about how they would want to let others know who they are with the robots, and what ways individual robots can have personalities (e.g., \textit{How would you want to let others know that one of the robots is ``your'' robot?}; \textit{How do you want to reflect your personality through your robot?}).\\
        \toprule
        \textbf{W4 \newline Communication}
        & Showing consent demos
        & Several live demos showing different ways participants could use the robots to indicate if they did or did not want to join in on a behaviour, such as lifting up and setting down their robot to have it participate with the others. Participants were asked follow-up discussion questions about how they could use the robots in participatory behaviours (e.g., \textit{Can you think of any better ways to either join in or say no if you don’t want your robot to participate?}).\\
        \bottomrule
        & Workshop series reflection activity
        & Participants were asked to place sticky notes on a whiteboard to respond to various questions about their workshop experience. Each question had a column for ``What I thought before'' and ``What I think now''. Examples questions include \textit{What you think the robots can communicate?}, and \textit{To what extent do the robots make you feel closer to your friends?} \\
        \bottomrule
        & Survey about workshop experience
        & Participants completed a paper survey with closed- and open-ended questions about their workshop experience. Example questions included: \textit{Did you learn anything new in the workshop series? If so, what?}; \textit{On a 5-point scale from 1 (not at all) to 5 (a lot) How much has your perception of robots changed now that you have completed the workshop series?} \\
        \bottomrule
        & Breakout group communication
        & In breakout groups (in two separate rooms), participants took turns using our prototype to send messages back and forth with the robots. This included simple phrases and emotions, as well as free time to `play around' with the prototype. Participants were asked to guess what they thought was being expressed with the robots, and respond back---the other group then guessed what the response meant. Participants were also asked follow-up discussion questions about how it felt to have the robots touch them, usability of the system, and how they felt about communicating with the robots (e.g., \textit{Did you feel like you could communicate well - did the other group understand you? Was it easy to respond back? Why or why not?}). \\
        \toprule
        \textbf{W5 \newline Design \newline Generalizability}
        & Demo and discussion: tangible control
        & New and returning participants were shown a live demo of a tangible control method (participants could drag one robot around, and the other robots would follow it). They were asked follow-up discussion questions about what they thought of the new method, as well as which overall method was their favourite (e.g., \textit{Were you able to get the robots to do what you wanted using this method? If not, why?}; \textit{Do you think there are specific situations where you would want to use this control method? Are there any situations you would not want to use it? Why?}).  \\
        \bottomrule
        & ESSbots for social wellbeing discussion 
        & Group discussion about participants' attitudes towards wellbeing with respect to social media, and how the robots might support them. Example questions included: \textit{What specific things about these robots would help you feel connected or supported by your remote friends, or help you support your friends?}; \textit{How is communicating through the robots similar and how is it different from other forms of online communication?} \\
        \bottomrule
    \end{tabular}

\end{table}


\newpage
\section{Appendix: Complete Data Analysis Codebook}
    \label{app:codebook}

\begin{table}[!htp]
\caption{Complete list of codes generated as part of our qualitative analysis and whether they appeared in each workshop.}

\Description{Table with seven columns, one for the pilot workshop and each of the five workshops in our workshop series. The table has 58 codes, one in each row. Under each workshop column, each code is labelled with ``yes" if it appeared in our qualitative analysis for that workshop.}
\label{tab:all_codes}

\centering
\footnotesize
\begin{tabular}{lrrrrrrr}\toprule
\textbf{Code} 
    &\textbf{Formative Workshops}
    &\textbf{Workshop 1} &\textbf{Workshop 2} &\textbf{Workshop 3} &\textbf{Workshop 4} &\textbf{Workshop 5} \\\midrule
behaviour.Interpretation & \checkmark &\checkmark & & &\checkmark &\checkmark \\
behaviour.Lights &\checkmark &\checkmark &\checkmark &\checkmark &\checkmark &\checkmark \\
behaviour.Mimic & &\checkmark & &\checkmark & &\checkmark \\
behaviour.Movement &\checkmark &\checkmark &\checkmark &\checkmark &\checkmark &\checkmark \\
behaviour.PositionInEnvironment &\checkmark & &\checkmark & &\checkmark &\checkmark \\
behaviour.Sound &\checkmark &\checkmark &\checkmark &\checkmark &\checkmark &\checkmark \\
behaviour.Speed &\checkmark & &\checkmark & & &\checkmark \\
behaviour.Swarm &\checkmark &\checkmark &\checkmark & & &\checkmark \\
behaviour.Touch &\checkmark & & & &\checkmark &\checkmark \\
behaviour.TurnOff & & & & &\checkmark & \\
behaviour.Vibration &\checkmark & & & & & \\
Buttons.Additions & & &\checkmark & & &\checkmark \\
Buttons.Benefits & & &\checkmark & &\checkmark &\checkmark \\
Buttons.Challenges & & &\checkmark & & &\checkmark \\
Communication.Benefits & & & &\checkmark &\checkmark &\checkmark \\
Communication.Content &\checkmark &\checkmark &\checkmark &\checkmark & &\checkmark \\
Communication.MessageComplexity &\checkmark &\checkmark &\checkmark &\checkmark &\checkmark &\checkmark \\
Communication.NonESSbotCommunication & &\checkmark &\checkmark &\checkmark &\checkmark &\checkmark \\
Communication.Quality &\checkmark & & &\checkmark &\checkmark &\checkmark \\
Communication.Responses & &\checkmark &\checkmark &\checkmark & & \\
Communication.Situational &\checkmark & &\checkmark &\checkmark &\checkmark &\checkmark \\
Communication.Touch &\checkmark & & &\checkmark &\checkmark &\checkmark \\
Communication.Challenges &\checkmark & & &\checkmark &\checkmark &\checkmark \\
Controller.Additions & & &\checkmark & & & \\
Controller.Benefits & & &\checkmark & & & \\
Controller.Challenges & & &\checkmark & & & \\
Customization &\checkmark &\checkmark & &\checkmark &\checkmark &\checkmark \\
Features.Additions &\checkmark &\checkmark &\checkmark &\checkmark &\checkmark &\checkmark \\
Features.RobotQualities &\checkmark & & & & &\checkmark \\
Features.SystemQualities &\checkmark &\checkmark & &\checkmark &\checkmark &\checkmark \\
Interaction.Benefits & &\checkmark & & &\checkmark &\checkmark \\
Interaction.Challenges & &\checkmark & &\checkmark &\checkmark & \\
Interaction.Consent & & & &\checkmark &\checkmark &\checkmark \\
Interaction.ControlLevel &\checkmark &\checkmark &\checkmark &\checkmark &\checkmark & \\
Interaction.Device & &\checkmark &\checkmark &\checkmark &\checkmark &\checkmark \\
Interaction.Generalizability & & & & & &\checkmark \\
Interaction.Motivation & & & & &\checkmark & \\
Interaction.Situational & & &\checkmark &\checkmark &\checkmark &\checkmark \\
Interface.Additions & & &\checkmark &\checkmark &\checkmark &\checkmark \\
Interface.Benefits & & & & &\checkmark &\checkmark \\
Interface.Challenges & & & & &\checkmark &\checkmark \\
ParticipantQuestion &\checkmark & & & & & \\
Perception.Ambivalent &\checkmark & &\checkmark &\checkmark &\checkmark &\checkmark \\
Perception.ConceptualConstruct &\checkmark &\checkmark &\checkmark &\checkmark &\checkmark &\checkmark \\
Perception.Negative & &\checkmark &\checkmark &\checkmark &\checkmark & \\
Perception.Positive &\checkmark & &\checkmark &\checkmark &\checkmark &\checkmark \\
SocialTechIntegration & &\checkmark & & &\checkmark &\checkmark \\
TouchControl.Additions & & &\checkmark & &\checkmark & \\
TouchControl.Benefits & & &\checkmark & & &\checkmark \\
TouchControl.Challenges & & & & & &\checkmark \\
Usage.GamesAndFun &\checkmark &\checkmark &\checkmark &\checkmark &\checkmark &\checkmark \\
Usage.GroupDynamics & & &\checkmark &\checkmark &\checkmark &\checkmark \\
Usage.PersonalUse &\checkmark &\checkmark & &\checkmark & & \\
Usage.SocialConnection & &\checkmark & &\checkmark &\checkmark & \\
Usage.UserGroup & &\checkmark &\checkmark & & & \\
VisualScripting.Additions & & &\checkmark & & &\checkmark \\
VisualScripting.Benefits & & &\checkmark & & &\checkmark \\
VisualScripting.Challenges & & &\checkmark & & & \\
\bottomrule
\vspace{0.2em}
\end{tabular}

\end{table}

\end{document}